\documentclass[twocolumn,aps,prl,groupedaddress,amsmath,amssymb]{revtex4-1}
\usepackage{url}
\usepackage[colorlinks=true, linkcolor=blue,urlcolor=blue,anchorcolor=blue,citecolor=blue,bookmarksnumbered]{hyperref}
\usepackage{graphicx}
\usepackage{dcolumn}
\usepackage{bm}
\usepackage{verbatim}
\usepackage{algorithmicx}
\usepackage{mathrsfs}
\usepackage{amsmath} 
\usepackage{color,xcolor}
\usepackage{booktabs}

\begin{document}
	
\title{Quantum parametric amplification of phonon-mediated magnon-spin interaction}
\author{Yan Wang$^{1,2}$}\author{Hui-Lai Zhang$^{1,2}$}\author{Jin-Lei Wu$^{3}$}\author{Jie Song$^{4}$}\author{Kun Yang$^{1,2}$}\author{Wei Qin$^{5}$}\author{Hui Jing$^{2,6\ast}$}\author{Le-Man Kuang$^{2,6\ast}$}
\affiliation{$^{1}$College of Physics and Electronic Engineering, Zhengzhou University of Light Industry, Zhengzhou 450001, China\\
$^{2}$ Academy for Quantum Science and Technology, Zhengzhou University of Light Industry, Zhengzhou 450001, China\\
$^{3}$ School of Physics and Microelectronics, Zhengzhou University, Zhengzhou 450001, China\\
$^{4}$ School of Physics, Harbin Institute of Technology, Harbin, 150001, China\\
$^{5}$ Theoretical Quantum Physics Laboratory, RIKEN Cluster for Pioneering Research, Wako-shi, Saitama 351-0198, Japan\\
$^{6}$ Key Laboratory of Low-Dimensional Quantum Structures and Quantum Control of Ministry of Education, Department of Physics and Synergetic Innovation Center for Quantum Effects and Applications, Hunan Normal University, Changsha 410081, China\\
$^{\ast}$Corresponding authors (Hui~Jing,~email:~jinghui73@foxmail.com;\\
Le-Man~Kuang,~email:~lmkuang@hunnu.edu.cn)}
	
	\begin{abstract}
		\noindent The recently developed hybrid magnonics provides new opportunities for advances in both the study of magnetism and the development of quantum information processing. However, engineering coherent quantum state transfer between magnons and specific information carriers, in particular, mechanical oscillators and solid-state spins, remains challenging due to the intrinsically weak interactions and the frequency mismatch between different components. Here, we show how to strongly couple the magnon modes in a nanomagnet to the quantized mechanical motion (phonons) of a micromechanical cantilever in a hybrid tripartite system. The coherent and enhanced magnon-phonon coupling is engineered by introducing the quantum parametric amplification of the mechanical motion. With experimentally feasible parameters, we show that the mechanical parametric drive can be adjusted to drive the system into the strong-coupling regime and even the ultrastrong-coupling regime. Furthermore, we show the coherent state transfer between the nanomagnet and a nitrogen-vacancy center in the dispersive-coupling regime, with the magnon-spin interaction mediated by the virtually-excited squeezed phonons. The amplified mechanical noise can hardly interrupt the coherent dynamics of the system even for low mechanical quality factors, which removes the requirement of applying additional engineered-reservoir techniques. Our work opens up prospects for developing novel quantum transducers, quantum memories and high-precision measurements.\\
		\\
		\noindent{\bf mechanical parametric amplification, strong-coupling regime, hybrid quantum system\\
			\\
			PACS number(s):}~~~{42.50.Lc, 75.30.Ds, 63.20.Ls}
	\end{abstract}
	\maketitle
	\noindent \textbf{\begin{large}1~~~{Introduction}\end{large}}\\
	\\
The studies on magnons, i.e., collective spin excitations of ordered magnets, have recently made remarkable progress in exploring fundamental quantum
physics and realizing active quantum components for applications in quantum science and technology \cite{YUAN20221,ZARERAMESHTI20221,Barman2021,doi10106350020277,doi10106350021099,doi10106350046202,LachanceQuirion2019}. In this context, the generation, manipulation and detection of magnons, i.e., the so called magnonics, attracts increasing research interests for the perspective of interfacing magnons with other quantum excitations. Examples include microwave and optical photons \cite{PhysRevLett.111.127003,PhysRevLett.113.083603,PhysRevLett.113.156401,PhysRevLett.116.223601,PhysRevLett.117.123605,PhysRevLett.123.107701,PhysRevLett.123.107702,PhysRevLett.125.237201,PhysRevLett.124.053602}, superconducting qubits \cite{tabuchi2015coherent,doi:10.1126/sciadv.1603150,doi:10.1126/science.aaz9236}, and acoustic phonons due to magnetostrictive interactions \cite{doi:10.1126/sciadv.1501286,PhysRevLett.121.203601,PhysRevLett.124.213604,PhysRevLett.129.243601}. The versatility of magnons opens new opportunities to further engineer coherent interactions between magnons and other physical systems with drastically different properties. 

Hybrid quantum systems (HQSs) combine different physical components with complementary functionalities, thus providing multitasking capabilities that individual components cannot offer \cite{Kurizki3866}. In particular, benefiting from the unprecedented tunability and compatibility, HQSs based on magnonics, or say hybrid magnonics, promote the development of novel quantum technologies, such as microwave-to-optical quantum transducers \cite{PhysRevB.93.174427,Zhu:20,Chai:22} and quantum detection and sensing \cite{PhysRevB.99.214415,PhysRevLett.125.147201,PhysRevLett.125.117701,doi:10.1063/5.0024369,PhysRevA.103.063708}. However, a current challenge of functionalizing hybrid magnonics is how to coherently couple the magnons to quantum systems that either weakly couple to magnons or have large energy difference from magnon modes. Specifically, the former is mainly caused by the inadequate spatial matching, thus making the coupling strengths lower than intrinsic losses and decoherence of both systems. For the latter, the coherent process of quantum state transfer can hardly take place if the energy cannot be conserved, even in the presence of strong interactions. A typical case is to coherently couple magnons to external mechanical oscillators (corresponding to mechanical phonons aside from acoustic phonons due to the deformation of the magnetic system itself). This is generally difficult to achieve due to their dissimilar excitation energies or frequency mismatch \cite{doi:10.1126/sciadv.1501286,PhysRevLett.124.093602,PhysRevB.101.125404}. This cannot enable the coherent information transfer between magnons and mechanical phonons, and thus may limit the development of practical applications based on hybrid magnon-phonon systems \cite{holanda2018detecting,doi:10.1063/10.0000872}, such as high-precision metrology of small magnetic fields and mechanical displacements. In addition, recent studies have found that magnons can strongly interact with solid-state spins such as nitrogen-vacancy centers and silicon-vacancy centers in diamond \cite{PhysRevLett.125.247702,doi:10.1021/acs.jpcc.0c11536,PhysRevApplied.16.064008,PRXQuantum.2.040314,PhysRevB.105.245310}. However, reaching the single magnon-spin strong-coupling regime (SCR) often requires the spins to be positioned very close to the magnet surface, which inevitably causes difficulties of manipulating single spins. 

In this work, we propose an experimentally feasible method for strongly coupling magnons to mechanical phonons and to single solid-state spin in a hybrid tripartite system. Our proposal is based on a hybrid device where a micromechanical cantilever, pumped by a detuned parametric drive, interacts simultaneously with a nanomagnet and a single nitrogen-vacancy (NV) center via magnetic field gradients. We show that the magnon mode can be coupled, in a coherent and tunable way, to the phonon mode by adjusting the mechanical parametric drive (MPD). With experimentally accessible parameters, the magnon-mechanical coupling can be exponentially enhanced, thus driving the initially weak-coupling system into the SCR and even the ultrastrong-coupling regime (UCR). The coupling enhancement can facilitate the extension of coherence time of the quantum system where the mechanical mode acts as a long-lived quantum memory. We also extend our protocol by coupling an NV spin to the mechanical phonon. In the dispersive-coupling regime, we realize coherent quantum state transfer between the nanomagnet and the NV center. Remarkably, the mechanical noise amplified by the MPD can hardly corrupt the coherent and strong-coupling dynamics of the proposed system without using additional engineered-reservoir techniques that are generally required for suppressing the amplified noise \cite{PhysRevLett.114.093602,PhysRevA.100.062501,PhysRevLett.120.093601,PhysRevLett.120.093602,PhysRevLett.126.023602,PhysRevLett.128.083604,PhysRevLett.125.153602,PhysRevApplied.17.024009}, which could simplify the experimental requirement and implementation.

Related forms of using the quantum parametric amplification (QPA) to significantly enhance and controllably manipulate the interaction between quantum
objects have been studied in various coupled systems ranging from optomechanics \cite{PhysRevLett.114.093602,lemonde2016enhanced,PhysRevA.100.062501,zhao2020weak,liu2023phase}, cavity or circuit QED \cite{PhysRevLett.120.093601,PhysRevLett.120.093602,PhysRevA.99.023833,PhysRevLett.124.073602,PhysRevLett.125.203601,PhysRevA.101.053826,PhysRevA.102.032601,PhysRevLett.127.093602,PhysRevLett.126.023602,PhysRevLett.128.083604,villiers2023dynamically}, trapped ions \cite{PhysRevLett.122.030501,burd2021quantum,PhysRevA.107.032425} and hybrid spin-mechanical systems \cite{PhysRevLett.125.153602,PhysRevApplied.17.024009,PhysRevA.107.023722}. Our work further extends and determines the capabilities of exploiting the QPA to enhance the magnon-phonon and the phonon-mediated magnon-spin interactions in a hybrid bipartite system. Notice that a recent study has shown that the QPA can be used to enhance the direct tripartite coupling among spins, magnons and phonons \cite{PhysRevLett.130.073602}, where markedly different physical model and coupling mechanism were considered in a different hybrid setup. Our findings in achieving the effective enhancement of the bipartite interaction between the magnon and an external phonon provide a versatile tool for various intriguing applications, such as the improvement of magnon-assisted ground-state cooling of the mechanical motion \cite{PhysRevLett.124.093602,ASJAD20233}, and the preparation of highly entangled cat states between the magnon and the solid spin in the weak-coupling regime \cite{PhysRevB.103.L100403,PhysRevLett.127.087203}. As for the enhancement of the bipartite interaction between the magnon and the NV spin, our strategy strengthens the ability for processing complicated quantum information tasks where both the NV spin and the magnons can act as collaborate quantum memories. Our scheme may also find inspiring applications to design quantum transducers, quantum memories and high-precision measurements. \\
	\\
	\noindent \textbf{\begin{large}2~~~{Quantum dynamics of hybrid magnon-phonon-spin system}\end{large}}\\\\
\begin{figure*}[htbp]
	\centering
	\includegraphics[width=2\columnwidth]{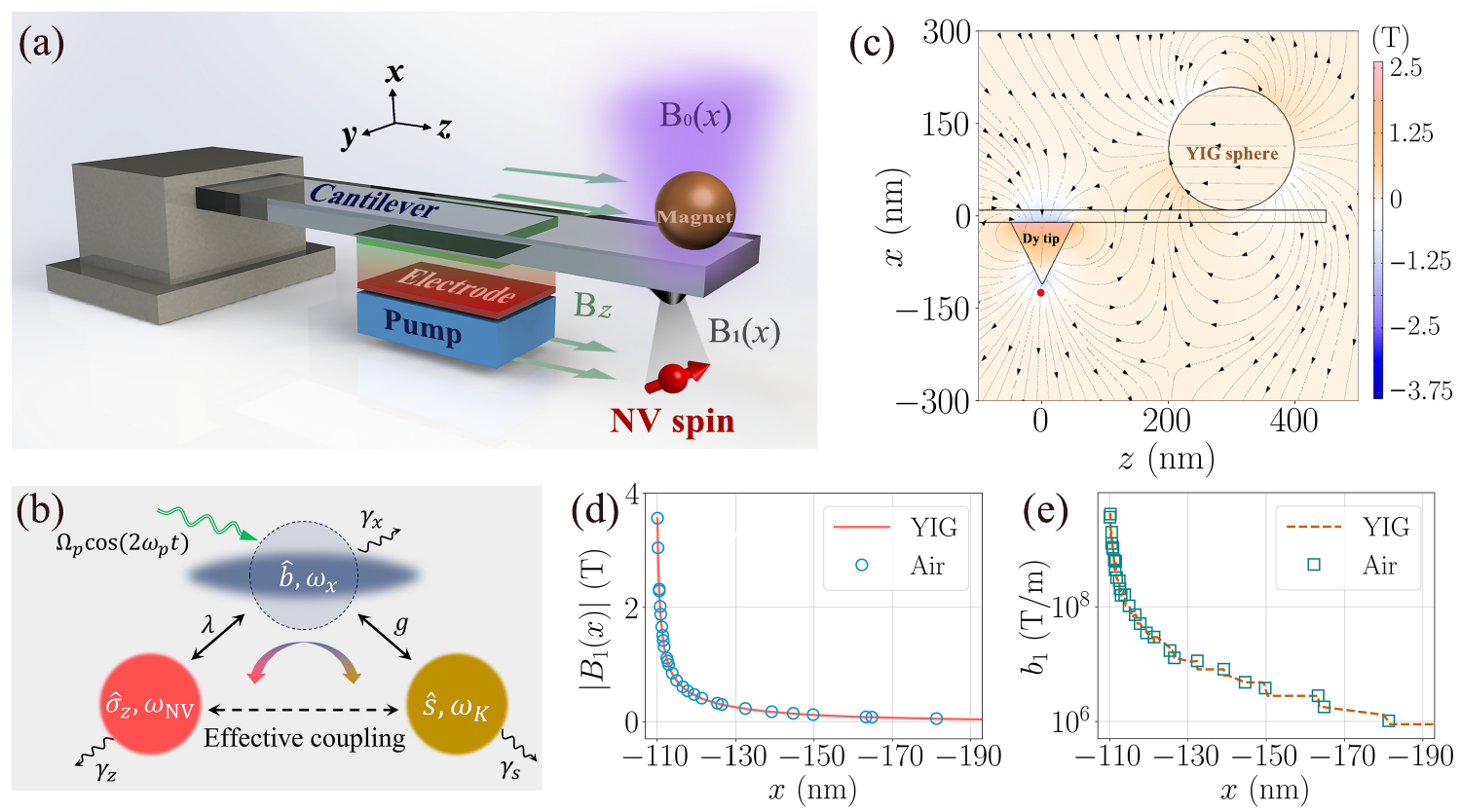}
	\caption{(a) Schematic illustration of the proposed hybrid quantum system. A micromechanical cantilever holding a spherical nanomagnet is parametrically driven by a pump source that electrically modulates its spring constant. The nanomagnet interacts with an external magnetic field, which has a homogeneous component $\textbf{B}_{z}$ and a linear gradient $\textbf{B}_{0}(x)$. A magnet tip attached underneath to the cantilever produces a strong magnetic gradient $\textbf{B}_{1}(x)$ that couples the motion of cantilever to a lower single NV center via the Zeeman effect. Dimensions of each components are not to scale for visual clarity. (b) Schematic of the physical model. The effective coupling between the spin qubit of frequency $\omega_{\mathrm{NV}}$ and the magnon mode of frequency $\omega_{K}$ is mediated by the mechanical mode of frequency $\omega_{x}$, on which a two-phonon (parametric) drive is applied to amplify its zero-point fluctuations. This results in an exponential enhancement of both the spin-mechanical coupling $\lambda$ and the magnon-mechanical coupling $g$, thereby amplifying the resulting effective magnon-spin interaction. (c) Finite-element simulations of the magnetic field lines and the strength distribution generated by a Dy tip of 100 nm underside diameter and height, and a YIG nanosphere of 100 nm radius, which are separated by a distance of 300 nm along the $z$ direction. A diamond cantilever with dimensions of $(\ell,w,t)=(4,0.1,0.02)$ $\mu$m is situated in the middle. Only the $x$-$z$ plane of the spacial magnetic distributions is illustrated here. Absolute values of the magnetic strength $|B_{1}(x)|$ (d) and calculated magnetic gradient $b_{1}$ (e) along the $x$ direction for $z=0$. Solid curves and hollow markers are results in the presence and absence of the YIG sphere. A large magnetic gradient $b_{1}\approx4.5\times10^7$ T/m is obtained at the position of the NV spin (marked with a red dot in (c)).}
	\label{fig1}
\end{figure*}
We consider a hybrid tripartite system, as schematically illustrated in Fig.~\ref{fig1}(a), where a spherical nanomagnet of radius $R$ is attached on top of a singly-clamped micromechanical cantilever of dimensions ($\ell,w,t$) \cite{vinante2011magnetic,burgess2013quantitative,PhysRevLett.111.207203,PhysRevB.92.235441}, and a single NV center is placed under a magnetic tip fixed at the bottom of the cantilever  \cite{kolkowitz2012coherent}. The nanomagnet interacts with an external magnetic field, which has a homogeneous component $\textbf{B}_{z}=B_{z}\textbf{e}_{z}$, and a linear gradient $\textbf{B}_{0}(x)=-b_{0}x\textbf{e}_{x}$, with $b_{0}$ the magnetic field gradient. The homogeneous field $\textbf{B}_{z}$ is introduced to align the magnetization of the nanomagnet to the $z$ direction, while the linear gradient $\textbf{B}_{0}(x)$ is used to couple the nanomagnet to the mechanical motion of the cantilever (see below). The motion of the magnetic tip (assumed to be only along the $x$ direction) produces a strong magnetic gradient $\textbf{B}_{1}(x)=-b_{1}x\textbf{e}_{x}$, which causes Zeeman shifts of the NV center, giving rise to the magnetic coupling between the NV spin and the mechanical motion of the cantilever (see below).
Note here that the magnetic gradient $\textbf{B}_{0}(x)$ has no effect on the spin-mechanical coupling because it does not change the local magnetic field sensed by the motionless NV spin. 
Additionally, we assume that the cantilever is electrically pumped by a periodic drive that modulates the mechanical spring constant in time {\cite{PhysRevLett.67.699,PhysRevLett.107.213603}, thereby amplifying the zero-point fluctuations of the mechanical motion. To be specific, a time-varying voltage originating from a tunable oscillating pump under the cantilever is applied to form a general capacitor between the two electrodes, one of which is coated on the lower surface of the cantilever and the other is placed on top of the pump, as illustrated in Fig.~\ref{fig1}(a).
	\\\\
    \noindent \textbf{2.1~~~{Free Hamiltonians for phonon, spin and magnon}}\\
    \\
    We begin with the Hamiltonian of the mechanical cantilever. By modeling the mechanical motion of the cantilever as a harmonic oscillator \cite{PhysRevB.79.041302,rabl2010quantum} and following the standard quantization procedure \cite{lemonde2016enhanced,PhysRevLett.125.153602,PhysRevApplied.17.024009}, the Hamiltonian of the cantilever oscillator is obtained as
    \begin{equation}\label{Hmec}
    \frac{\hat{H}_{\mathrm{mec}}}{\hbar}=\omega_{x} \hat{b}^{\dagger} \hat{b}-\Omega_{p}\cos(2\omega_{p}t)(\hat{b}^{2}+\hat{b}^{\dagger2}),
    \end{equation} 
    where $\hat{b}$ ($\hat{b}^{\dagger}$) is the annihilation (creation) operator of the motional quanta, i.e., phonon, with frequency $\omega_{x}$, and $\omega_{p}$, $\Omega_{p}$ are the pump frequency and amplitude, respectively. Equation (\ref{Hmec}) shows that a two-phonon (i.e., parametric) drive of the mechanical mode can be achieved by applying the time-dependent pump to the mechanical oscillator. This detuned parametric drive, as detailed below, modifies the eigenstates of the mechanical Hamiltonian to be squeezed Fock states, whose amplified fluctuations in the anti-squeezed quadrature yield larger interactions with both the magnon and the NV spin [see Fig.~\ref{fig1}(b)].
    In addition, working with the typical mechanical zero-point fluctuations of hundreds of femtometers, the proper distances between the two electrode plates, e.g., tens of nanometers, and the accessible parameters with respect to the electric drive module, the nonlinear amplitude $\Omega_{p}$ of up to GHz could be realized \cite{PhysRevLett.125.153602}.

    For the single NV center, its electronic ground state is a $S=1$ spin triplet with basis states $\left| m_s\right\rangle$, where $m_s=0,\pm1$. 
    The zero-field splitting between the degenerate sublevels $\left| m_s=\pm1\right\rangle$ and $\left| m_s=0\right\rangle$ is $D = 2\pi \times 2.87$ GHz \cite{childress2006coherent,kolkowitz2012coherent,taylor2008high}. We assume that the NV symmetry axis is along the $z$ direction. The magnetic field $\textbf{B}_{z}$ lifts the degeneracy of the states $|m_s=\pm1\rangle$, causing a Zeeman splitting $\delta=2 g_{e} \mu_{B} B_{z}/\hbar$, where $g_{e} \simeq 2$ is the Land$\acute{\mathrm{e}}$ factor and $\mu_{B}$ the Bohr magneton. Taking the state $|0\rangle$ to be the energy reference, the free
    Hamiltonian of the NV center has the form
    \begin{equation}\label{H_NV}
    \frac{\hat{H}_{\mathrm{nv}}}{\hbar}=\sum_{j=\pm 1} \delta_j|j\rangle\langle j|,
    \end{equation}
    where $\delta_j\equiv D\pm\delta/2$. By choosing a proper magnetic strength $B_{z}$ such that $\left|\delta_{+1}-\omega_{x}\right| \gg\left|\delta_{-1}-\omega_{x}\right|$, the contribution $\propto$ $\left(\left|+1\rangle\langle 0\right| +\left|0\rangle\langle +1\right| \right) (\hat{b}^{\dagger}+\hat{b}) $ oscillates rapidly and thus can be excluded from the dynamics. The Hamiltonian (\ref{H_NV}) can then be rewritten as
    \begin{equation}\label{H_NV1}
    \frac{\hat{H}_{\mathrm{NV}}}{\hbar}=\frac{1}{2}\omega_{\mathrm{NV}}\hat{\sigma}_{z},
    \end{equation}
    where $\omega_{\mathrm{NV}}\equiv\delta_{-1}$ and $\hat{\sigma}_{z}=|-1\rangle\langle-1|-|0\rangle\langle0|$. 
    
    Next, we turn our attention to the spherical nanomagnet that supports the magnetization-wave or spin-wave (magnon) modes \cite{ZARERAMESHTI20221}. Under the dipolar, isotropic, and magnetostatic approximations \cite{PhysRevB.101.125404}, one can obtain the  magnetization eigenmodes, known as the magnetostatic dipolar spin waves, or Walker modes for spherical magnets \cite{PhysRev.105.390,doi:10.1063/1.1723117,https://doi.org/10.1002/pssb.2220820102}. The eigenfrequencies of the Walker modes, $\omega_{\beta}\equiv\omega_{lmn}$, depend on the external homogeneous field $B_{z}\equiv\mu_{0}H_{z}$, where $l=1,2,3,\dots$, $m\in[-l,l]$, $n=0,1,2,\dots,n_{\mathrm{max}}(l)$, and $\mu_{0}$ is the vacuum permeability. These Walker modes can be further quantized \cite{MILLS200616}, resulting in the free magnon Hamiltonian 
    \begin{equation}
    \frac{\hat{H}_{\mathrm{m}}}{\hbar}=\sum_\beta \omega_\beta \hat{s}_\beta^{\dagger} \hat{s}_\beta,
    \end{equation}
    with the bosonic ladder operators $[\hat{s}_\beta, \hat{s}_{\beta^{\prime}}^{\dagger}]=\delta_{\beta \beta^{\prime}}$. Correspondingly, the spin-wave magnetization operator in the Schr$\mathrm{\ddot{o}}$dinger picture is given by
    \begin{equation}
    \hat{\mathbf{m}}(\mathbf{r})=\sum_\beta \mathcal{M}_{0 \beta}[\tilde{\mathbf{m}}_\beta(\mathbf{r}) \hat{s}_\beta+\text {H.c.}],
    \end{equation}
    where $\tilde{\mathbf{m}}_\beta(\mathbf{r})$ is the mode function of the spin wave, and $\mathcal{M}_{0\beta}\equiv\sqrt{\hbar|\gamma| M_S/\tilde{\Lambda}_\beta}$ is the zero-point magnetization, with $|\gamma|=g_{e} \mu_{B}/\hbar$ the absolute value of the gyromagnetic ratio, $M_S$ the saturation magnetization and $\tilde{\Lambda}_\beta \equiv 2 \operatorname{Im} \int d V \tilde{m}_x^*(\mathbf{r}) \tilde{m}_y(\mathbf{r})$ the normalization constant. 
    
    In the present work, we focus only on the simplest Walker mode, i.e., the spatially uniform Kittel mode \cite{PhysRev.73.155}, $\beta\equiv\{110\} \equiv K$ . In this case, the free Hamiltonian of the Kittel magnon mode is given by
    \begin{equation}
    \hat{H}_{\mathrm{mag}}/\hbar=\omega_{K} \hat{s}^{\dagger} \hat{s},
    \end{equation}
    where $\omega_{K}=|\gamma|B_{z}$ is the eigenfrequency of the Kittel mode, and the mode index $K$ has been omitted for simplicity. Considering that the magnetized mode function of the Kittel mode takes a homogeneous form, $\tilde{\mathbf{m}}_{\mathrm{K}}(\mathbf{r})=\mathbf{e}_x+i \mathbf{e}_y$, the zero-point magnetization becomes $\mathcal{M}_{K}=\sqrt{\hbar|\gamma| M_S/2V}$, with $V$ the nanomagnet volume. We remark that there is no $z$ component in the magnetized mode function. The corresponding magnetization operator for the Kittel mode can then be obtained as
    \begin{equation}\label{mr}
    \hat{\mathbf{m}}_{K}(\mathbf{r})=\mathcal{M}_{K}[(\mathbf{e}_x+i \mathbf{e}_y)\hat{s}+\text {H.c.}].
    \end{equation}	
	\\
	\noindent \textbf{2.2~~~{Interaction between magnon and phonon or spin and phonon}}\\\\
	In order to couple the magnon mode of the nanomagnet to the mechanical motion of the cantilever, we apply a spatially inhomogeneous magnetic field $\textbf{B}_{0}(x)$. For a general driving magnetic field $\mathbf{H}_d\left(\mathbf{r},t\right)$, it can increase the total energy of the nanomagnet by \cite{PhysRevB.101.125404} 
	\begin{equation}\label{Delta_E(t)}
	\Delta E(t)=-\frac{\mu_0}{2}\int \text{d}V\left[M_{S}\textbf{e}_{z}+\mathbf{m}(\mathbf{r}, t)\right] \cdot \mathbf{H}_d\left(\mathbf{r}, t\right).
	\end{equation}
	Here, the first and second terms describe the interactions of the driving field with the static magnetization $M_{S}\textbf{e}_{z}$ and the magnetization of the spin waves $\mathbf{m}(\mathbf{r}, t)$, respectively. To obtain the magnon-mechanical interaction Hamiltonian of our system in the Schr$\mathrm{\ddot{o}}$dinger picture, we substitute the dynamical variable $\mathbf{m}(\mathbf{r}, t)$ in Eq.~(\ref{Delta_E(t)}) by its quantum operator $\hat{\mathbf{m}}_{K}(\mathbf{r})$ given in Eq.~(\ref{mr}), the variable $\mathbf{H}_d\left(\mathbf{r}, t\right)$ by $\mathbf{H}_{0}(x)\equiv\mathbf{B}_{0}(x)/\mu_{0}$, and the variation in magnetic energy $\Delta E(t)$ by operator $\hat{V}$. The resulting interaction Hamiltonian is given by
	\begin{equation}
	\hat{V}=\frac{b_0V\mathcal{M}_{K}}{2}(\hat{s}^\dagger+\hat{s})\hat{x},
	\end{equation}
	which describes the coupling between the Kittel magnon and the mechanical motion along the $x$ direction. By further rewriting the position operator of the mechanical oscillator as $\hat{x}=x_{\mathrm{zpf}}(\hat{b}^{\dagger}+\hat{b})$, where $x_{\mathrm{zpf}}=\sqrt{\hbar / 2 M_{\mathrm{eff}} \omega_{x}}$ is the zero-point fluctuation amplitude, with $M_{\mathrm{eff}}=M+M_{m}+M_{t}$ covering the effective masses of the cantilever ($M$), the nanomagnet ($M_{m}$) and the magnetic tip ($M_{t}$), the magnon-mechanical interaction Hamiltonian becomes
	\begin{equation}
	\hat{H}_{\mathrm{m}\text{-}\mathrm{m}}/\hbar=g(\hat{s}^\dagger+\hat{s})(\hat{b}^{\dagger}+\hat{b}),
	\end{equation}
	where $g\equiv b_0x_{\mathrm{zpf}}\mathcal{M}_{K}V/2\hbar$ is the magnon-mechanical coupling rate.
	
	For the proposed hybrid system in Fig.~\ref{fig1}(a), we desire to provide a mechanism where the spin-phonon coupling is determined only by the magnetic gradient $\textbf{B}_{1}(x)$ originating from the vibration of magnetic tip, while the inhomogeneous external field $\textbf{B}_{0}(x)$ is responsible only for the magnon-phonon coupling. To this end, we first select a magnetic tip with a size of $\sim$100 nm, and the magnetic substance is selected as Dysprosium (Dy) due to a high saturation magnetization of up to 3.7 T \cite{doi:10.1063/1.3673910}, which is larger than that of yttrium iron garnet (YIG) with $\mu_{0}M_{S}\approx0.74$ T. This allows us to obtain a strong magnetic gradient exceeding $10^{7}$ T/m at short distances for the NV center, as demonstrated in the simulation below. Moreover, we consider a specific arrangement of magnets such that the magnetic tip is positioned, along the $z$ axis, at a distance of hundreds of nanometers away from the nanomagnet. This can avoid the interplay of the magnetic fields generated by the magnetic tip and the nanomagnet due to an adequate spatial separation. To verify this, we numerically simulate the relevant magnetic fields and compute the value of magnetic gradient using the finite-element analysis method, and the results are shown in Figs.~\ref{fig1}(c)-\ref{fig1}(e). Figure \ref{fig1}(c) is an example of the simulated magnetic field lines and strength distribution for a Dy tip with underside diameter and height of 100 nm, and a YIG sphere with radius of 100 nm, which are separated by 300 nm in the direction of $z$-axis. One can see in this configuration that the field distribution around the Dy tip can hardly be affected by the YIG sphere and vice versa. Moreover, Figs.~\ref{fig1}(d) and \ref{fig1}(e) show values of the magnetic strength $|B_{1}(x)|$ and the gradient $b_{1}$ along the $x$ direction for $z=0$. Here, the effect of the YIG sphere on the Dy tip is evaluated by comparing the values of $B_{1}(x)$ and $b_{1}$ after removing the YIG sphere (labeled with ``Air'' in the plots). We find that both the magnetic strength and the magnetic gradient remain unchanged in the presence or absence of the YIG sphere, revealing the negligible influence of the YIG sphere on the magnetic tip in the present configuration.

	The oscillations of the mechanical cantilever as well as the Dy tip produce a strong magnetic gradient that causes the Zeeman shifts of the NV center \cite{PhysRevLett.117.015502,PhysRevLett.121.123604,Zhou:22}, thereby coupling the spin and mechanical degrees of freedom. The magnetic interaction Hamiltonian can be written as
	\begin{equation}
	\hat{H}_{\mathrm{int}}=\frac{\mu_{B} g_{e}}{\hbar} \hat{\textbf{S}}\cdot \textbf{B}(\textbf{r}_0),
	\end{equation}
	where $\hat{\textbf{S}}$ denotes the spin operator and $\textbf{r}_0$ the position of the NV center. In our setup, the bending motion of the cantilever (along the $x$ direction) modulates the local magnetic field sensed by the NV center, giving rise to the spin-mechanical coupling with the interaction Hamiltonian
	\begin{equation}
	\hat{H}_{\mathrm{s}\text{-}\mathrm{m}}=-\frac{\mu_{B} g_{e} b_{1}}{\hbar}\hat{S}_{x}\hat{x}.
	\end{equation}
	After using again $\hat{x}=x_{\mathrm{zpf}}(\hat{b}^{\dagger}+\hat{b})$, this interaction Hamiltonian becomes
	\begin{equation}
	\hat{H}_{\mathrm{s}\text{-}\mathrm{m}}/\hbar=\lambda(\hat{\sigma}_{+}+\hat{\sigma}_{-})(\hat{b}^{\dagger}+\hat{b}),
	\end{equation}
	where $\lambda\equiv-\mu_{B} g_{e} b_{1}x_{\mathrm{zpf}}/\hbar$ denotes the spin-mechanical coupling rate and the spin operators $\hat{\sigma}_{+}=\hat{\sigma}_{-}^{\dagger}=|-1\rangle\langle0|$. This spin-mechanical coupling depends on the magnetic gradient $b_{1}$, which can be adjusted by positioning the NV spin at the suitable distances from the Dy tip. For instance, a large gradient $b_{1}\approx4.5\times10^7$ T/m can be obtained for a distance of about 10 nm between the Dy tip and the NV spin, as shown in Fig.~\ref{fig1}(e).
	
	To sum up, we obtain the total Hamiltonian for the proposed hybrid system
	\begin{equation} \label{H_hybrid}
	\begin{split}
	\frac{\hat{H}_{\mathrm{hybrid}}}{\hbar}=&\hat{H}_{\mathrm{mec}}+\hat{H}_{\mathrm{NV}}+\hat{H}_{\mathrm{mag}}+\hat{H}_{\mathrm{m}\text{-}\mathrm{m}}+\hat{H}_{\mathrm{s}\text{-}\mathrm{m}}\\
	=&\omega_{x} \hat{b}^{\dagger} \hat{b}+\frac{1}{2}\omega_{\mathrm{NV}}\hat{\sigma}_{z}+\omega_{K} \hat{s}^{\dagger} \hat{s}
	\\
	&+g(\hat{s}^\dagger+\hat{s})(\hat{b}^{\dagger}+\hat{b})+\lambda(\hat{\sigma}_{+}+\hat{\sigma}_{-})(\hat{b}^{\dagger}+\hat{b})\\
	&-\Omega_{p}\cos(2\omega_{p}t)(\hat{b}^{2}+\hat{b}^{\dagger2}).
	\end{split}
	\end{equation}
	\\
	\noindent \textbf{\begin{large}3~~~{Giant enhancement of phonon-mediated magnon-spin interaction}\end{large}}\\
	\\
	In this section, we show how to engineer coherent and enhanced magnon-mechanical interaction and the consequent magnon-spin interaction mediated by a squeezed phonon. Our approach is based on the simultaneous enhancement of the magnon-mechanical and spin-mechanical couplings with the squeezed phonons being virtually excited under some proper parameters. Since the direct enhancement of the spin-mechanical coupling via the parametric amplification of the mechanical motion has been discussed in detail elsewhere \cite{PhysRevLett.125.153602,PhysRevApplied.17.024009}, in the following we first show our new results of extending the previous scheme  \cite{PhysRevLett.125.153602,PhysRevApplied.17.024009} to involve the magnonic degree of freedom, and on this basis we derive the enhanced magnon-spin coupling in the dispersive regime.
	\\\\
	\noindent \textbf{3.1~~~{Exponentially enhanced magnon-phonon coupling in the squeezed frame}}
	\\\\
	We first consider the coupling between the nanomagnet and the mechanical oscillator by excluding the spin degree of freedom. In this case, the magnon-mechanical Hamiltonian, in the frame rotating at $\omega_p$ and after dropping the high-frequency oscillation terms, is given by
	\begin{equation}\label{H_mm}
	\begin{split}
	\frac{\hat{H}_{\mathrm{mm}}}{\hbar}=&\Delta_{x} \hat{b}^{\dagger} \hat{b}+\Delta_{K} \hat{s}^{\dagger} \hat{s}-\frac{\Omega_{p}}{2}(\hat{b}^{2}+\hat{b}^{\dagger2})
	\\
	&+g(\hat{b}\hat{s}^\dagger+\hat{b}^\dagger\hat{s}^\dagger e^{\mathrm{i}2\omega_pt}+\mathrm{H.c.}),
	\end{split}
	\end{equation}
	where $\Delta_{x}=\omega_{x}-\omega_{p}$ and $\Delta_{K}=\omega_{K}-\omega_{p}$. To see the effect of the MPD clearly, it is convenient to transform the Hamiltonian (\ref{H_mm}) to the squeezed frame by performing a unitary squeezing transformation, $\hat{U}_{s}=\exp [r_p(\hat{b}^{2}-\hat{b}^{\dagger 2}) / 2]$, where the squeezing parameter $r_p$ is defined via $r_{p}=(1 / 2) \operatorname{arctanh}\left(\Omega_{p} / \Delta_{x}\right)$, yielding 
	\begin{equation}\label{H_mms}
	\begin{split}
	\frac{\hat{H}_{\mathrm{mm}}^{\mathrm{S}}}{\hbar}=&\Delta_{s} \hat{b}^{\dagger}_{s} \hat{b}_{s}+\Delta_{K} \hat{s}^{\dagger} \hat{s}+\big[g_{r}(\hat{b}_{s}\hat{s}^\dagger+\hat{b}_{s}^\dagger \hat{s}^\dagger e^{\mathrm{i}2\omega_pt})\\
	&-g_{c}(\hat{b}_{s}^\dagger\hat{s}^\dagger+\hat{b}_{s} \hat{s}^\dagger e^{\mathrm{i}2\omega_pt})+\mathrm{H.c.}\big],
	\end{split}
	\end{equation}
	where $\Delta_{s}=\Delta_{x}/\cosh(2r_p)$ is the squeezed-phonon frequency, $g_{r}=g\cosh(r_p)$ and $g_{c}=g\sinh(r_p)$ are the enhanced magnon-mechanical coupling strengths. Such an enhancement is a consequence of the MPD, which modifies the eigenstates of the mechanical Hamiltonian and generates the squeezed phonons with the amplified fluctuations responsible for the larger interaction with the magnon. In addition, Eq.~(\ref{H_mms}) implies that different quantum dynamical processes can be achieved under different parameter regimes. As an example, we discuss the magnon-mechanical interaction in the regime of $\Delta_{s}=\Delta_{K}$, where the magnon is in resonance with the mechanical phonon (see Appendix A for more detailed discussions). The unitary dynamics of such a system is governed by a Jaynes-Cummings-type Hamiltonian
	\begin{equation}\label{H_mmsr}
	\begin{split}
	\frac{\hat{H}_{\mathrm{mm,r}}^{\mathrm{S}}}{\hbar}\simeq&\Delta_{s} \hat{b}^{\dagger}_{s} \hat{b}_{s}+\Delta_{K} \hat{s}^{\dagger} \hat{s}+g_{r}(\hat{b}_{s}\hat{s}^\dagger+\mathrm{H.c.}),
	\end{split}
	\end{equation}
	which suggests a coherent exchange of the phonon excitation and the magnon, with the coupling being exponentially enhanced. The direct realization of such a coherent coupling between the magnon and the mechanical phonons could be versatile for various applications, e.g., magnon-assisted ground-state cooling of the mechanical motion \cite{PhysRevLett.124.093602}, and quantum control and detection of the magnon excitations by measuring the mechanical motion \cite{PRXQuantum.2.040344}.
	
	In addition to the coherent dynamics described by Eq.~(\ref{H_mmsr}), we take into account the couplings of the mechanical oscillator and the nanomagnet to the Markovian baths, which cause the mechanics to be damped at a rate $\gamma_{x}$ and the magnon at a rate $\gamma_{s}$. In the presence of the MPD, the noise coming from the mechanical bath is also amplified. In the weak mechanical dissipation limit $\gamma_{x}\ll\Delta_{s}$, the mechanical bath with a thermal occupation $\bar{n}_{x}$ corresponds to a bath for the $\hat{b}_{s}$ mode with a
	thermal occupation $\bar{n}_{x}^{s}=\bar{n}_{x}\cosh(2r_{p})+\sinh^{2}(r_{p})$ (see Appendix B for details). That is, the $\hat{b}_{s}$ mode can be considered as being coupled, with an unchanged vacuum damping rate $\gamma_{x}$, to a thermal bath but with thermal occupation $\bar{n}_{x}^{s}$. The system dynamics under dissipation can then be described by the Lindblad master equation
	\begin{equation} \label{mas equ}
	\dot{\hat{\rho}}=-\frac{\mathrm{i}}{\hbar}[\hat{H}_{\mathrm{mm}}^{\mathrm{S}},\hat{\rho}]+\mathcal{L}_{x}[\hat{\rho}]+\mathcal{L}_{s}[\hat{\rho}].
	\end{equation}
	Here, $\hat{H}_{\mathrm{mm}}^{\mathrm{S}}$ is given by Eq.~(\ref{H_mms}), $\hat{\rho}$ is the density operator, $\mathcal{L}_{x}[\hat{\rho}]=\gamma_x[\left(\bar{n}_{x}^{s}+1\right) L_{\hat{b}_s}\hat{\rho}+\bar{n}_{x}^{s} L_{\hat{b}_s^{\dagger}}\hat{\rho}]$ and  $\mathcal{L}_{s}[\hat{\rho}]=\gamma_s[\left(\bar{n}_s+1\right) L_{\hat{s}}\hat{\rho}+\bar{n}_s L_{\hat{s}^{\dagger}}\hat{\rho}]$, where $L_{\hat{o}} \hat{\rho}\equiv\hat{o} \hat{\rho} \hat{o}^{\dagger}-[\hat{o}^{\dagger}\hat{o} ,\hat{\rho}]/2$. Moreover, $\bar{n}_{j}=\left[\exp \left(\hbar \omega_{j} / k_{B} T_{j}\right)-1\right]^{-1}$ ($j=x,s$), with $k_{B}$ the Boltzmann constant and $T_{j}$ the temperature of the thermal bath of each degree of freedom. We assume that all modes couple with the thermal reservoirs at a common temperature $T$. It is also convenient to define the mechanical quality ($Q$) factor as
	$Q_{x}=\omega_x/\gamma_{x}$. The experimental values of $Q_{x}$ exceeding $10^8$ have been demonstrated recently in nanomechanical resonators \cite{PhysRevLett.116.147202,tsaturyan2017ultracoherent,ghadimi2018elastic}. Regarding the magnon linewidth, we consider $\gamma_{s}=2\pi\times1$ MHz for the YIG material \cite{PhysRevLett.113.156401,tabuchi2015coherent,PhysRevLett.129.243601}. 
	
	Next, we explore the physical regimes where our
	proposed magnon-mechanical system can be achieved using realistic parameters. We assume that the dimensions of the micromechanical cantilever are $(\ell,w,t)=(4,0.1,0.02)$ $\mu$m. Then, the frequency and the effective mass of this mechanical oscillator are estimated, respectively, as $\omega_{x} \approx 3.516 \times\left(t / \ell ^{2}\right) \sqrt{E / 12 \rho}\approx 2\pi \times 3.8$ MHz and $ M = \rho \ell w t /4 \approx 7 \times 10^{-18} $ kg, with the Young’s modulus $E \approx 1.22 \times 10^{12} \mathrm{~Pa}$ and the mass density $\rho \approx 3.52 \times 10^{3} \mathrm{~kg/m^3}$ for diamond. For the YIG nanomagnet, the sizes in the range of 10 $\le R\le$ 100 nm are considered hereafter, in accordance with the approximations for deriving magnon eigenmodes \cite{PhysRevB.101.125404}. The magnon frequency $\omega_{K}=|\gamma|B_{z}$ can be tuned by the external magnetic field $B_{z}$, and we apply a field $B_{z}\approx0.084$ T such that the spin Hamiltonian (\ref{H_NV1}) is valid. This leads to $\omega_{K}\approx2\pi \times 2.35$ GHz. In addition, we consider a large field gradient $b_0\approx10^7$ T/m for the nanomagnet \cite{mamin2007nuclear,doi:10.1063/1.3673910,Poggio_2010}. Once the parameters $b_0$, $R$ and $x_{\mathrm{zpf}}$ are given, the magnon-mechanical coupling can be determined via  $g=b_0x_{\mathrm{zpf}}\mathcal{M}_{K}V/2\hbar$.  We list the main system parameters used for our numerical calculations in Table \ref{tab1} (see Appendix C).
		
	\begin{figure*}
		\includegraphics[width=2\columnwidth]{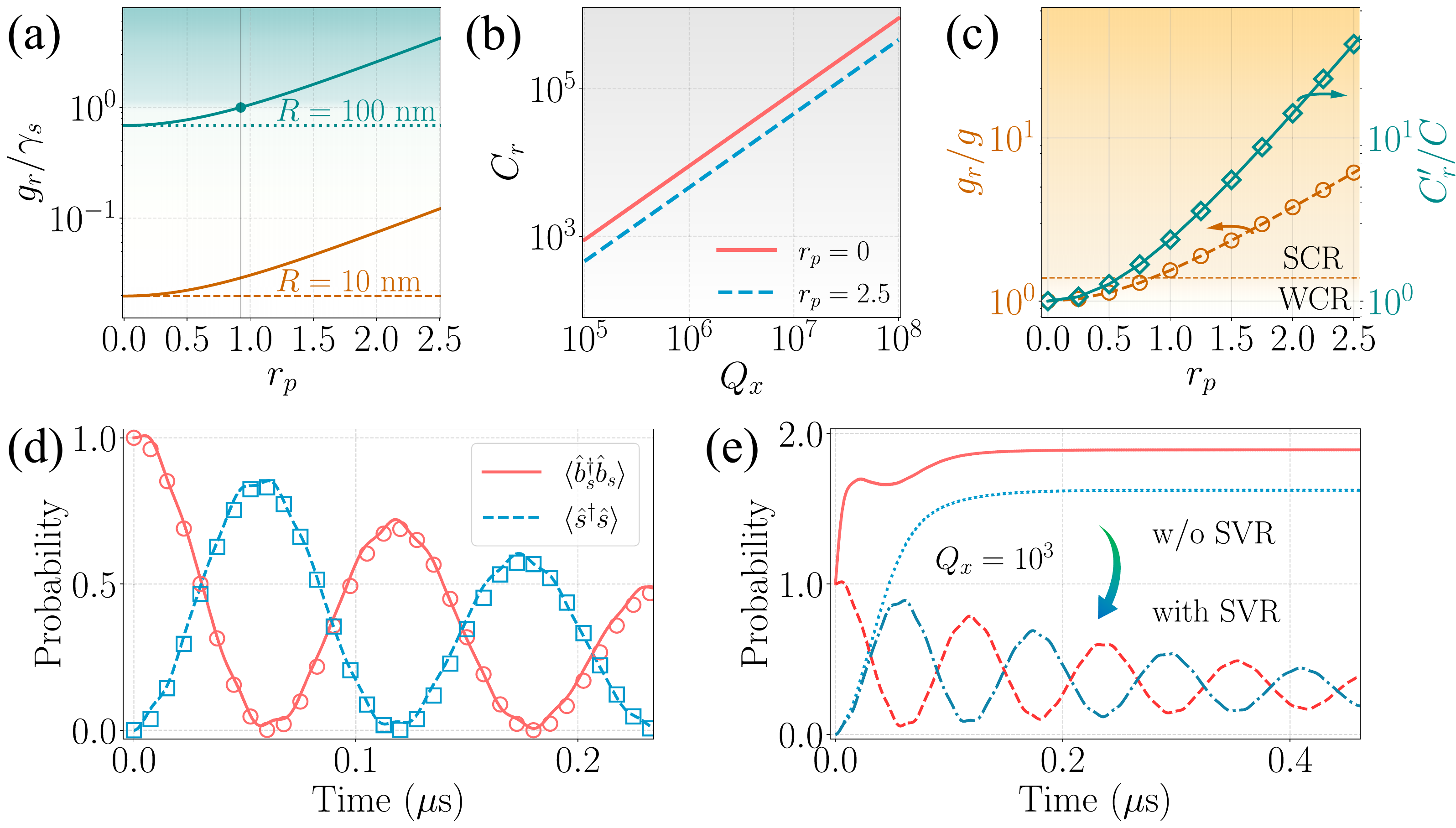}
		\caption{{Relevant parameters and quantum dynamics of the magnon-mechanical hybrid system. (a) Coupling-dissipation ratio $g_{r}/\gamma_{s}$ versus the squeezing parameter $r_p$ for the YIG spheres with radii $R=10$ and 100 nm. The dashed and dotted horizontal lines correspond to the bare coupling rates $g$ for $R=10$ and 100 nm, respectively. (b) Cooperativity $C_{r}$ versus the mechanical $Q$ factor $Q_{x}$ for $r_p=0$ (without the MPD) and 2.5. (c) Enhancement of the coupling, $g_{r}/g$, and the cooperativity, $C_{r}^{\prime}/C$, versus $r_{p}$. The shaded areas in (a) and (d) indicate the SCR with $g_{r}/\gamma_{s}>1$. (d) Time evolution of the occupation probabilities of the mean phonon number $\langle \hat{b}_{s}^\dagger\hat{b}_{s}\rangle$ and magnon number $\langle \hat{s}^\dagger\hat{s}\rangle$. The solid curves and hollow marks are obtained by numerically solving the master equation (\ref{mas equ}) with the total Hamiltonian (\ref{H_mms}) and the effective Hamiltonian (\ref{H_mmsr}), respectively. (e) Time evolution of the occupation probabilities of $\langle \hat{b}_{s}^\dagger\hat{b}_{s}\rangle$ (solid and dashed curves) and $\langle \hat{s}^\dagger\hat{s}\rangle$ (dotted and dashed-dotted curves) when considering a low mechanical $Q$ factor of $Q_{x}=10^{3}$. The solid and dotted curves are obtained without using an auxiliary squeezed-vacuum reservoir (SVR), while the dashed and dashed-dotted curves are obtained after introducing an auxiliary SVR with matching parameters $r_{e}=r_{p}$ and $\theta_{e}=\pi$. The coherent state transfer, which is corrupted by the amplified mechanical noise, can be recovered by introducing the auxiliary SVR that eliminates completely the amplified noise. Other parameters used are given in the main text.}} 
		\label{fig2}
	\end{figure*}

In Fig.~\ref{fig2}(a), we plot the ratio $g_{r}/\gamma_{s}$ versus $r_p$ for two values of the nanomagnet radius $R$. Because we always have $\gamma_{x}\ll\gamma_{s}$ for typical values of $Q_{x}$, the ratio $g_{r}/\gamma_{s}>1$ can thus be seen as an indication of the system entering the SCR, whereas $g_{r}/\gamma_{s}\le1$ corresponds to the weak-coupling regime (WCR). One can see from Fig.~\ref{fig2}(a) that $g_{r}/\gamma_{s}$ increases gradually as a function of $r_p$ due to the effect of the MPD. For a relatively small $R$ of 10 nm, the current system cannot reach the SCR even for $r_p=2.5$, whereas for a larger $R$ of 100 nm, the system can be enhanced into the SCR as long as $r_p>0.93$. Remarkably, one can obtain $g_{r}/|\Delta_{s}|>0.1$ for $r_p\ge2.42$, indicating that the system enters the UCR.
Figure \ref{fig2}(b) shows the cooperativity $C_{r}=4g_{r}^2/[\gamma_{x}\gamma_{s}(1+\bar{n}_{x}^{s})]$ versus the mechanical $Q$ factor $Q_{x}$ ranging from $10^5$ to $10^8$ for different $r_{p}$, considering $R=100$ nm and a cryogenic temperature $T=10$ mK \cite{tsaturyan2017ultracoherent}. The system resides in the high-cooperativity ($C>1$) regime even at a moderate $Q_{x}\sim10^5$, and $C_{r}$ increases significantly with increasing $Q_{x}$. In contrast, increasing $r_{p}$ leads to the decrease in $C_{r}$, due to the amplified mechanical noise responsible for the rapidly accumulated thermal phonons $\bar{n}_{x}^{s}$ with increasing the parametric drive. Note that this is generally a prominent problem for realizing coherent energy transfer in previous schemes, however it can be circumvented in our proposal as verified in the following. 
	
By numerically solving the master equation (\ref{mas equ}), the time evolution of the occupation probabilities of the mean phonon and magnon numbers can be evaluated. Figure \ref{fig2}(d) shows $\langle \hat{b}_{s}^\dagger\hat{b}_{s}\rangle$ (solid curve) and $\langle \hat{s}^\dagger\hat{s}\rangle$ (dashed curve) versus the evolution time for $r_{p}=2.5$ and $Q_{x}=10^8$. The hollow marks are obtained by solving the master equation (\ref{mas equ}) with the effective Hamiltonian (\ref{H_mmsr}), and are in good agreement with the exact numerical results (solid and dashed curves). The typical vacuum Rabi oscillations occur for multiple periods, revealing the SCR that the system resides. Notably, we also find that the coherent dynamics can hardly be interrupted in the presence of the amplified mechanical noise, and this is shown to be valid at least for $Q_{x}\ge10^5$. For larger mechanical damping, e.g., $Q_{x}=10^3$, the coherent dynamics of the system are corrupted drastically by the amplified thermal noise, with the state transfer between the mechanical phonon and the magnon being completely decoupled, as depicted in Fig.~\ref{fig2}(e) (see the top two curves).
To be more specific, the enhancement of the mechanical noise induced by the phonon squeezing can be effectively evaluated via $\gamma_x'/\gamma_x=\cosh(2r_p)$, where $\gamma_x'$ is the effective damping rate of the $\hat{b}_{s}$ mode. The enhancement of the magnon-mechanical coupling is $g_{r}/g=\cosh(r_{p})$. For $Q_{x}\ge10^5$ or $\gamma_x\le2\pi\times38$ Hz, the enhanced damping rate $\gamma_x'\le2\pi\times2.8$ kHz, which is at least three orders of magnitude smaller than the parametrically-enhanced coupling $g_r\approx2\pi\times4.23$ MHz. This allows the observation of the coherent state transfer in the system. In contrast, for relatively low $Q_x=10^{3}$ or $\gamma_x=2\pi\times3.8$ kHz, the damping rate is effectively enhanced by the phonon squeezing to $\gamma_x'=2\pi\times0.28$ MHz. In this case, $g_{r}$ is just about 15 times larger than $\gamma_x'$, and therefore, the amplified noise corrupts the coherent state transfer in the system. This may lead to a certain limitation of the proposed scheme in case of the mechanical resonator possessing a extremely-low $Q$. To eliminate completely the amplified noise from the mechanical bath, a possible strategy is to inject an orthogonally squeezed vacuum, such that the bath of the $\hat{b}_{s}$ mode remains in vacuum (see Appendix B for more details). The coherent dynamics of the system can then be recovered, as verified in Fig.~\ref{fig2}(e) (see the bottom two curves). Remarkably, the cooperativity $C_{r}$ after introducing the auxiliary reservoir is renewed as $C_{r}^\prime\equiv4g_{r}^2/[\gamma_{x}\gamma_{s}(1+\bar{n}_{x})]$, which can be parametrically-enhanced by increasing $r_{p}$, as shown Fig.~\ref{fig2}(c). 
	\\\\
	\noindent \textbf{3.2~~~{Quantum dynamics with enhanced magnon-spin coupling}}
	\\\\
	Now let us integrate the NV spin into the abovementioned magnon-mechanical system. The Hamiltonian describing the hybrid tripartite system, in the rotating frame with $\omega_p$, is given by
	\begin{equation}\label{H_mmnv}
	\begin{split}
	\frac{\hat{H}_{\mathrm{Hybrid}}}{\hbar}=&\Delta_{x} \hat{b}^{\dagger} \hat{b}+\Delta_{K} \hat{s}^{\dagger} \hat{s}+\frac{1}{2}\Delta_{\mathrm{NV}}\hat{\sigma}_{z}
	-\frac{\Omega_{p}}{2}(\hat{b}^{2}+\hat{b}^{\dagger2})
	\\
	&+\big[g(\hat{b}\hat{s}^\dagger+\hat{b}^\dagger\hat{s}^\dagger e^{\mathrm{i}2\omega_pt})+\lambda(\hat{b}\hat{\sigma}_{+}+\hat{b}^\dagger\hat{\sigma}_{+} e^{\mathrm{i}2\omega_pt})\\
	&+\mathrm{H.c.}\big],
	\end{split}
	\end{equation}
	with $\Delta_{\mathrm{NV}}=\omega_{\mathrm{NV}}-\omega_{p}$. We use the squeezing transformation $\hat{U}_{s}$ again to diagonalize the mechanics-only parts in $\hat{H}_{\mathrm{Hybrid}}$, and the resulting Hamiltonian reads
	\begin{equation}\label{H_mmnvs}
	\begin{split}
	\frac{\hat{H}_{\mathrm{Hybrid}}^{\mathrm{S}}}{\hbar}=&\Delta_{s} \hat{b}^{\dagger}_{s} \hat{b}_{s}+\Delta_{K} \hat{s}^{\dagger} \hat{s}+\frac{1}{2}\Delta_{\mathrm{NV}}\hat{\sigma}_{z}
	\\
	&+\big[g_{r}(\hat{b}_{s}\hat{s}^\dagger+\hat{b}_{s}^\dagger \hat{s}^\dagger e^{\mathrm{i}2\omega_pt})\\
	&-g_{c}(\hat{b}_{s}^\dagger\hat{s}^\dagger+\hat{b}_{s} \hat{s}^\dagger e^{\mathrm{i}2\omega_pt})\\
	&+\lambda_{r}(\hat{b}_{s}\hat{\sigma}_{+}+\hat{b}_{s}^\dagger\hat{\sigma}_{+}e^{\mathrm{i}2\omega_pt})\\
	&-\lambda_{c}(\hat{b}_{s}^\dagger\hat{\sigma}_{+}+\hat{b}_{s}\hat{\sigma}_{+}e^{\mathrm{i}2\omega_pt})+\mathrm{H.c.}\big],
	\end{split}
	\end{equation}
	where $\lambda_{r}=\lambda\cosh(r_p)$ and $\lambda_{c}=\lambda\sinh(r_p)$ are the enhanced spin-mechanical couplings by the MPD. In the dispersive regime with properly large detunings (see Appendix D), the squeezed phonon mode $\hat{b}_{s}$ is only virtually excited, which allows us to adiabatically eliminate the mechanical degree of freedom from the dynamics. We further consider that the magnon frequency is very close to the NV spin frequency, namely $\omega_{K}\approx\omega_{\mathrm{NV}}$, which can be achieved by introducing an additional static magnetic field (with the opposite direction to $\textbf{B}_{z}$) for the nanomagnet (see relevant realistic parameters below). Furthermore, the magnetic gradient $b_{1}$ is adjusted such that the bare spin-mechanical coupling $\lambda$ is nearly equal to the bare magnon-mechanical coupling $g$. Given these settings, the effective Hamiltonian describing the phonon-mediated magnon-spin interactions can then be written as (see Appendix D)
	\begin{equation}\label{H_msr}
	\frac{\hat{H}_{\mathrm{ms}}^{\mathrm{S}}}{\hbar}=\Delta_{k}\hat{s}^\dagger\hat{s}+\Delta_{n}\hat{\sigma}_{+}\hat{\sigma}_{-}+g_{ms}(\hat{s}^\dagger\hat{\sigma}_{-}+ \hat{s}\hat{\sigma}_{+}),
	\end{equation}
	where $\Delta_{k}=g_{r}^2/(\Delta_{K}-\Delta_{s})$, $\Delta_{n}=\lambda_{r}^2/(\Delta_{\mathrm{NV}}-\Delta_{s})$ are the effective detunings, and $g_{ms}=g_{r}\lambda_{r}/[\Delta_{K(\mathrm{NV})}-\Delta_{s}]$ is the effective magnon-spin coupling rate. The Hamiltonian (\ref{H_msr}) indicates that we have achieved an effective two-mode system where the magnon is resonantly coupled to the NV spin with an enhanced coupling rate. 
	
	Under a realistic consideration, the dissipative dynamics of the tripartite system is governed by the master equation
	\begin{equation} \label{mas equ1}
	\dot{\hat{\rho}}=-\frac{\mathrm{i}}{\hbar}[\hat{H}_{\mathrm{Hybrid}}^{\mathrm{S}},\hat{\rho}]+\mathcal{L}_{x}[\hat{\rho}]+\mathcal{L}_{K}[\hat{\rho}]+\mathcal{L}_{z}[\hat{\rho}],
	\end{equation}
	where $\hat{H}_{\mathrm{Hybrid}}^{\mathrm{S}}$ is given in Eq.~(\ref{H_mmnvs}), and $\mathcal{L}_{z}[\hat{\rho}]=\gamma_zL_{\hat{\sigma}_{z}}\hat{\rho}$ with $\gamma_{z}$ the single spin dephasing rate of the NV center. Experimentally, the NV dephasing time $T_{2}=1/\gamma_{z}$ can readily reach several hundreds of kilohertz at both room temperature and cryogenic temperature \cite{ishikawa2012optical,mamin2013nanoscale,bar2013solid,ovartchaiyapong2014dynamic}, and hereafter we consider $\gamma_{z}\approx2\pi\times1$ kHz. As for other experimental parameters, we recall the scenario considered in Sec.~3.1, where relevant experimental parameters are $\omega_{x}\approx2\pi\times3.8$ MHz, $Q_{x}=10^8$, and $g\approx2\pi\times0.69$ MHz for $R=100$ nm. For the NV spin, we apply a static magnetic field $B_{z}\approx0.068$ T, which gives $\omega_{\mathrm{NV}}\approx2\pi\times0.96$ GHz. Furthermore, an additional local field $B_{z}'\approx0.034$ T is introduced to adjust the frequency of the nanomagnet, such that $\omega_{K}=|\gamma|(B_{z}-B_{z}')\approx\omega_{\mathrm{NV}}$ \cite{PhysRevLett.113.156401}. Regarding the bare coupling between the NV spin and the mechanical oscillator, we consider $b_{1}\approx4.56\times10^7$ T/m, resulting $\lambda\approx2\pi\times0.69$ MHz, namely $\lambda\approx g$. Then, the system dynamics 
	can be evaluated by exactly solving the master equation (\ref{mas equ1}).
	
	\begin{figure*}
		\includegraphics[width=1.8\columnwidth]{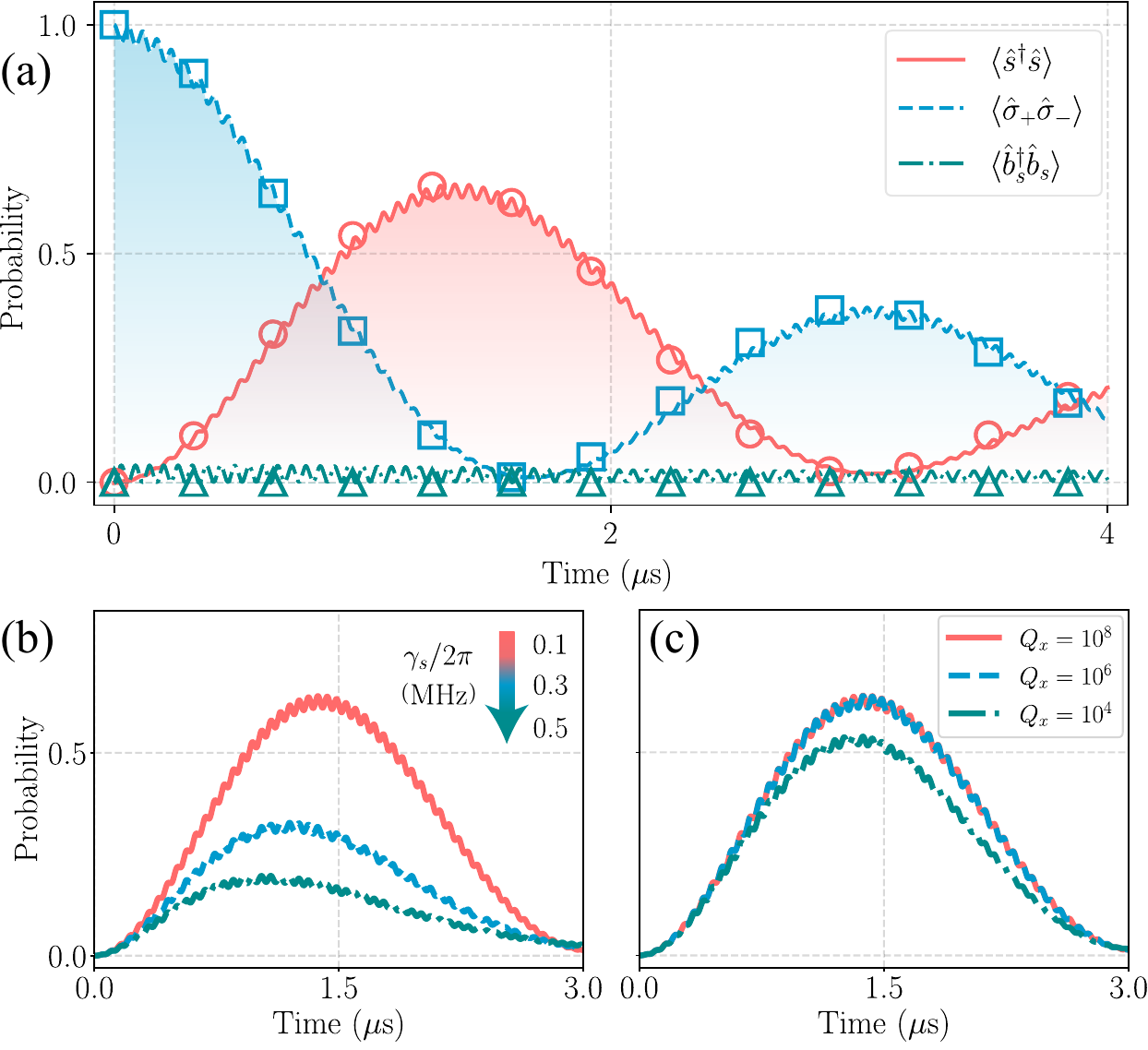}
		\caption{{Quantum dynamics of the hybrid tripartite system in the dispersive regime. (a) Time evolution of the occupation probabilities of the mean phonon number, the mean magnon numbers and the spin excited state. The solid curves and hollow marks are obtained by numerically solving the master equation (\ref{mas equ1}) with the total Hamiltonian (\ref{H_mmnvs}) and the effective Hamiltonian (\ref{H_msr}), respectively. (b),(c) Time evolution of the occupation probability of the spin excited state for different magnon dissipation rate $\gamma_{s}$ and mechanical $Q$ factor $Q_{x}$. In (b) and (c), $Q_{x}$ and $\gamma_{s}$ are fixed, respectively, to be $10^{8}$ and $2\pi\times0.1$ MHz. Other specific parameters used here are given in the main text.}}
		\label{fig3}
	\end{figure*}
	
	The dynamics of the hybrid system is examined for a proper MPD and the suitably large detunings, for example, $r_{p}=1.54$, $\Delta_{s}=-55g_{r}$ and $\Delta_{K}=-45g_{r}$. The numerical results are shown in Fig.~\ref{fig3}(a). Here, we assume that the NV spin is initially in the excited state, while the nanomagnet and the mechanical oscillator are in their vacuum states. The evident vacuum Rabi oscillations between the nanomagnet and the NV spin with excited-state population $\langle\hat{\sigma}_{+}\hat{\sigma}_{-}\rangle$ can be observed clearly, whereas the mechanical oscillator is nearly unexcited (namely, virtually excited) with $\langle\hat{b}_{s}^\dagger\hat{b}_{s}\rangle\approx0$ during the evolution. This implies that the quantum state initially stored in the NV spin can be transferred to the nanomagnet through the mechanical oscillator serving as a quantum bus, and vice versa. In our simulations, the decay rate of the magnon mode is set to be $\gamma_{s}=2\pi\times0.1$ MHz. Experimentally, a narrow magnon linewidth $<2\pi\times0.6$ MHz has been demonstrated in a YIG sphere \cite{doi:10.1126/sciadv.1501286}, which could be more smaller when considering an ultrapure YIG \cite{rameshti2022cavity}. For the relatively strong magnon dissipation, the state transfer becomes more inefficient with increasing $\gamma_{s}$,  while it exhibits strong robustness against the mechanical dissipation for $Q_{x}$ as low as $10^{4}$, as shown in Figs.~\ref{fig3}(b) and \ref{fig3}(c).
	\\\\
	\noindent \textbf{\begin{large}4~~{Conclusion}\end{large}}
	\\\\
	In conclusion, we have presented an experimentally feasible method to realize the strong magnon-mechanical and phonon-mediated magnon-spin couplings in a hybrid tripartite system. We have shown that the magnon mode supported by the nanomagnet can be coupled, in a strong and tunable way, to the phonon mode of the micromechanical cantilever by introducing the parametric amplification of the mechanical motion. With experimentally accessible parameters, this magnon-mechanical coupling can readily reach the strong-coupling regime, thus removing the obstacle to achieving the strong coupling between the two quantum systems with a large energy difference. By further coupling an NV spin to the mechanical motion and working in the dispersive regime, we have engineered an efficient strategy for quantum state transfer between the nanomagnet and the NV center. This provides a versatile platform for processing complicated quantum information tasks where the NV spin and the magnon can both act as collaborate quantum memories. Our proposal could also find other potential applications such as quantum control and measurement of magnon excitations, and the design of quantum transducers and sensors, etc.
	\\
	\\
	\begin{acknowledgements}
		\begin{small}{\it This work was supported by the National Natural Science Foundation of China (Grant Nos. 12205256, 11935006, 11774086, 12247105, and 1217050862), the Henan Provincial Science and Technology Research Project (Grant Nos. 232102221001, 
		232102210175), HNQSTIT project (Grant No. 2022112), and the Fundamental Research Funds for the Central Universities (Grant No. 2023FRFK06012).}\end{small}
	\end{acknowledgements}

\bibliography{apssamp}

\begin{thebibliography}{100}%
\makeatletter
\providecommand \@ifxundefined [1]{%
 \@ifx{#1\undefined}
}%
\providecommand \@ifnum [1]{%
 \ifnum #1\expandafter \@firstoftwo
 \else \expandafter \@secondoftwo
 \fi
}%
\providecommand \@ifx [1]{%
 \ifx #1\expandafter \@firstoftwo
 \else \expandafter \@secondoftwo
 \fi
}%
\providecommand \natexlab [1]{#1}%
\providecommand \enquote  [1]{``#1''}%
\providecommand \bibnamefont  [1]{#1}%
\providecommand \bibfnamefont [1]{#1}%
\providecommand \citenamefont [1]{#1}%
\providecommand \href@noop [0]{\@secondoftwo}%
\providecommand \href [0]{\begingroup \@sanitize@url \@href}%
\providecommand \@href[1]{\@@startlink{#1}\@@href}%
\providecommand \@@href[1]{\endgroup#1\@@endlink}%
\providecommand \@sanitize@url [0]{\catcode `\\12\catcode `\$12\catcode
  `\&12\catcode `\#12\catcode `\^12\catcode `\_12\catcode `\%12\relax}%
\providecommand \@@startlink[1]{}%
\providecommand \@@endlink[0]{}%
\providecommand \url  [0]{\begingroup\@sanitize@url \@url }%
\providecommand \@url [1]{\endgroup\@href {#1}{\urlprefix }}%
\providecommand \urlprefix  [0]{URL }%
\providecommand \Eprint [0]{\href }%
\providecommand \doibase [0]{http://dx.doi.org/}%
\providecommand \selectlanguage [0]{\@gobble}%
\providecommand \bibinfo  [0]{\@secondoftwo}%
\providecommand \bibfield  [0]{\@secondoftwo}%
\providecommand \translation [1]{[#1]}%
\providecommand \BibitemOpen [0]{}%
\providecommand \bibitemStop [0]{}%
\providecommand \bibitemNoStop [0]{.\EOS\space}%
\providecommand \EOS [0]{\spacefactor3000\relax}%
\providecommand \BibitemShut  [1]{\csname bibitem#1\endcsname}%
\let\auto@bib@innerbib\@empty
\bibitem [{\citenamefont {Yuan}\ \emph {et~al.}(2022)\citenamefont {Yuan},
  \citenamefont {Cao}, \citenamefont {Kamra}, \citenamefont {Duine},\ and\
  \citenamefont {Yan}}]{YUAN20221}%
  \BibitemOpen
  \bibfield  {author} {\bibinfo {author} {\bibfnamefont {H.}~\bibnamefont
  {Yuan}}, \bibinfo {author} {\bibfnamefont {Y.}~\bibnamefont {Cao}}, \bibinfo
  {author} {\bibfnamefont {A.}~\bibnamefont {Kamra}}, \bibinfo {author}
  {\bibfnamefont {R.~A.}\ \bibnamefont {Duine}}, \ and\ \bibinfo {author}
  {\bibfnamefont {P.}~\bibnamefont {Yan}},\ }\href {\doibase
  https://doi.org/10.1016/j.physrep.2022.03.002} {\bibfield  {journal}
  {\bibinfo  {journal} {Phys. Rep.}\ }\textbf {\bibinfo {volume} {965}},\
  \bibinfo {pages} {1} (\bibinfo {year} {2022})}\BibitemShut {NoStop}%
\bibitem [{\citenamefont {{Zare Rameshti}}\ \emph {et~al.}(2022)\citenamefont
  {{Zare Rameshti}}, \citenamefont {{Viola Kusminskiy}}, \citenamefont {Haigh},
  \citenamefont {Usami}, \citenamefont {Lachance-Quirion}, \citenamefont
  {Nakamura}, \citenamefont {Hu}, \citenamefont {Tang}, \citenamefont {Bauer},\
  and\ \citenamefont {Blanter}}]{ZARERAMESHTI20221}%
  \BibitemOpen
  \bibfield  {author} {\bibinfo {author} {\bibfnamefont {B.}~\bibnamefont
  {{Zare Rameshti}}}, \bibinfo {author} {\bibfnamefont {S.}~\bibnamefont
  {{Viola Kusminskiy}}}, \bibinfo {author} {\bibfnamefont {J.~A.}\ \bibnamefont
  {Haigh}}, \bibinfo {author} {\bibfnamefont {K.}~\bibnamefont {Usami}},
  \bibinfo {author} {\bibfnamefont {D.}~\bibnamefont {Lachance-Quirion}},
  \bibinfo {author} {\bibfnamefont {Y.}~\bibnamefont {Nakamura}}, \bibinfo
  {author} {\bibfnamefont {C.-M.}\ \bibnamefont {Hu}}, \bibinfo {author}
  {\bibfnamefont {H.~X.}\ \bibnamefont {Tang}}, \bibinfo {author}
  {\bibfnamefont {G.~E.}\ \bibnamefont {Bauer}}, \ and\ \bibinfo {author}
  {\bibfnamefont {Y.~M.}\ \bibnamefont {Blanter}},\ }\href {\doibase
  https://doi.org/10.1016/j.physrep.2022.06.001} {\bibfield  {journal}
  {\bibinfo  {journal} {Phys. Rep.}\ }\textbf {\bibinfo {volume} {979}},\
  \bibinfo {pages} {1} (\bibinfo {year} {2022})}\BibitemShut {NoStop}%
\bibitem [{\citenamefont {Barman}\ \emph {et~al.}(2021)\citenamefont {Barman},
  \citenamefont {Gubbiotti}, \citenamefont {Ladak},\ and\ \citenamefont
  {et~al.}}]{Barman2021}%
  \BibitemOpen
  \bibfield  {author} {\bibinfo {author} {\bibfnamefont {A.}~\bibnamefont
  {Barman}}, \bibinfo {author} {\bibfnamefont {G.}~\bibnamefont {Gubbiotti}},
  \bibinfo {author} {\bibfnamefont {S.}~\bibnamefont {Ladak}}, \ and\ \bibinfo
  {author} {\bibnamefont {et~al.}},\ }\href {\doibase 10.1088/1361-648X/abec1a}
  {\bibfield  {journal} {\bibinfo  {journal} {J. Phys.: Condens. Matter}\
  }\textbf {\bibinfo {volume} {33}},\ \bibinfo {pages} {413001} (\bibinfo
  {year} {2021})}\BibitemShut {NoStop}%
\bibitem [{\citenamefont {Li}\ \emph {et~al.}(2020{\natexlab{a}})\citenamefont
  {Li}, \citenamefont {Zhang}, \citenamefont {Tyberkevych}, \citenamefont
  {Kwok}, \citenamefont {Hoffmann},\ and\ \citenamefont
  {Novosad}}]{doi10106350020277}%
  \BibitemOpen
  \bibfield  {author} {\bibinfo {author} {\bibfnamefont {Y.}~\bibnamefont
  {Li}}, \bibinfo {author} {\bibfnamefont {W.}~\bibnamefont {Zhang}}, \bibinfo
  {author} {\bibfnamefont {V.}~\bibnamefont {Tyberkevych}}, \bibinfo {author}
  {\bibfnamefont {W.-K.}\ \bibnamefont {Kwok}}, \bibinfo {author}
  {\bibfnamefont {A.}~\bibnamefont {Hoffmann}}, \ and\ \bibinfo {author}
  {\bibfnamefont {V.}~\bibnamefont {Novosad}},\ }\href {\doibase
  10.1063/5.0020277} {\bibfield  {journal} {\bibinfo  {journal} {J. Appl.
  Phys.}\ }\textbf {\bibinfo {volume} {128}},\ \bibinfo {pages} {130902}
  (\bibinfo {year} {2020}{\natexlab{a}})}\BibitemShut {NoStop}%
\bibitem [{\citenamefont {Kamra}\ \emph {et~al.}(2020)\citenamefont {Kamra},
  \citenamefont {Belzig},\ and\ \citenamefont {Brataas}}]{doi10106350021099}%
  \BibitemOpen
  \bibfield  {author} {\bibinfo {author} {\bibfnamefont {A.}~\bibnamefont
  {Kamra}}, \bibinfo {author} {\bibfnamefont {W.}~\bibnamefont {Belzig}}, \
  and\ \bibinfo {author} {\bibfnamefont {A.}~\bibnamefont {Brataas}},\ }\href
  {\doibase 10.1063/5.0021099} {\bibfield  {journal} {\bibinfo  {journal}
  {Appl. Phys. Lett.}\ }\textbf {\bibinfo {volume} {117}},\ \bibinfo {pages}
  {090501} (\bibinfo {year} {2020})}\BibitemShut {NoStop}%
\bibitem [{\citenamefont {Harder}\ \emph {et~al.}(2021)\citenamefont {Harder},
  \citenamefont {Yao}, \citenamefont {Gui},\ and\ \citenamefont
  {Hu}}]{doi10106350046202}%
  \BibitemOpen
  \bibfield  {author} {\bibinfo {author} {\bibfnamefont {M.}~\bibnamefont
  {Harder}}, \bibinfo {author} {\bibfnamefont {B.~M.}\ \bibnamefont {Yao}},
  \bibinfo {author} {\bibfnamefont {Y.~S.}\ \bibnamefont {Gui}}, \ and\
  \bibinfo {author} {\bibfnamefont {C.-M.}\ \bibnamefont {Hu}},\ }\href
  {\doibase 10.1063/5.0046202} {\bibfield  {journal} {\bibinfo  {journal} {J.
  Appl. Phys.}\ }\textbf {\bibinfo {volume} {129}},\ \bibinfo {pages} {201101}
  (\bibinfo {year} {2021})}\BibitemShut {NoStop}%
\bibitem [{\citenamefont {Lachance-Quirion}\ \emph {et~al.}(2019)\citenamefont
  {Lachance-Quirion}, \citenamefont {Tabuchi}, \citenamefont {Gloppe},
  \citenamefont {Usami},\ and\ \citenamefont {Nakamura}}]{LachanceQuirion2019}%
  \BibitemOpen
  \bibfield  {author} {\bibinfo {author} {\bibfnamefont {D.}~\bibnamefont
  {Lachance-Quirion}}, \bibinfo {author} {\bibfnamefont {Y.}~\bibnamefont
  {Tabuchi}}, \bibinfo {author} {\bibfnamefont {A.}~\bibnamefont {Gloppe}},
  \bibinfo {author} {\bibfnamefont {K.}~\bibnamefont {Usami}}, \ and\ \bibinfo
  {author} {\bibfnamefont {Y.}~\bibnamefont {Nakamura}},\ }\href {\doibase
  10.7567/1882-0786/ab248d} {\bibfield  {journal} {\bibinfo  {journal} {Appl.
  Phys. Express}\ }\textbf {\bibinfo {volume} {12}},\ \bibinfo {pages} {070101}
  (\bibinfo {year} {2019})}\BibitemShut {NoStop}%
\bibitem [{\citenamefont {Huebl}\ \emph {et~al.}(2013)\citenamefont {Huebl},
  \citenamefont {Zollitsch}, \citenamefont {Lotze}, \citenamefont {Hocke},
  \citenamefont {Greifenstein}, \citenamefont {Marx}, \citenamefont {Gross},\
  and\ \citenamefont {Goennenwein}}]{PhysRevLett.111.127003}%
  \BibitemOpen
  \bibfield  {author} {\bibinfo {author} {\bibfnamefont {H.}~\bibnamefont
  {Huebl}}, \bibinfo {author} {\bibfnamefont {C.~W.}\ \bibnamefont
  {Zollitsch}}, \bibinfo {author} {\bibfnamefont {J.}~\bibnamefont {Lotze}},
  \bibinfo {author} {\bibfnamefont {F.}~\bibnamefont {Hocke}}, \bibinfo
  {author} {\bibfnamefont {M.}~\bibnamefont {Greifenstein}}, \bibinfo {author}
  {\bibfnamefont {A.}~\bibnamefont {Marx}}, \bibinfo {author} {\bibfnamefont
  {R.}~\bibnamefont {Gross}}, \ and\ \bibinfo {author} {\bibfnamefont
  {S.~T.~B.}\ \bibnamefont {Goennenwein}},\ }\href {\doibase
  10.1103/PhysRevLett.111.127003} {\bibfield  {journal} {\bibinfo  {journal}
  {Phys. Rev. Lett.}\ }\textbf {\bibinfo {volume} {111}},\ \bibinfo {pages}
  {127003} (\bibinfo {year} {2013})}\BibitemShut {NoStop}%
\bibitem [{\citenamefont {Tabuchi}\ \emph {et~al.}(2014)\citenamefont
  {Tabuchi}, \citenamefont {Ishino}, \citenamefont {Ishikawa}, \citenamefont
  {Yamazaki}, \citenamefont {Usami},\ and\ \citenamefont
  {Nakamura}}]{PhysRevLett.113.083603}%
  \BibitemOpen
  \bibfield  {author} {\bibinfo {author} {\bibfnamefont {Y.}~\bibnamefont
  {Tabuchi}}, \bibinfo {author} {\bibfnamefont {S.}~\bibnamefont {Ishino}},
  \bibinfo {author} {\bibfnamefont {T.}~\bibnamefont {Ishikawa}}, \bibinfo
  {author} {\bibfnamefont {R.}~\bibnamefont {Yamazaki}}, \bibinfo {author}
  {\bibfnamefont {K.}~\bibnamefont {Usami}}, \ and\ \bibinfo {author}
  {\bibfnamefont {Y.}~\bibnamefont {Nakamura}},\ }\href {\doibase
  10.1103/PhysRevLett.113.083603} {\bibfield  {journal} {\bibinfo  {journal}
  {Phys. Rev. Lett.}\ }\textbf {\bibinfo {volume} {113}},\ \bibinfo {pages}
  {083603} (\bibinfo {year} {2014})}\BibitemShut {NoStop}%
\bibitem [{\citenamefont {Zhang}\ \emph {et~al.}(2014)\citenamefont {Zhang},
  \citenamefont {Zou}, \citenamefont {Jiang},\ and\ \citenamefont
  {Tang}}]{PhysRevLett.113.156401}%
  \BibitemOpen
  \bibfield  {author} {\bibinfo {author} {\bibfnamefont {X.}~\bibnamefont
  {Zhang}}, \bibinfo {author} {\bibfnamefont {C.-L.}\ \bibnamefont {Zou}},
  \bibinfo {author} {\bibfnamefont {L.}~\bibnamefont {Jiang}}, \ and\ \bibinfo
  {author} {\bibfnamefont {H.~X.}\ \bibnamefont {Tang}},\ }\href {\doibase
  10.1103/PhysRevLett.113.156401} {\bibfield  {journal} {\bibinfo  {journal}
  {Phys. Rev. Lett.}\ }\textbf {\bibinfo {volume} {113}},\ \bibinfo {pages}
  {156401} (\bibinfo {year} {2014})}\BibitemShut {NoStop}%
\bibitem [{\citenamefont {Osada}\ \emph {et~al.}(2016)\citenamefont {Osada},
  \citenamefont {Hisatomi}, \citenamefont {Noguchi}, \citenamefont {Tabuchi},
  \citenamefont {Yamazaki}, \citenamefont {Usami}, \citenamefont {Sadgrove},
  \citenamefont {Yalla}, \citenamefont {Nomura},\ and\ \citenamefont
  {Nakamura}}]{PhysRevLett.116.223601}%
  \BibitemOpen
  \bibfield  {author} {\bibinfo {author} {\bibfnamefont {A.}~\bibnamefont
  {Osada}}, \bibinfo {author} {\bibfnamefont {R.}~\bibnamefont {Hisatomi}},
  \bibinfo {author} {\bibfnamefont {A.}~\bibnamefont {Noguchi}}, \bibinfo
  {author} {\bibfnamefont {Y.}~\bibnamefont {Tabuchi}}, \bibinfo {author}
  {\bibfnamefont {R.}~\bibnamefont {Yamazaki}}, \bibinfo {author}
  {\bibfnamefont {K.}~\bibnamefont {Usami}}, \bibinfo {author} {\bibfnamefont
  {M.}~\bibnamefont {Sadgrove}}, \bibinfo {author} {\bibfnamefont
  {R.}~\bibnamefont {Yalla}}, \bibinfo {author} {\bibfnamefont
  {M.}~\bibnamefont {Nomura}}, \ and\ \bibinfo {author} {\bibfnamefont
  {Y.}~\bibnamefont {Nakamura}},\ }\href {\doibase
  10.1103/PhysRevLett.116.223601} {\bibfield  {journal} {\bibinfo  {journal}
  {Phys. Rev. Lett.}\ }\textbf {\bibinfo {volume} {116}},\ \bibinfo {pages}
  {223601} (\bibinfo {year} {2016})}\BibitemShut {NoStop}%
\bibitem [{\citenamefont {Zhang}\ \emph
  {et~al.}(2016{\natexlab{a}})\citenamefont {Zhang}, \citenamefont {Zhu},
  \citenamefont {Zou},\ and\ \citenamefont {Tang}}]{PhysRevLett.117.123605}%
  \BibitemOpen
  \bibfield  {author} {\bibinfo {author} {\bibfnamefont {X.}~\bibnamefont
  {Zhang}}, \bibinfo {author} {\bibfnamefont {N.}~\bibnamefont {Zhu}}, \bibinfo
  {author} {\bibfnamefont {C.-L.}\ \bibnamefont {Zou}}, \ and\ \bibinfo
  {author} {\bibfnamefont {H.~X.}\ \bibnamefont {Tang}},\ }\href {\doibase
  10.1103/PhysRevLett.117.123605} {\bibfield  {journal} {\bibinfo  {journal}
  {Phys. Rev. Lett.}\ }\textbf {\bibinfo {volume} {117}},\ \bibinfo {pages}
  {123605} (\bibinfo {year} {2016}{\natexlab{a}})}\BibitemShut {NoStop}%
\bibitem [{\citenamefont {Li}\ \emph {et~al.}(2019)\citenamefont {Li},
  \citenamefont {Polakovic}, \citenamefont {Wang}, \citenamefont {Xu},
  \citenamefont {Lendinez}, \citenamefont {Zhang}, \citenamefont {Ding},
  \citenamefont {Khaire}, \citenamefont {Saglam}, \citenamefont {Divan},
  \citenamefont {Pearson}, \citenamefont {Kwok}, \citenamefont {Xiao},
  \citenamefont {Novosad}, \citenamefont {Hoffmann},\ and\ \citenamefont
  {Zhang}}]{PhysRevLett.123.107701}%
  \BibitemOpen
  \bibfield  {author} {\bibinfo {author} {\bibfnamefont {Y.}~\bibnamefont
  {Li}}, \bibinfo {author} {\bibfnamefont {T.}~\bibnamefont {Polakovic}},
  \bibinfo {author} {\bibfnamefont {Y.-L.}\ \bibnamefont {Wang}}, \bibinfo
  {author} {\bibfnamefont {J.}~\bibnamefont {Xu}}, \bibinfo {author}
  {\bibfnamefont {S.}~\bibnamefont {Lendinez}}, \bibinfo {author}
  {\bibfnamefont {Z.}~\bibnamefont {Zhang}}, \bibinfo {author} {\bibfnamefont
  {J.}~\bibnamefont {Ding}}, \bibinfo {author} {\bibfnamefont {T.}~\bibnamefont
  {Khaire}}, \bibinfo {author} {\bibfnamefont {H.}~\bibnamefont {Saglam}},
  \bibinfo {author} {\bibfnamefont {R.}~\bibnamefont {Divan}}, \bibinfo
  {author} {\bibfnamefont {J.}~\bibnamefont {Pearson}}, \bibinfo {author}
  {\bibfnamefont {W.-K.}\ \bibnamefont {Kwok}}, \bibinfo {author}
  {\bibfnamefont {Z.}~\bibnamefont {Xiao}}, \bibinfo {author} {\bibfnamefont
  {V.}~\bibnamefont {Novosad}}, \bibinfo {author} {\bibfnamefont
  {A.}~\bibnamefont {Hoffmann}}, \ and\ \bibinfo {author} {\bibfnamefont
  {W.}~\bibnamefont {Zhang}},\ }\href {\doibase 10.1103/PhysRevLett.123.107701}
  {\bibfield  {journal} {\bibinfo  {journal} {Phys. Rev. Lett.}\ }\textbf
  {\bibinfo {volume} {123}},\ \bibinfo {pages} {107701} (\bibinfo {year}
  {2019})}\BibitemShut {NoStop}%
\bibitem [{\citenamefont {Hou}\ and\ \citenamefont
  {Liu}(2019)}]{PhysRevLett.123.107702}%
  \BibitemOpen
  \bibfield  {author} {\bibinfo {author} {\bibfnamefont {J.~T.}\ \bibnamefont
  {Hou}}\ and\ \bibinfo {author} {\bibfnamefont {L.}~\bibnamefont {Liu}},\
  }\href {\doibase 10.1103/PhysRevLett.123.107702} {\bibfield  {journal}
  {\bibinfo  {journal} {Phys. Rev. Lett.}\ }\textbf {\bibinfo {volume} {123}},\
  \bibinfo {pages} {107702} (\bibinfo {year} {2019})}\BibitemShut {NoStop}%
\bibitem [{\citenamefont {Xu}\ \emph {et~al.}(2020)\citenamefont {Xu},
  \citenamefont {Zhong}, \citenamefont {Han}, \citenamefont {Jin},
  \citenamefont {Jiang},\ and\ \citenamefont {Zhang}}]{PhysRevLett.125.237201}%
  \BibitemOpen
  \bibfield  {author} {\bibinfo {author} {\bibfnamefont {J.}~\bibnamefont
  {Xu}}, \bibinfo {author} {\bibfnamefont {C.}~\bibnamefont {Zhong}}, \bibinfo
  {author} {\bibfnamefont {X.}~\bibnamefont {Han}}, \bibinfo {author}
  {\bibfnamefont {D.}~\bibnamefont {Jin}}, \bibinfo {author} {\bibfnamefont
  {L.}~\bibnamefont {Jiang}}, \ and\ \bibinfo {author} {\bibfnamefont
  {X.}~\bibnamefont {Zhang}},\ }\href {\doibase 10.1103/PhysRevLett.125.237201}
  {\bibfield  {journal} {\bibinfo  {journal} {Phys. Rev. Lett.}\ }\textbf
  {\bibinfo {volume} {125}},\ \bibinfo {pages} {237201} (\bibinfo {year}
  {2020})}\BibitemShut {NoStop}%
\bibitem [{\citenamefont {Yuan}\ \emph {et~al.}(2020)\citenamefont {Yuan},
  \citenamefont {Yan}, \citenamefont {Zheng}, \citenamefont {He}, \citenamefont
  {Xia},\ and\ \citenamefont {Yung}}]{PhysRevLett.124.053602}%
  \BibitemOpen
  \bibfield  {author} {\bibinfo {author} {\bibfnamefont {H.~Y.}\ \bibnamefont
  {Yuan}}, \bibinfo {author} {\bibfnamefont {P.}~\bibnamefont {Yan}}, \bibinfo
  {author} {\bibfnamefont {S.}~\bibnamefont {Zheng}}, \bibinfo {author}
  {\bibfnamefont {Q.~Y.}\ \bibnamefont {He}}, \bibinfo {author} {\bibfnamefont
  {K.}~\bibnamefont {Xia}}, \ and\ \bibinfo {author} {\bibfnamefont {M.-H.}\
  \bibnamefont {Yung}},\ }\href {\doibase 10.1103/PhysRevLett.124.053602}
  {\bibfield  {journal} {\bibinfo  {journal} {Phys. Rev. Lett.}\ }\textbf
  {\bibinfo {volume} {124}},\ \bibinfo {pages} {053602} (\bibinfo {year}
  {2020})}\BibitemShut {NoStop}%
\bibitem [{\citenamefont {Tabuchi}\ \emph {et~al.}(2015)\citenamefont
  {Tabuchi}, \citenamefont {Ishino}, \citenamefont {Noguchi}, \citenamefont
  {Ishikawa}, \citenamefont {Yamazaki}, \citenamefont {Usami},\ and\
  \citenamefont {Nakamura}}]{tabuchi2015coherent}%
  \BibitemOpen
  \bibfield  {author} {\bibinfo {author} {\bibfnamefont {Y.}~\bibnamefont
  {Tabuchi}}, \bibinfo {author} {\bibfnamefont {S.}~\bibnamefont {Ishino}},
  \bibinfo {author} {\bibfnamefont {A.}~\bibnamefont {Noguchi}}, \bibinfo
  {author} {\bibfnamefont {T.}~\bibnamefont {Ishikawa}}, \bibinfo {author}
  {\bibfnamefont {R.}~\bibnamefont {Yamazaki}}, \bibinfo {author}
  {\bibfnamefont {K.}~\bibnamefont {Usami}}, \ and\ \bibinfo {author}
  {\bibfnamefont {Y.}~\bibnamefont {Nakamura}},\ }\href
  {https://www.science.org/doi/full/10.1126/science.aaa3693} {\bibfield
  {journal} {\bibinfo  {journal} {Science}\ }\textbf {\bibinfo {volume}
  {349}},\ \bibinfo {pages} {405} (\bibinfo {year} {2015})}\BibitemShut
  {NoStop}%
\bibitem [{\citenamefont {Lachance-Quirion}\ \emph {et~al.}(2017)\citenamefont
  {Lachance-Quirion}, \citenamefont {Tabuchi}, \citenamefont {Ishino},
  \citenamefont {Noguchi}, \citenamefont {Ishikawa}, \citenamefont {Yamazaki},\
  and\ \citenamefont {Nakamura}}]{doi:10.1126/sciadv.1603150}%
  \BibitemOpen
  \bibfield  {author} {\bibinfo {author} {\bibfnamefont {D.}~\bibnamefont
  {Lachance-Quirion}}, \bibinfo {author} {\bibfnamefont {Y.}~\bibnamefont
  {Tabuchi}}, \bibinfo {author} {\bibfnamefont {S.}~\bibnamefont {Ishino}},
  \bibinfo {author} {\bibfnamefont {A.}~\bibnamefont {Noguchi}}, \bibinfo
  {author} {\bibfnamefont {T.}~\bibnamefont {Ishikawa}}, \bibinfo {author}
  {\bibfnamefont {R.}~\bibnamefont {Yamazaki}}, \ and\ \bibinfo {author}
  {\bibfnamefont {Y.}~\bibnamefont {Nakamura}},\ }\href {\doibase
  10.1126/sciadv.1603150} {\bibfield  {journal} {\bibinfo  {journal} {Sci.
  Adv.}\ }\textbf {\bibinfo {volume} {3}},\ \bibinfo {pages} {e1603150}
  (\bibinfo {year} {2017})}\BibitemShut {NoStop}%
\bibitem [{\citenamefont {Lachance-Quirion}\ \emph {et~al.}(2020)\citenamefont
  {Lachance-Quirion}, \citenamefont {Wolski}, \citenamefont {Tabuchi},
  \citenamefont {Kono}, \citenamefont {Usami},\ and\ \citenamefont
  {Nakamura}}]{doi:10.1126/science.aaz9236}%
  \BibitemOpen
  \bibfield  {author} {\bibinfo {author} {\bibfnamefont {D.}~\bibnamefont
  {Lachance-Quirion}}, \bibinfo {author} {\bibfnamefont {S.~P.}\ \bibnamefont
  {Wolski}}, \bibinfo {author} {\bibfnamefont {Y.}~\bibnamefont {Tabuchi}},
  \bibinfo {author} {\bibfnamefont {S.}~\bibnamefont {Kono}}, \bibinfo {author}
  {\bibfnamefont {K.}~\bibnamefont {Usami}}, \ and\ \bibinfo {author}
  {\bibfnamefont {Y.}~\bibnamefont {Nakamura}},\ }\href {\doibase
  10.1126/science.aaz9236} {\bibfield  {journal} {\bibinfo  {journal}
  {Science}\ }\textbf {\bibinfo {volume} {367}},\ \bibinfo {pages} {425}
  (\bibinfo {year} {2020})}\BibitemShut {NoStop}%
\bibitem [{\citenamefont {Zhang}\ \emph
  {et~al.}(2016{\natexlab{b}})\citenamefont {Zhang}, \citenamefont {Zou},
  \citenamefont {Jiang},\ and\ \citenamefont
  {Tang}}]{doi:10.1126/sciadv.1501286}%
  \BibitemOpen
  \bibfield  {author} {\bibinfo {author} {\bibfnamefont {X.}~\bibnamefont
  {Zhang}}, \bibinfo {author} {\bibfnamefont {C.-L.}\ \bibnamefont {Zou}},
  \bibinfo {author} {\bibfnamefont {L.}~\bibnamefont {Jiang}}, \ and\ \bibinfo
  {author} {\bibfnamefont {H.~X.}\ \bibnamefont {Tang}},\ }\href {\doibase
  10.1126/sciadv.1501286} {\bibfield  {journal} {\bibinfo  {journal} {Sci.
  Adv.}\ }\textbf {\bibinfo {volume} {2}},\ \bibinfo {pages} {e1501286}
  (\bibinfo {year} {2016}{\natexlab{b}})}\BibitemShut {NoStop}%
\bibitem [{\citenamefont {Li}\ \emph {et~al.}(2018)\citenamefont {Li},
  \citenamefont {Zhu},\ and\ \citenamefont {Agarwal}}]{PhysRevLett.121.203601}%
  \BibitemOpen
  \bibfield  {author} {\bibinfo {author} {\bibfnamefont {J.}~\bibnamefont
  {Li}}, \bibinfo {author} {\bibfnamefont {S.-Y.}\ \bibnamefont {Zhu}}, \ and\
  \bibinfo {author} {\bibfnamefont {G.~S.}\ \bibnamefont {Agarwal}},\ }\href
  {\doibase 10.1103/PhysRevLett.121.203601} {\bibfield  {journal} {\bibinfo
  {journal} {Phys. Rev. Lett.}\ }\textbf {\bibinfo {volume} {121}},\ \bibinfo
  {pages} {203601} (\bibinfo {year} {2018})}\BibitemShut {NoStop}%
\bibitem [{\citenamefont {Yu}\ \emph {et~al.}(2020)\citenamefont {Yu},
  \citenamefont {Shen},\ and\ \citenamefont {Li}}]{PhysRevLett.124.213604}%
  \BibitemOpen
  \bibfield  {author} {\bibinfo {author} {\bibfnamefont {M.}~\bibnamefont
  {Yu}}, \bibinfo {author} {\bibfnamefont {H.}~\bibnamefont {Shen}}, \ and\
  \bibinfo {author} {\bibfnamefont {J.}~\bibnamefont {Li}},\ }\href {\doibase
  10.1103/PhysRevLett.124.213604} {\bibfield  {journal} {\bibinfo  {journal}
  {Phys. Rev. Lett.}\ }\textbf {\bibinfo {volume} {124}},\ \bibinfo {pages}
  {213604} (\bibinfo {year} {2020})}\BibitemShut {NoStop}%
\bibitem [{\citenamefont {Shen}\ \emph {et~al.}(2022)\citenamefont {Shen},
  \citenamefont {Xu}, \citenamefont {Zhang}, \citenamefont {Zhang},
  \citenamefont {Wang}, \citenamefont {Chai}, \citenamefont {Zou},
  \citenamefont {Guo},\ and\ \citenamefont {Dong}}]{PhysRevLett.129.243601}%
  \BibitemOpen
  \bibfield  {author} {\bibinfo {author} {\bibfnamefont {Z.}~\bibnamefont
  {Shen}}, \bibinfo {author} {\bibfnamefont {G.-T.}\ \bibnamefont {Xu}},
  \bibinfo {author} {\bibfnamefont {M.}~\bibnamefont {Zhang}}, \bibinfo
  {author} {\bibfnamefont {Y.-L.}\ \bibnamefont {Zhang}}, \bibinfo {author}
  {\bibfnamefont {Y.}~\bibnamefont {Wang}}, \bibinfo {author} {\bibfnamefont
  {C.-Z.}\ \bibnamefont {Chai}}, \bibinfo {author} {\bibfnamefont {C.-L.}\
  \bibnamefont {Zou}}, \bibinfo {author} {\bibfnamefont {G.-C.}\ \bibnamefont
  {Guo}}, \ and\ \bibinfo {author} {\bibfnamefont {C.-H.}\ \bibnamefont
  {Dong}},\ }\href {\doibase 10.1103/PhysRevLett.129.243601} {\bibfield
  {journal} {\bibinfo  {journal} {Phys. Rev. Lett.}\ }\textbf {\bibinfo
  {volume} {129}},\ \bibinfo {pages} {243601} (\bibinfo {year}
  {2022})}\BibitemShut {NoStop}%
\bibitem [{\citenamefont {Kurizki}\ \emph {et~al.}(2015)\citenamefont
  {Kurizki}, \citenamefont {Bertet}, \citenamefont {Kubo}, \citenamefont
  {M{\o}lmer}, \citenamefont {Petrosyan}, \citenamefont {Rabl},\ and\
  \citenamefont {Schmiedmayer}}]{Kurizki3866}%
  \BibitemOpen
  \bibfield  {author} {\bibinfo {author} {\bibfnamefont {G.}~\bibnamefont
  {Kurizki}}, \bibinfo {author} {\bibfnamefont {P.}~\bibnamefont {Bertet}},
  \bibinfo {author} {\bibfnamefont {Y.}~\bibnamefont {Kubo}}, \bibinfo {author}
  {\bibfnamefont {K.}~\bibnamefont {M{\o}lmer}}, \bibinfo {author}
  {\bibfnamefont {D.}~\bibnamefont {Petrosyan}}, \bibinfo {author}
  {\bibfnamefont {P.}~\bibnamefont {Rabl}}, \ and\ \bibinfo {author}
  {\bibfnamefont {J.}~\bibnamefont {Schmiedmayer}},\ }\href {\doibase
  10.1073/pnas.1419326112} {\bibfield  {journal} {\bibinfo  {journal} {Proc.
  Natl Acad. Sci. U.S.A.}\ }\textbf {\bibinfo {volume} {112}},\ \bibinfo
  {pages} {3866} (\bibinfo {year} {2015})}\BibitemShut {NoStop}%
\bibitem [{\citenamefont {Hisatomi}\ \emph {et~al.}(2016)\citenamefont
  {Hisatomi}, \citenamefont {Osada}, \citenamefont {Tabuchi}, \citenamefont
  {Ishikawa}, \citenamefont {Noguchi}, \citenamefont {Yamazaki}, \citenamefont
  {Usami},\ and\ \citenamefont {Nakamura}}]{PhysRevB.93.174427}%
  \BibitemOpen
  \bibfield  {author} {\bibinfo {author} {\bibfnamefont {R.}~\bibnamefont
  {Hisatomi}}, \bibinfo {author} {\bibfnamefont {A.}~\bibnamefont {Osada}},
  \bibinfo {author} {\bibfnamefont {Y.}~\bibnamefont {Tabuchi}}, \bibinfo
  {author} {\bibfnamefont {T.}~\bibnamefont {Ishikawa}}, \bibinfo {author}
  {\bibfnamefont {A.}~\bibnamefont {Noguchi}}, \bibinfo {author} {\bibfnamefont
  {R.}~\bibnamefont {Yamazaki}}, \bibinfo {author} {\bibfnamefont
  {K.}~\bibnamefont {Usami}}, \ and\ \bibinfo {author} {\bibfnamefont
  {Y.}~\bibnamefont {Nakamura}},\ }\href {\doibase 10.1103/PhysRevB.93.174427}
  {\bibfield  {journal} {\bibinfo  {journal} {Phys. Rev. B}\ }\textbf {\bibinfo
  {volume} {93}},\ \bibinfo {pages} {174427} (\bibinfo {year}
  {2016})}\BibitemShut {NoStop}%
\bibitem [{\citenamefont {Zhu}\ \emph {et~al.}(2020{\natexlab{a}})\citenamefont
  {Zhu}, \citenamefont {Zhang}, \citenamefont {Han}, \citenamefont {Zou},
  \citenamefont {Zhong}, \citenamefont {Wang}, \citenamefont {Jiang},\ and\
  \citenamefont {Tang}}]{Zhu:20}%
  \BibitemOpen
  \bibfield  {author} {\bibinfo {author} {\bibfnamefont {N.}~\bibnamefont
  {Zhu}}, \bibinfo {author} {\bibfnamefont {X.}~\bibnamefont {Zhang}}, \bibinfo
  {author} {\bibfnamefont {X.}~\bibnamefont {Han}}, \bibinfo {author}
  {\bibfnamefont {C.-L.}\ \bibnamefont {Zou}}, \bibinfo {author} {\bibfnamefont
  {C.}~\bibnamefont {Zhong}}, \bibinfo {author} {\bibfnamefont {C.-H.}\
  \bibnamefont {Wang}}, \bibinfo {author} {\bibfnamefont {L.}~\bibnamefont
  {Jiang}}, \ and\ \bibinfo {author} {\bibfnamefont {H.~X.}\ \bibnamefont
  {Tang}},\ }\href {\doibase 10.1364/OPTICA.397967} {\bibfield  {journal}
  {\bibinfo  {journal} {Optica}\ }\textbf {\bibinfo {volume} {7}},\ \bibinfo
  {pages} {1291} (\bibinfo {year} {2020}{\natexlab{a}})}\BibitemShut {NoStop}%
\bibitem [{\citenamefont {Chai}\ \emph {et~al.}(2022)\citenamefont {Chai},
  \citenamefont {Shen}, \citenamefont {Zhang}, \citenamefont {Zhao},
  \citenamefont {Guo}, \citenamefont {Zou},\ and\ \citenamefont
  {Dong}}]{Chai:22}%
  \BibitemOpen
  \bibfield  {author} {\bibinfo {author} {\bibfnamefont {C.-Z.}\ \bibnamefont
  {Chai}}, \bibinfo {author} {\bibfnamefont {Z.}~\bibnamefont {Shen}}, \bibinfo
  {author} {\bibfnamefont {Y.-L.}\ \bibnamefont {Zhang}}, \bibinfo {author}
  {\bibfnamefont {H.-Q.}\ \bibnamefont {Zhao}}, \bibinfo {author}
  {\bibfnamefont {G.-C.}\ \bibnamefont {Guo}}, \bibinfo {author} {\bibfnamefont
  {C.-L.}\ \bibnamefont {Zou}}, \ and\ \bibinfo {author} {\bibfnamefont
  {C.-H.}\ \bibnamefont {Dong}},\ }\href {\doibase 10.1364/PRJ.446226}
  {\bibfield  {journal} {\bibinfo  {journal} {Photon. Res.}\ }\textbf {\bibinfo
  {volume} {10}},\ \bibinfo {pages} {820} (\bibinfo {year} {2022})}\BibitemShut
  {NoStop}%
\bibitem [{\citenamefont {Cao}\ and\ \citenamefont
  {Yan}(2019)}]{PhysRevB.99.214415}%
  \BibitemOpen
  \bibfield  {author} {\bibinfo {author} {\bibfnamefont {Y.}~\bibnamefont
  {Cao}}\ and\ \bibinfo {author} {\bibfnamefont {P.}~\bibnamefont {Yan}},\
  }\href {\doibase 10.1103/PhysRevB.99.214415} {\bibfield  {journal} {\bibinfo
  {journal} {Phys. Rev. B}\ }\textbf {\bibinfo {volume} {99}},\ \bibinfo
  {pages} {214415} (\bibinfo {year} {2019})}\BibitemShut {NoStop}%
\bibitem [{\citenamefont {Colombano}\ \emph {et~al.}(2020)\citenamefont
  {Colombano}, \citenamefont {Arregui}, \citenamefont {Bonell}, \citenamefont
  {Capuj}, \citenamefont {Chavez-Angel}, \citenamefont {Pitanti}, \citenamefont
  {Valenzuela}, \citenamefont {Sotomayor-Torres}, \citenamefont
  {Navarro-Urrios},\ and\ \citenamefont {Costache}}]{PhysRevLett.125.147201}%
  \BibitemOpen
  \bibfield  {author} {\bibinfo {author} {\bibfnamefont {M.~F.}\ \bibnamefont
  {Colombano}}, \bibinfo {author} {\bibfnamefont {G.}~\bibnamefont {Arregui}},
  \bibinfo {author} {\bibfnamefont {F.}~\bibnamefont {Bonell}}, \bibinfo
  {author} {\bibfnamefont {N.~E.}\ \bibnamefont {Capuj}}, \bibinfo {author}
  {\bibfnamefont {E.}~\bibnamefont {Chavez-Angel}}, \bibinfo {author}
  {\bibfnamefont {A.}~\bibnamefont {Pitanti}}, \bibinfo {author} {\bibfnamefont
  {S.~O.}\ \bibnamefont {Valenzuela}}, \bibinfo {author} {\bibfnamefont
  {C.~M.}\ \bibnamefont {Sotomayor-Torres}}, \bibinfo {author} {\bibfnamefont
  {D.}~\bibnamefont {Navarro-Urrios}}, \ and\ \bibinfo {author} {\bibfnamefont
  {M.~V.}\ \bibnamefont {Costache}},\ }\href {\doibase
  10.1103/PhysRevLett.125.147201} {\bibfield  {journal} {\bibinfo  {journal}
  {Phys. Rev. Lett.}\ }\textbf {\bibinfo {volume} {125}},\ \bibinfo {pages}
  {147201} (\bibinfo {year} {2020})}\BibitemShut {NoStop}%
\bibitem [{\citenamefont {Wolski}\ \emph {et~al.}(2020)\citenamefont {Wolski},
  \citenamefont {Lachance-Quirion}, \citenamefont {Tabuchi}, \citenamefont
  {Kono}, \citenamefont {Noguchi}, \citenamefont {Usami},\ and\ \citenamefont
  {Nakamura}}]{PhysRevLett.125.117701}%
  \BibitemOpen
  \bibfield  {author} {\bibinfo {author} {\bibfnamefont {S.~P.}\ \bibnamefont
  {Wolski}}, \bibinfo {author} {\bibfnamefont {D.}~\bibnamefont
  {Lachance-Quirion}}, \bibinfo {author} {\bibfnamefont {Y.}~\bibnamefont
  {Tabuchi}}, \bibinfo {author} {\bibfnamefont {S.}~\bibnamefont {Kono}},
  \bibinfo {author} {\bibfnamefont {A.}~\bibnamefont {Noguchi}}, \bibinfo
  {author} {\bibfnamefont {K.}~\bibnamefont {Usami}}, \ and\ \bibinfo {author}
  {\bibfnamefont {Y.}~\bibnamefont {Nakamura}},\ }\href {\doibase
  10.1103/PhysRevLett.125.117701} {\bibfield  {journal} {\bibinfo  {journal}
  {Phys. Rev. Lett.}\ }\textbf {\bibinfo {volume} {125}},\ \bibinfo {pages}
  {117701} (\bibinfo {year} {2020})}\BibitemShut {NoStop}%
\bibitem [{\citenamefont {Crescini}\ \emph {et~al.}(2020)\citenamefont
  {Crescini}, \citenamefont {Braggio}, \citenamefont {Carugno}, \citenamefont
  {Ortolan},\ and\ \citenamefont {Ruoso}}]{doi:10.1063/5.0024369}%
  \BibitemOpen
  \bibfield  {author} {\bibinfo {author} {\bibfnamefont {N.}~\bibnamefont
  {Crescini}}, \bibinfo {author} {\bibfnamefont {C.}~\bibnamefont {Braggio}},
  \bibinfo {author} {\bibfnamefont {G.}~\bibnamefont {Carugno}}, \bibinfo
  {author} {\bibfnamefont {A.}~\bibnamefont {Ortolan}}, \ and\ \bibinfo
  {author} {\bibfnamefont {G.}~\bibnamefont {Ruoso}},\ }\href {\doibase
  10.1063/5.0024369} {\bibfield  {journal} {\bibinfo  {journal} {Appl. Phys.
  Lett.}\ }\textbf {\bibinfo {volume} {117}},\ \bibinfo {pages} {144001}
  (\bibinfo {year} {2020})}\BibitemShut {NoStop}%
\bibitem [{\citenamefont {Lu}\ \emph {et~al.}(2021)\citenamefont {Lu},
  \citenamefont {Zhang}, \citenamefont {Zhang},\ and\ \citenamefont
  {Jing}}]{PhysRevA.103.063708}%
  \BibitemOpen
  \bibfield  {author} {\bibinfo {author} {\bibfnamefont {T.-X.}\ \bibnamefont
  {Lu}}, \bibinfo {author} {\bibfnamefont {H.}~\bibnamefont {Zhang}}, \bibinfo
  {author} {\bibfnamefont {Q.}~\bibnamefont {Zhang}}, \ and\ \bibinfo {author}
  {\bibfnamefont {H.}~\bibnamefont {Jing}},\ }\href {\doibase
  10.1103/PhysRevA.103.063708} {\bibfield  {journal} {\bibinfo  {journal}
  {Phys. Rev. A}\ }\textbf {\bibinfo {volume} {103}},\ \bibinfo {pages}
  {063708} (\bibinfo {year} {2021})}\BibitemShut {NoStop}%
\bibitem [{\citenamefont {Gonzalez-Ballestero}\ \emph
  {et~al.}(2020{\natexlab{a}})\citenamefont {Gonzalez-Ballestero},
  \citenamefont {Gieseler},\ and\ \citenamefont
  {Romero-Isart}}]{PhysRevLett.124.093602}%
  \BibitemOpen
  \bibfield  {author} {\bibinfo {author} {\bibfnamefont {C.}~\bibnamefont
  {Gonzalez-Ballestero}}, \bibinfo {author} {\bibfnamefont {J.}~\bibnamefont
  {Gieseler}}, \ and\ \bibinfo {author} {\bibfnamefont {O.}~\bibnamefont
  {Romero-Isart}},\ }\href {\doibase 10.1103/PhysRevLett.124.093602} {\bibfield
   {journal} {\bibinfo  {journal} {Phys. Rev. Lett.}\ }\textbf {\bibinfo
  {volume} {124}},\ \bibinfo {pages} {093602} (\bibinfo {year}
  {2020}{\natexlab{a}})}\BibitemShut {NoStop}%
\bibitem [{\citenamefont {Gonzalez-Ballestero}\ \emph
  {et~al.}(2020{\natexlab{b}})\citenamefont {Gonzalez-Ballestero},
  \citenamefont {H\"ummer}, \citenamefont {Gieseler},\ and\ \citenamefont
  {Romero-Isart}}]{PhysRevB.101.125404}%
  \BibitemOpen
  \bibfield  {author} {\bibinfo {author} {\bibfnamefont {C.}~\bibnamefont
  {Gonzalez-Ballestero}}, \bibinfo {author} {\bibfnamefont {D.}~\bibnamefont
  {H\"ummer}}, \bibinfo {author} {\bibfnamefont {J.}~\bibnamefont {Gieseler}},
  \ and\ \bibinfo {author} {\bibfnamefont {O.}~\bibnamefont {Romero-Isart}},\
  }\href {\doibase 10.1103/PhysRevB.101.125404} {\bibfield  {journal} {\bibinfo
   {journal} {Phys. Rev. B}\ }\textbf {\bibinfo {volume} {101}},\ \bibinfo
  {pages} {125404} (\bibinfo {year} {2020}{\natexlab{b}})}\BibitemShut
  {NoStop}%
\bibitem [{\citenamefont {Holanda}\ \emph {et~al.}(2018)\citenamefont
  {Holanda}, \citenamefont {Maior}, \citenamefont {Azevedo},\ and\
  \citenamefont {Rezende}}]{holanda2018detecting}%
  \BibitemOpen
  \bibfield  {author} {\bibinfo {author} {\bibfnamefont {J.}~\bibnamefont
  {Holanda}}, \bibinfo {author} {\bibfnamefont {D.}~\bibnamefont {Maior}},
  \bibinfo {author} {\bibfnamefont {A.}~\bibnamefont {Azevedo}}, \ and\
  \bibinfo {author} {\bibfnamefont {S.}~\bibnamefont {Rezende}},\ }\href
  {https://www.nature.com/articles/s41567-018-0079-y} {\bibfield  {journal}
  {\bibinfo  {journal} {Nature Phys.}\ }\textbf {\bibinfo {volume} {14}},\
  \bibinfo {pages} {500} (\bibinfo {year} {2018})}\BibitemShut {NoStop}%
\bibitem [{\citenamefont {Bozhko}\ \emph {et~al.}(2020)\citenamefont {Bozhko},
  \citenamefont {Vasyuchka}, \citenamefont {Chumak},\ and\ \citenamefont
  {Serga}}]{doi:10.1063/10.0000872}%
  \BibitemOpen
  \bibfield  {author} {\bibinfo {author} {\bibfnamefont {D.~A.}\ \bibnamefont
  {Bozhko}}, \bibinfo {author} {\bibfnamefont {V.~I.}\ \bibnamefont
  {Vasyuchka}}, \bibinfo {author} {\bibfnamefont {A.~V.}\ \bibnamefont
  {Chumak}}, \ and\ \bibinfo {author} {\bibfnamefont {A.~A.}\ \bibnamefont
  {Serga}},\ }\href {\doibase 10.1063/10.0000872} {\bibfield  {journal}
  {\bibinfo  {journal} {Low Temp. Phys.}\ }\textbf {\bibinfo {volume} {46}},\
  \bibinfo {pages} {383} (\bibinfo {year} {2020})}\BibitemShut {NoStop}%
\bibitem [{\citenamefont {Neuman}\ \emph {et~al.}(2020)\citenamefont {Neuman},
  \citenamefont {Wang},\ and\ \citenamefont {Narang}}]{PhysRevLett.125.247702}%
  \BibitemOpen
  \bibfield  {author} {\bibinfo {author} {\bibfnamefont {T.~c.~v.}\
  \bibnamefont {Neuman}}, \bibinfo {author} {\bibfnamefont {D.~S.}\
  \bibnamefont {Wang}}, \ and\ \bibinfo {author} {\bibfnamefont
  {P.}~\bibnamefont {Narang}},\ }\href {\doibase
  10.1103/PhysRevLett.125.247702} {\bibfield  {journal} {\bibinfo  {journal}
  {Phys. Rev. Lett.}\ }\textbf {\bibinfo {volume} {125}},\ \bibinfo {pages}
  {247702} (\bibinfo {year} {2020})}\BibitemShut {NoStop}%
\bibitem [{\citenamefont {Wang}\ \emph {et~al.}(2021)\citenamefont {Wang},
  \citenamefont {Neuman},\ and\ \citenamefont
  {Narang}}]{doi:10.1021/acs.jpcc.0c11536}%
  \BibitemOpen
  \bibfield  {author} {\bibinfo {author} {\bibfnamefont {D.~S.}\ \bibnamefont
  {Wang}}, \bibinfo {author} {\bibfnamefont {T.}~\bibnamefont {Neuman}}, \ and\
  \bibinfo {author} {\bibfnamefont {P.}~\bibnamefont {Narang}},\ }\href
  {\doibase 10.1021/acs.jpcc.0c11536} {\bibfield  {journal} {\bibinfo
  {journal} {J. Phys. Chem. C}\ }\textbf {\bibinfo {volume} {125}},\ \bibinfo
  {pages} {6222} (\bibinfo {year} {2021})}\BibitemShut {NoStop}%
\bibitem [{\citenamefont {Skogvoll}\ \emph {et~al.}(2021)\citenamefont
  {Skogvoll}, \citenamefont {Lidal}, \citenamefont {Danon},\ and\ \citenamefont
  {Kamra}}]{PhysRevApplied.16.064008}%
  \BibitemOpen
  \bibfield  {author} {\bibinfo {author} {\bibfnamefont {I.~C.}\ \bibnamefont
  {Skogvoll}}, \bibinfo {author} {\bibfnamefont {J.}~\bibnamefont {Lidal}},
  \bibinfo {author} {\bibfnamefont {J.}~\bibnamefont {Danon}}, \ and\ \bibinfo
  {author} {\bibfnamefont {A.}~\bibnamefont {Kamra}},\ }\href {\doibase
  10.1103/PhysRevApplied.16.064008} {\bibfield  {journal} {\bibinfo  {journal}
  {Phys. Rev. Appl.}\ }\textbf {\bibinfo {volume} {16}},\ \bibinfo {pages}
  {064008} (\bibinfo {year} {2021})}\BibitemShut {NoStop}%
\bibitem [{\citenamefont {Fukami}\ \emph {et~al.}(2021)\citenamefont {Fukami},
  \citenamefont {Candido}, \citenamefont {Awschalom},\ and\ \citenamefont
  {Flatt\'e}}]{PRXQuantum.2.040314}%
  \BibitemOpen
  \bibfield  {author} {\bibinfo {author} {\bibfnamefont {M.}~\bibnamefont
  {Fukami}}, \bibinfo {author} {\bibfnamefont {D.~R.}\ \bibnamefont {Candido}},
  \bibinfo {author} {\bibfnamefont {D.~D.}\ \bibnamefont {Awschalom}}, \ and\
  \bibinfo {author} {\bibfnamefont {M.~E.}\ \bibnamefont {Flatt\'e}},\ }\href
  {\doibase 10.1103/PRXQuantum.2.040314} {\bibfield  {journal} {\bibinfo
  {journal} {PRX Quantum}\ }\textbf {\bibinfo {volume} {2}},\ \bibinfo {pages}
  {040314} (\bibinfo {year} {2021})}\BibitemShut {NoStop}%
\bibitem [{\citenamefont {Xiong}\ \emph {et~al.}(2022)\citenamefont {Xiong},
  \citenamefont {Tian}, \citenamefont {Zhang},\ and\ \citenamefont
  {You}}]{PhysRevB.105.245310}%
  \BibitemOpen
  \bibfield  {author} {\bibinfo {author} {\bibfnamefont {W.}~\bibnamefont
  {Xiong}}, \bibinfo {author} {\bibfnamefont {M.}~\bibnamefont {Tian}},
  \bibinfo {author} {\bibfnamefont {G.-Q.}\ \bibnamefont {Zhang}}, \ and\
  \bibinfo {author} {\bibfnamefont {J.~Q.}\ \bibnamefont {You}},\ }\href
  {\doibase 10.1103/PhysRevB.105.245310} {\bibfield  {journal} {\bibinfo
  {journal} {Phys. Rev. B}\ }\textbf {\bibinfo {volume} {105}},\ \bibinfo
  {pages} {245310} (\bibinfo {year} {2022})}\BibitemShut {NoStop}%
\bibitem [{\citenamefont {L\"u}\ \emph {et~al.}(2015)\citenamefont {L\"u},
  \citenamefont {Wu}, \citenamefont {Johansson}, \citenamefont {Jing},
  \citenamefont {Zhang},\ and\ \citenamefont {Nori}}]{PhysRevLett.114.093602}%
  \BibitemOpen
  \bibfield  {author} {\bibinfo {author} {\bibfnamefont {X.-Y.}\ \bibnamefont
  {L\"u}}, \bibinfo {author} {\bibfnamefont {Y.}~\bibnamefont {Wu}}, \bibinfo
  {author} {\bibfnamefont {J.~R.}\ \bibnamefont {Johansson}}, \bibinfo {author}
  {\bibfnamefont {H.}~\bibnamefont {Jing}}, \bibinfo {author} {\bibfnamefont
  {J.}~\bibnamefont {Zhang}}, \ and\ \bibinfo {author} {\bibfnamefont
  {F.}~\bibnamefont {Nori}},\ }\href {\doibase 10.1103/PhysRevLett.114.093602}
  {\bibfield  {journal} {\bibinfo  {journal} {Phys. Rev. Lett.}\ }\textbf
  {\bibinfo {volume} {114}},\ \bibinfo {pages} {093602} (\bibinfo {year}
  {2015})}\BibitemShut {NoStop}%
\bibitem [{\citenamefont {Qin}\ \emph {et~al.}(2019)\citenamefont {Qin},
  \citenamefont {Macr\`{\i}}, \citenamefont {Miranowicz}, \citenamefont
  {Savasta},\ and\ \citenamefont {Nori}}]{PhysRevA.100.062501}%
  \BibitemOpen
  \bibfield  {author} {\bibinfo {author} {\bibfnamefont {W.}~\bibnamefont
  {Qin}}, \bibinfo {author} {\bibfnamefont {V.}~\bibnamefont {Macr\`{\i}}},
  \bibinfo {author} {\bibfnamefont {A.}~\bibnamefont {Miranowicz}}, \bibinfo
  {author} {\bibfnamefont {S.}~\bibnamefont {Savasta}}, \ and\ \bibinfo
  {author} {\bibfnamefont {F.}~\bibnamefont {Nori}},\ }\href {\doibase
  10.1103/PhysRevA.100.062501} {\bibfield  {journal} {\bibinfo  {journal}
  {Phys. Rev. A}\ }\textbf {\bibinfo {volume} {100}},\ \bibinfo {pages}
  {062501} (\bibinfo {year} {2019})}\BibitemShut {NoStop}%
\bibitem [{\citenamefont {Qin}\ \emph {et~al.}(2018)\citenamefont {Qin},
  \citenamefont {Miranowicz}, \citenamefont {Li}, \citenamefont {L\"u},
  \citenamefont {You},\ and\ \citenamefont {Nori}}]{PhysRevLett.120.093601}%
  \BibitemOpen
  \bibfield  {author} {\bibinfo {author} {\bibfnamefont {W.}~\bibnamefont
  {Qin}}, \bibinfo {author} {\bibfnamefont {A.}~\bibnamefont {Miranowicz}},
  \bibinfo {author} {\bibfnamefont {P.-B.}\ \bibnamefont {Li}}, \bibinfo
  {author} {\bibfnamefont {X.-Y.}\ \bibnamefont {L\"u}}, \bibinfo {author}
  {\bibfnamefont {J.~Q.}\ \bibnamefont {You}}, \ and\ \bibinfo {author}
  {\bibfnamefont {F.}~\bibnamefont {Nori}},\ }\href {\doibase
  10.1103/PhysRevLett.120.093601} {\bibfield  {journal} {\bibinfo  {journal}
  {Phys. Rev. Lett.}\ }\textbf {\bibinfo {volume} {120}},\ \bibinfo {pages}
  {093601} (\bibinfo {year} {2018})}\BibitemShut {NoStop}%
\bibitem [{\citenamefont {Leroux}\ \emph {et~al.}(2018)\citenamefont {Leroux},
  \citenamefont {Govia},\ and\ \citenamefont {Clerk}}]{PhysRevLett.120.093602}%
  \BibitemOpen
  \bibfield  {author} {\bibinfo {author} {\bibfnamefont {C.}~\bibnamefont
  {Leroux}}, \bibinfo {author} {\bibfnamefont {L.~C.~G.}\ \bibnamefont
  {Govia}}, \ and\ \bibinfo {author} {\bibfnamefont {A.~A.}\ \bibnamefont
  {Clerk}},\ }\href {\doibase 10.1103/PhysRevLett.120.093602} {\bibfield
  {journal} {\bibinfo  {journal} {Phys. Rev. Lett.}\ }\textbf {\bibinfo
  {volume} {120}},\ \bibinfo {pages} {093602} (\bibinfo {year}
  {2018})}\BibitemShut {NoStop}%
\bibitem [{\citenamefont {Chen}\ \emph {et~al.}(2021)\citenamefont {Chen},
  \citenamefont {Qin}, \citenamefont {Wang}, \citenamefont {Miranowicz},\ and\
  \citenamefont {Nori}}]{PhysRevLett.126.023602}%
  \BibitemOpen
  \bibfield  {author} {\bibinfo {author} {\bibfnamefont {Y.-H.}\ \bibnamefont
  {Chen}}, \bibinfo {author} {\bibfnamefont {W.}~\bibnamefont {Qin}}, \bibinfo
  {author} {\bibfnamefont {X.}~\bibnamefont {Wang}}, \bibinfo {author}
  {\bibfnamefont {A.}~\bibnamefont {Miranowicz}}, \ and\ \bibinfo {author}
  {\bibfnamefont {F.}~\bibnamefont {Nori}},\ }\href {\doibase
  10.1103/PhysRevLett.126.023602} {\bibfield  {journal} {\bibinfo  {journal}
  {Phys. Rev. Lett.}\ }\textbf {\bibinfo {volume} {126}},\ \bibinfo {pages}
  {023602} (\bibinfo {year} {2021})}\BibitemShut {NoStop}%
\bibitem [{\citenamefont {Tang}\ \emph {et~al.}(2022)\citenamefont {Tang},
  \citenamefont {Tang}, \citenamefont {Chen}, \citenamefont {Nori},
  \citenamefont {Xiao},\ and\ \citenamefont {Xia}}]{PhysRevLett.128.083604}%
  \BibitemOpen
  \bibfield  {author} {\bibinfo {author} {\bibfnamefont {L.}~\bibnamefont
  {Tang}}, \bibinfo {author} {\bibfnamefont {J.}~\bibnamefont {Tang}}, \bibinfo
  {author} {\bibfnamefont {M.}~\bibnamefont {Chen}}, \bibinfo {author}
  {\bibfnamefont {F.}~\bibnamefont {Nori}}, \bibinfo {author} {\bibfnamefont
  {M.}~\bibnamefont {Xiao}}, \ and\ \bibinfo {author} {\bibfnamefont
  {K.}~\bibnamefont {Xia}},\ }\href {\doibase 10.1103/PhysRevLett.128.083604}
  {\bibfield  {journal} {\bibinfo  {journal} {Phys. Rev. Lett.}\ }\textbf
  {\bibinfo {volume} {128}},\ \bibinfo {pages} {083604} (\bibinfo {year}
  {2022})}\BibitemShut {NoStop}%
\bibitem [{\citenamefont {Li}\ \emph {et~al.}(2020{\natexlab{b}})\citenamefont
  {Li}, \citenamefont {Zhou}, \citenamefont {Gao},\ and\ \citenamefont
  {Nori}}]{PhysRevLett.125.153602}%
  \BibitemOpen
  \bibfield  {author} {\bibinfo {author} {\bibfnamefont {P.-B.}\ \bibnamefont
  {Li}}, \bibinfo {author} {\bibfnamefont {Y.}~\bibnamefont {Zhou}}, \bibinfo
  {author} {\bibfnamefont {W.-B.}\ \bibnamefont {Gao}}, \ and\ \bibinfo
  {author} {\bibfnamefont {F.}~\bibnamefont {Nori}},\ }\href {\doibase
  10.1103/PhysRevLett.125.153602} {\bibfield  {journal} {\bibinfo  {journal}
  {Phys. Rev. Lett.}\ }\textbf {\bibinfo {volume} {125}},\ \bibinfo {pages}
  {153602} (\bibinfo {year} {2020}{\natexlab{b}})}\BibitemShut {NoStop}%
\bibitem [{\citenamefont {Wang}\ \emph {et~al.}(2022)\citenamefont {Wang},
  \citenamefont {Wu}, \citenamefont {Han}, \citenamefont {Xia}, \citenamefont
  {Jiang},\ and\ \citenamefont {Song}}]{PhysRevApplied.17.024009}%
  \BibitemOpen
  \bibfield  {author} {\bibinfo {author} {\bibfnamefont {Y.}~\bibnamefont
  {Wang}}, \bibinfo {author} {\bibfnamefont {J.-L.}\ \bibnamefont {Wu}},
  \bibinfo {author} {\bibfnamefont {J.-X.}\ \bibnamefont {Han}}, \bibinfo
  {author} {\bibfnamefont {Y.}~\bibnamefont {Xia}}, \bibinfo {author}
  {\bibfnamefont {Y.-Y.}\ \bibnamefont {Jiang}}, \ and\ \bibinfo {author}
  {\bibfnamefont {J.}~\bibnamefont {Song}},\ }\href {\doibase
  10.1103/PhysRevApplied.17.024009} {\bibfield  {journal} {\bibinfo  {journal}
  {Phys. Rev. Applied}\ }\textbf {\bibinfo {volume} {17}},\ \bibinfo {pages}
  {024009} (\bibinfo {year} {2022})}\BibitemShut {NoStop}%
\bibitem [{\citenamefont {Lemonde}\ \emph {et~al.}(2016)\citenamefont
  {Lemonde}, \citenamefont {Didier},\ and\ \citenamefont
  {Clerk}}]{lemonde2016enhanced}%
  \BibitemOpen
  \bibfield  {author} {\bibinfo {author} {\bibfnamefont {M.-A.}\ \bibnamefont
  {Lemonde}}, \bibinfo {author} {\bibfnamefont {N.}~\bibnamefont {Didier}}, \
  and\ \bibinfo {author} {\bibfnamefont {A.~A.}\ \bibnamefont {Clerk}},\ }\href
  {https://www.nature.com/articles/ncomms11338/} {\bibfield  {journal}
  {\bibinfo  {journal} {Nat. Commun.}\ }\textbf {\bibinfo {volume} {7}},\
  \bibinfo {pages} {1} (\bibinfo {year} {2016})}\BibitemShut {NoStop}%
\bibitem [{\citenamefont {Zhao}\ \emph {et~al.}(2020)\citenamefont {Zhao},
  \citenamefont {Zhang}, \citenamefont {Miranowicz},\ and\ \citenamefont
  {Jing}}]{zhao2020weak}%
  \BibitemOpen
  \bibfield  {author} {\bibinfo {author} {\bibfnamefont {W.}~\bibnamefont
  {Zhao}}, \bibinfo {author} {\bibfnamefont {S.-D.}\ \bibnamefont {Zhang}},
  \bibinfo {author} {\bibfnamefont {A.}~\bibnamefont {Miranowicz}}, \ and\
  \bibinfo {author} {\bibfnamefont {H.}~\bibnamefont {Jing}},\ }\href
  {https://link.springer.com/article/10.1007/s11433-019-9451-3} {\bibfield
  {journal} {\bibinfo  {journal} {Sci. China Phys. Mech. Astron.}\ }\textbf
  {\bibinfo {volume} {63}},\ \bibinfo {pages} {224211} (\bibinfo {year}
  {2020})}\BibitemShut {NoStop}%
\bibitem [{\citenamefont {Liu}\ \emph {et~al.}(2023)\citenamefont {Liu},
  \citenamefont {Jiao}, \citenamefont {Li}, \citenamefont {Xu}, \citenamefont
  {He},\ and\ \citenamefont {Jing}}]{liu2023phase}%
  \BibitemOpen
  \bibfield  {author} {\bibinfo {author} {\bibfnamefont {J.-X.}\ \bibnamefont
  {Liu}}, \bibinfo {author} {\bibfnamefont {Y.-F.}\ \bibnamefont {Jiao}},
  \bibinfo {author} {\bibfnamefont {Y.}~\bibnamefont {Li}}, \bibinfo {author}
  {\bibfnamefont {X.-W.}\ \bibnamefont {Xu}}, \bibinfo {author} {\bibfnamefont
  {Q.-Y.}\ \bibnamefont {He}}, \ and\ \bibinfo {author} {\bibfnamefont
  {H.}~\bibnamefont {Jing}},\ }\href
  {https://link.springer.com/article/10.1007/s11433-022-2043-3#citeas}
  {\bibfield  {journal} {\bibinfo  {journal} {Sci. China Phys. Mech. Astron.}\
  }\textbf {\bibinfo {volume} {66}},\ \bibinfo {pages} {230312} (\bibinfo
  {year} {2023})}\BibitemShut {NoStop}%
\bibitem [{\citenamefont {Wang}\ \emph {et~al.}(2019)\citenamefont {Wang},
  \citenamefont {Li}, \citenamefont {Sampuli}, \citenamefont {Song},
  \citenamefont {Jiang},\ and\ \citenamefont {Xia}}]{PhysRevA.99.023833}%
  \BibitemOpen
  \bibfield  {author} {\bibinfo {author} {\bibfnamefont {Y.}~\bibnamefont
  {Wang}}, \bibinfo {author} {\bibfnamefont {C.}~\bibnamefont {Li}}, \bibinfo
  {author} {\bibfnamefont {E.~M.}\ \bibnamefont {Sampuli}}, \bibinfo {author}
  {\bibfnamefont {J.}~\bibnamefont {Song}}, \bibinfo {author} {\bibfnamefont
  {Y.}~\bibnamefont {Jiang}}, \ and\ \bibinfo {author} {\bibfnamefont
  {Y.}~\bibnamefont {Xia}},\ }\href {\doibase 10.1103/PhysRevA.99.023833}
  {\bibfield  {journal} {\bibinfo  {journal} {Phys. Rev. A}\ }\textbf {\bibinfo
  {volume} {99}},\ \bibinfo {pages} {023833} (\bibinfo {year}
  {2019})}\BibitemShut {NoStop}%
\bibitem [{\citenamefont {Zhu}\ \emph {et~al.}(2020{\natexlab{b}})\citenamefont
  {Zhu}, \citenamefont {Ping}, \citenamefont {Yang},\ and\ \citenamefont
  {Agarwal}}]{PhysRevLett.124.073602}%
  \BibitemOpen
  \bibfield  {author} {\bibinfo {author} {\bibfnamefont {C.~J.}\ \bibnamefont
  {Zhu}}, \bibinfo {author} {\bibfnamefont {L.~L.}\ \bibnamefont {Ping}},
  \bibinfo {author} {\bibfnamefont {Y.~P.}\ \bibnamefont {Yang}}, \ and\
  \bibinfo {author} {\bibfnamefont {G.~S.}\ \bibnamefont {Agarwal}},\ }\href
  {\doibase 10.1103/PhysRevLett.124.073602} {\bibfield  {journal} {\bibinfo
  {journal} {Phys. Rev. Lett.}\ }\textbf {\bibinfo {volume} {124}},\ \bibinfo
  {pages} {073602} (\bibinfo {year} {2020}{\natexlab{b}})}\BibitemShut
  {NoStop}%
\bibitem [{\citenamefont {Groszkowski}\ \emph {et~al.}(2020)\citenamefont
  {Groszkowski}, \citenamefont {Lau}, \citenamefont {Leroux}, \citenamefont
  {Govia},\ and\ \citenamefont {Clerk}}]{PhysRevLett.125.203601}%
  \BibitemOpen
  \bibfield  {author} {\bibinfo {author} {\bibfnamefont {P.}~\bibnamefont
  {Groszkowski}}, \bibinfo {author} {\bibfnamefont {H.-K.}\ \bibnamefont
  {Lau}}, \bibinfo {author} {\bibfnamefont {C.}~\bibnamefont {Leroux}},
  \bibinfo {author} {\bibfnamefont {L.~C.~G.}\ \bibnamefont {Govia}}, \ and\
  \bibinfo {author} {\bibfnamefont {A.~A.}\ \bibnamefont {Clerk}},\ }\href
  {\doibase 10.1103/PhysRevLett.125.203601} {\bibfield  {journal} {\bibinfo
  {journal} {Phys. Rev. Lett.}\ }\textbf {\bibinfo {volume} {125}},\ \bibinfo
  {pages} {203601} (\bibinfo {year} {2020})}\BibitemShut {NoStop}%
\bibitem [{\citenamefont {Wang}\ \emph
  {et~al.}(2020{\natexlab{a}})\citenamefont {Wang}, \citenamefont {Wu},
  \citenamefont {Song}, \citenamefont {Zhang}, \citenamefont {Jiang},\ and\
  \citenamefont {Xia}}]{PhysRevA.101.053826}%
  \BibitemOpen
  \bibfield  {author} {\bibinfo {author} {\bibfnamefont {Y.}~\bibnamefont
  {Wang}}, \bibinfo {author} {\bibfnamefont {J.-L.}\ \bibnamefont {Wu}},
  \bibinfo {author} {\bibfnamefont {J.}~\bibnamefont {Song}}, \bibinfo {author}
  {\bibfnamefont {Z.-J.}\ \bibnamefont {Zhang}}, \bibinfo {author}
  {\bibfnamefont {Y.-Y.}\ \bibnamefont {Jiang}}, \ and\ \bibinfo {author}
  {\bibfnamefont {Y.}~\bibnamefont {Xia}},\ }\href {\doibase
  10.1103/PhysRevA.101.053826} {\bibfield  {journal} {\bibinfo  {journal}
  {Phys. Rev. A}\ }\textbf {\bibinfo {volume} {101}},\ \bibinfo {pages}
  {053826} (\bibinfo {year} {2020}{\natexlab{a}})}\BibitemShut {NoStop}%
\bibitem [{\citenamefont {Wang}\ \emph
  {et~al.}(2020{\natexlab{b}})\citenamefont {Wang}, \citenamefont {Wu},
  \citenamefont {Han}, \citenamefont {Jiang}, \citenamefont {Xia},\ and\
  \citenamefont {Song}}]{PhysRevA.102.032601}%
  \BibitemOpen
  \bibfield  {author} {\bibinfo {author} {\bibfnamefont {Y.}~\bibnamefont
  {Wang}}, \bibinfo {author} {\bibfnamefont {J.-L.}\ \bibnamefont {Wu}},
  \bibinfo {author} {\bibfnamefont {J.-X.}\ \bibnamefont {Han}}, \bibinfo
  {author} {\bibfnamefont {Y.-Y.}\ \bibnamefont {Jiang}}, \bibinfo {author}
  {\bibfnamefont {Y.}~\bibnamefont {Xia}}, \ and\ \bibinfo {author}
  {\bibfnamefont {J.}~\bibnamefont {Song}},\ }\href {\doibase
  10.1103/PhysRevA.102.032601} {\bibfield  {journal} {\bibinfo  {journal}
  {Phys. Rev. A}\ }\textbf {\bibinfo {volume} {102}},\ \bibinfo {pages}
  {032601} (\bibinfo {year} {2020}{\natexlab{b}})}\BibitemShut {NoStop}%
\bibitem [{\citenamefont {Qin}\ \emph {et~al.}(2021)\citenamefont {Qin},
  \citenamefont {Miranowicz}, \citenamefont {Jing},\ and\ \citenamefont
  {Nori}}]{PhysRevLett.127.093602}%
  \BibitemOpen
  \bibfield  {author} {\bibinfo {author} {\bibfnamefont {W.}~\bibnamefont
  {Qin}}, \bibinfo {author} {\bibfnamefont {A.}~\bibnamefont {Miranowicz}},
  \bibinfo {author} {\bibfnamefont {H.}~\bibnamefont {Jing}}, \ and\ \bibinfo
  {author} {\bibfnamefont {F.}~\bibnamefont {Nori}},\ }\href {\doibase
  10.1103/PhysRevLett.127.093602} {\bibfield  {journal} {\bibinfo  {journal}
  {Phys. Rev. Lett.}\ }\textbf {\bibinfo {volume} {127}},\ \bibinfo {pages}
  {093602} (\bibinfo {year} {2021})}\BibitemShut {NoStop}%
\bibitem [{\citenamefont {Villiers}\ \emph {et~al.}()\citenamefont {Villiers},
  \citenamefont {Smith}, \citenamefont {Petrescu}, \citenamefont {Borgognoni},
  \citenamefont {Delbecq}, \citenamefont {Sarlette}, \citenamefont {Mirrahimi},
  \citenamefont {Campagne-Ibarcq}, \citenamefont {Kontos},\ and\ \citenamefont
  {Leghtas}}]{villiers2023dynamically}%
  \BibitemOpen
  \bibfield  {author} {\bibinfo {author} {\bibfnamefont {M.}~\bibnamefont
  {Villiers}}, \bibinfo {author} {\bibfnamefont {W.~C.}\ \bibnamefont {Smith}},
  \bibinfo {author} {\bibfnamefont {A.}~\bibnamefont {Petrescu}}, \bibinfo
  {author} {\bibfnamefont {A.}~\bibnamefont {Borgognoni}}, \bibinfo {author}
  {\bibfnamefont {M.}~\bibnamefont {Delbecq}}, \bibinfo {author} {\bibfnamefont
  {A.}~\bibnamefont {Sarlette}}, \bibinfo {author} {\bibfnamefont
  {M.}~\bibnamefont {Mirrahimi}}, \bibinfo {author} {\bibfnamefont
  {P.}~\bibnamefont {Campagne-Ibarcq}}, \bibinfo {author} {\bibfnamefont
  {T.}~\bibnamefont {Kontos}}, \ and\ \bibinfo {author} {\bibfnamefont
  {Z.}~\bibnamefont {Leghtas}},\ }\href {https://arxiv.org/abs/2212.04991}
  {}\Eprint {http://arxiv.org/abs/2212.04991} {arXiv:2212.04991} \BibitemShut
  {NoStop}%
\bibitem [{\citenamefont {Ge}\ \emph {et~al.}(2019)\citenamefont {Ge},
  \citenamefont {Sawyer}, \citenamefont {Britton}, \citenamefont {Jacobs},
  \citenamefont {Bollinger},\ and\ \citenamefont
  {Foss-Feig}}]{PhysRevLett.122.030501}%
  \BibitemOpen
  \bibfield  {author} {\bibinfo {author} {\bibfnamefont {W.}~\bibnamefont
  {Ge}}, \bibinfo {author} {\bibfnamefont {B.~C.}\ \bibnamefont {Sawyer}},
  \bibinfo {author} {\bibfnamefont {J.~W.}\ \bibnamefont {Britton}}, \bibinfo
  {author} {\bibfnamefont {K.}~\bibnamefont {Jacobs}}, \bibinfo {author}
  {\bibfnamefont {J.~J.}\ \bibnamefont {Bollinger}}, \ and\ \bibinfo {author}
  {\bibfnamefont {M.}~\bibnamefont {Foss-Feig}},\ }\href {\doibase
  10.1103/PhysRevLett.122.030501} {\bibfield  {journal} {\bibinfo  {journal}
  {Phys. Rev. Lett.}\ }\textbf {\bibinfo {volume} {122}},\ \bibinfo {pages}
  {030501} (\bibinfo {year} {2019})}\BibitemShut {NoStop}%
\bibitem [{\citenamefont {Burd}\ \emph {et~al.}(2021)\citenamefont {Burd},
  \citenamefont {Srinivas}, \citenamefont {Knaack}, \citenamefont {Ge},
  \citenamefont {Wilson}, \citenamefont {Wineland}, \citenamefont {Leibfried},
  \citenamefont {Bollinger}, \citenamefont {Allcock},\ and\ \citenamefont
  {Slichter}}]{burd2021quantum}%
  \BibitemOpen
  \bibfield  {author} {\bibinfo {author} {\bibfnamefont {S.~C.}\ \bibnamefont
  {Burd}}, \bibinfo {author} {\bibfnamefont {R.}~\bibnamefont {Srinivas}},
  \bibinfo {author} {\bibfnamefont {H.~M.}\ \bibnamefont {Knaack}}, \bibinfo
  {author} {\bibfnamefont {W.}~\bibnamefont {Ge}}, \bibinfo {author}
  {\bibfnamefont {A.~C.}\ \bibnamefont {Wilson}}, \bibinfo {author}
  {\bibfnamefont {D.~J.}\ \bibnamefont {Wineland}}, \bibinfo {author}
  {\bibfnamefont {D.}~\bibnamefont {Leibfried}}, \bibinfo {author}
  {\bibfnamefont {J.~J.}\ \bibnamefont {Bollinger}}, \bibinfo {author}
  {\bibfnamefont {D.}~\bibnamefont {Allcock}}, \ and\ \bibinfo {author}
  {\bibfnamefont {D.}~\bibnamefont {Slichter}},\ }\href
  {https://www.nature.com/articles/s41567-021-01237-9} {\bibfield  {journal}
  {\bibinfo  {journal} {Nat. Phys.}\ }\textbf {\bibinfo {volume} {17}},\
  \bibinfo {pages} {898} (\bibinfo {year} {2021})}\BibitemShut {NoStop}%
\bibitem [{\citenamefont {Affolter}\ \emph {et~al.}(2023)\citenamefont
  {Affolter}, \citenamefont {Ge}, \citenamefont {Bullock}, \citenamefont
  {Burd}, \citenamefont {Gilmore}, \citenamefont {Lilieholm}, \citenamefont
  {Carter},\ and\ \citenamefont {Bollinger}}]{PhysRevA.107.032425}%
  \BibitemOpen
  \bibfield  {author} {\bibinfo {author} {\bibfnamefont {M.}~\bibnamefont
  {Affolter}}, \bibinfo {author} {\bibfnamefont {W.}~\bibnamefont {Ge}},
  \bibinfo {author} {\bibfnamefont {B.}~\bibnamefont {Bullock}}, \bibinfo
  {author} {\bibfnamefont {S.~C.}\ \bibnamefont {Burd}}, \bibinfo {author}
  {\bibfnamefont {K.~A.}\ \bibnamefont {Gilmore}}, \bibinfo {author}
  {\bibfnamefont {J.~F.}\ \bibnamefont {Lilieholm}}, \bibinfo {author}
  {\bibfnamefont {A.~L.}\ \bibnamefont {Carter}}, \ and\ \bibinfo {author}
  {\bibfnamefont {J.~J.}\ \bibnamefont {Bollinger}},\ }\href {\doibase
  10.1103/PhysRevA.107.032425} {\bibfield  {journal} {\bibinfo  {journal}
  {Phys. Rev. A}\ }\textbf {\bibinfo {volume} {107}},\ \bibinfo {pages}
  {032425} (\bibinfo {year} {2023})}\BibitemShut {NoStop}%
\bibitem [{\citenamefont {Pan}\ \emph {et~al.}(2023)\citenamefont {Pan},
  \citenamefont {Hei}, \citenamefont {Dong}, \citenamefont {Chen},
  \citenamefont {Shen}, \citenamefont {Ali},\ and\ \citenamefont
  {Li}}]{PhysRevA.107.023722}%
  \BibitemOpen
  \bibfield  {author} {\bibinfo {author} {\bibfnamefont {X.-F.}\ \bibnamefont
  {Pan}}, \bibinfo {author} {\bibfnamefont {X.-L.}\ \bibnamefont {Hei}},
  \bibinfo {author} {\bibfnamefont {X.-L.}\ \bibnamefont {Dong}}, \bibinfo
  {author} {\bibfnamefont {J.-Q.}\ \bibnamefont {Chen}}, \bibinfo {author}
  {\bibfnamefont {C.-P.}\ \bibnamefont {Shen}}, \bibinfo {author}
  {\bibfnamefont {H.}~\bibnamefont {Ali}}, \ and\ \bibinfo {author}
  {\bibfnamefont {P.-B.}\ \bibnamefont {Li}},\ }\href {\doibase
  10.1103/PhysRevA.107.023722} {\bibfield  {journal} {\bibinfo  {journal}
  {Phys. Rev. A}\ }\textbf {\bibinfo {volume} {107}},\ \bibinfo {pages}
  {023722} (\bibinfo {year} {2023})}\BibitemShut {NoStop}%
\bibitem [{\citenamefont {Hei}\ \emph {et~al.}(2023)\citenamefont {Hei},
  \citenamefont {Li}, \citenamefont {Pan},\ and\ \citenamefont
  {Nori}}]{PhysRevLett.130.073602}%
  \BibitemOpen
  \bibfield  {author} {\bibinfo {author} {\bibfnamefont {X.-L.}\ \bibnamefont
  {Hei}}, \bibinfo {author} {\bibfnamefont {P.-B.}\ \bibnamefont {Li}},
  \bibinfo {author} {\bibfnamefont {X.-F.}\ \bibnamefont {Pan}}, \ and\
  \bibinfo {author} {\bibfnamefont {F.}~\bibnamefont {Nori}},\ }\href {\doibase
  10.1103/PhysRevLett.130.073602} {\bibfield  {journal} {\bibinfo  {journal}
  {Phys. Rev. Lett.}\ }\textbf {\bibinfo {volume} {130}},\ \bibinfo {pages}
  {073602} (\bibinfo {year} {2023})}\BibitemShut {NoStop}%
\bibitem [{\citenamefont {Asjad}\ \emph {et~al.}(2023)\citenamefont {Asjad},
  \citenamefont {Li}, \citenamefont {Zhu},\ and\ \citenamefont
  {You}}]{ASJAD20233}%
  \BibitemOpen
  \bibfield  {author} {\bibinfo {author} {\bibfnamefont {M.}~\bibnamefont
  {Asjad}}, \bibinfo {author} {\bibfnamefont {J.}~\bibnamefont {Li}}, \bibinfo
  {author} {\bibfnamefont {S.-Y.}\ \bibnamefont {Zhu}}, \ and\ \bibinfo
  {author} {\bibfnamefont {J.}~\bibnamefont {You}},\ }\href {\doibase
  https://doi.org/10.1016/j.fmre.2022.07.006} {\bibfield  {journal} {\bibinfo
  {journal} {Fundam. Res.}\ }\textbf {\bibinfo {volume} {3}},\ \bibinfo {pages}
  {3} (\bibinfo {year} {2023})}\BibitemShut {NoStop}%
\bibitem [{\citenamefont {Sharma}\ \emph {et~al.}(2021)\citenamefont {Sharma},
  \citenamefont {Bittencourt}, \citenamefont {Karenowska},\ and\ \citenamefont
  {Kusminskiy}}]{PhysRevB.103.L100403}%
  \BibitemOpen
  \bibfield  {author} {\bibinfo {author} {\bibfnamefont {S.}~\bibnamefont
  {Sharma}}, \bibinfo {author} {\bibfnamefont {V.~A. S.~V.}\ \bibnamefont
  {Bittencourt}}, \bibinfo {author} {\bibfnamefont {A.~D.}\ \bibnamefont
  {Karenowska}}, \ and\ \bibinfo {author} {\bibfnamefont {S.~V.}\ \bibnamefont
  {Kusminskiy}},\ }\href {\doibase 10.1103/PhysRevB.103.L100403} {\bibfield
  {journal} {\bibinfo  {journal} {Phys. Rev. B}\ }\textbf {\bibinfo {volume}
  {103}},\ \bibinfo {pages} {L100403} (\bibinfo {year} {2021})}\BibitemShut
  {NoStop}%
\bibitem [{\citenamefont {Sun}\ \emph {et~al.}(2021)\citenamefont {Sun},
  \citenamefont {Zheng}, \citenamefont {Xiao}, \citenamefont {Gong},
  \citenamefont {He},\ and\ \citenamefont {Xia}}]{PhysRevLett.127.087203}%
  \BibitemOpen
  \bibfield  {author} {\bibinfo {author} {\bibfnamefont {F.-X.}\ \bibnamefont
  {Sun}}, \bibinfo {author} {\bibfnamefont {S.-S.}\ \bibnamefont {Zheng}},
  \bibinfo {author} {\bibfnamefont {Y.}~\bibnamefont {Xiao}}, \bibinfo {author}
  {\bibfnamefont {Q.}~\bibnamefont {Gong}}, \bibinfo {author} {\bibfnamefont
  {Q.}~\bibnamefont {He}}, \ and\ \bibinfo {author} {\bibfnamefont
  {K.}~\bibnamefont {Xia}},\ }\href {\doibase 10.1103/PhysRevLett.127.087203}
  {\bibfield  {journal} {\bibinfo  {journal} {Phys. Rev. Lett.}\ }\textbf
  {\bibinfo {volume} {127}},\ \bibinfo {pages} {087203} (\bibinfo {year}
  {2021})}\BibitemShut {NoStop}%
\bibitem [{\citenamefont {Vinante}\ \emph {et~al.}(2011)\citenamefont
  {Vinante}, \citenamefont {Wijts}, \citenamefont {Usenko}, \citenamefont
  {Schinkelshoek},\ and\ \citenamefont {Oosterkamp}}]{vinante2011magnetic}%
  \BibitemOpen
  \bibfield  {author} {\bibinfo {author} {\bibfnamefont {A.}~\bibnamefont
  {Vinante}}, \bibinfo {author} {\bibfnamefont {G.}~\bibnamefont {Wijts}},
  \bibinfo {author} {\bibfnamefont {O.}~\bibnamefont {Usenko}}, \bibinfo
  {author} {\bibfnamefont {L.}~\bibnamefont {Schinkelshoek}}, \ and\ \bibinfo
  {author} {\bibfnamefont {T.}~\bibnamefont {Oosterkamp}},\ }\href
  {https://www.nature.com/articles/ncomms1581} {\bibfield  {journal} {\bibinfo
  {journal} {Nat. Commun.}\ }\textbf {\bibinfo {volume} {2}},\ \bibinfo {pages}
  {1} (\bibinfo {year} {2011})}\BibitemShut {NoStop}%
\bibitem [{\citenamefont {Burgess}\ \emph {et~al.}(2013)\citenamefont
  {Burgess}, \citenamefont {Fraser}, \citenamefont {Sani}, \citenamefont
  {Vick}, \citenamefont {Hauer}, \citenamefont {Davis},\ and\ \citenamefont
  {Freeman}}]{burgess2013quantitative}%
  \BibitemOpen
  \bibfield  {author} {\bibinfo {author} {\bibfnamefont {J.}~\bibnamefont
  {Burgess}}, \bibinfo {author} {\bibfnamefont {A.}~\bibnamefont {Fraser}},
  \bibinfo {author} {\bibfnamefont {F.~F.}\ \bibnamefont {Sani}}, \bibinfo
  {author} {\bibfnamefont {D.}~\bibnamefont {Vick}}, \bibinfo {author}
  {\bibfnamefont {B.}~\bibnamefont {Hauer}}, \bibinfo {author} {\bibfnamefont
  {J.}~\bibnamefont {Davis}}, \ and\ \bibinfo {author} {\bibfnamefont
  {M.}~\bibnamefont {Freeman}},\ }\href
  {https://www.science.org/doi/full/10.1126/science.1231390} {\bibfield
  {journal} {\bibinfo  {journal} {Science}\ }\textbf {\bibinfo {volume}
  {339}},\ \bibinfo {pages} {1051} (\bibinfo {year} {2013})}\BibitemShut
  {NoStop}%
\bibitem [{\citenamefont {Vinante}\ and\ \citenamefont
  {Falferi}(2013)}]{PhysRevLett.111.207203}%
  \BibitemOpen
  \bibfield  {author} {\bibinfo {author} {\bibfnamefont {A.}~\bibnamefont
  {Vinante}}\ and\ \bibinfo {author} {\bibfnamefont {P.}~\bibnamefont
  {Falferi}},\ }\href {\doibase 10.1103/PhysRevLett.111.207203} {\bibfield
  {journal} {\bibinfo  {journal} {Phys. Rev. Lett.}\ }\textbf {\bibinfo
  {volume} {111}},\ \bibinfo {pages} {207203} (\bibinfo {year}
  {2013})}\BibitemShut {NoStop}%
\bibitem [{\citenamefont {den Haan}\ \emph {et~al.}(2015)\citenamefont {den
  Haan}, \citenamefont {Wagenaar}, \citenamefont {de~Voogd}, \citenamefont
  {Koning},\ and\ \citenamefont {Oosterkamp}}]{PhysRevB.92.235441}%
  \BibitemOpen
  \bibfield  {author} {\bibinfo {author} {\bibfnamefont {A.~M.~J.}\
  \bibnamefont {den Haan}}, \bibinfo {author} {\bibfnamefont {J.~J.~T.}\
  \bibnamefont {Wagenaar}}, \bibinfo {author} {\bibfnamefont {J.~M.}\
  \bibnamefont {de~Voogd}}, \bibinfo {author} {\bibfnamefont {G.}~\bibnamefont
  {Koning}}, \ and\ \bibinfo {author} {\bibfnamefont {T.~H.}\ \bibnamefont
  {Oosterkamp}},\ }\href {\doibase 10.1103/PhysRevB.92.235441} {\bibfield
  {journal} {\bibinfo  {journal} {Phys. Rev. B}\ }\textbf {\bibinfo {volume}
  {92}},\ \bibinfo {pages} {235441} (\bibinfo {year} {2015})}\BibitemShut
  {NoStop}%
\bibitem [{\citenamefont {Kolkowitz}\ \emph {et~al.}(2012)\citenamefont
  {Kolkowitz}, \citenamefont {Jayich}, \citenamefont {Unterreithmeier},
  \citenamefont {Bennett}, \citenamefont {Rabl}, \citenamefont {Harris},\ and\
  \citenamefont {Lukin}}]{kolkowitz2012coherent}%
  \BibitemOpen
  \bibfield  {author} {\bibinfo {author} {\bibfnamefont {S.}~\bibnamefont
  {Kolkowitz}}, \bibinfo {author} {\bibfnamefont {A.~C.~B.}\ \bibnamefont
  {Jayich}}, \bibinfo {author} {\bibfnamefont {Q.~P.}\ \bibnamefont
  {Unterreithmeier}}, \bibinfo {author} {\bibfnamefont {S.~D.}\ \bibnamefont
  {Bennett}}, \bibinfo {author} {\bibfnamefont {P.}~\bibnamefont {Rabl}},
  \bibinfo {author} {\bibfnamefont {J.}~\bibnamefont {Harris}}, \ and\ \bibinfo
  {author} {\bibfnamefont {M.~D.}\ \bibnamefont {Lukin}},\ }\href
  {https://science.sciencemag.org/content/335/6076/1603.abstract} {\bibfield
  {journal} {\bibinfo  {journal} {Science}\ }\textbf {\bibinfo {volume}
  {335}},\ \bibinfo {pages} {1603} (\bibinfo {year} {2012})}\BibitemShut
  {NoStop}%
\bibitem [{\citenamefont {Rugar}\ and\ \citenamefont
  {Gr\"utter}(1991)}]{PhysRevLett.67.699}%
  \BibitemOpen
  \bibfield  {author} {\bibinfo {author} {\bibfnamefont {D.}~\bibnamefont
  {Rugar}}\ and\ \bibinfo {author} {\bibfnamefont {P.}~\bibnamefont
  {Gr\"utter}},\ }\href {\doibase 10.1103/PhysRevLett.67.699} {\bibfield
  {journal} {\bibinfo  {journal} {Phys. Rev. Lett.}\ }\textbf {\bibinfo
  {volume} {67}},\ \bibinfo {pages} {699} (\bibinfo {year} {1991})}\BibitemShut
  {NoStop}%
\bibitem [{\citenamefont {Szorkovszky}\ \emph {et~al.}(2011)\citenamefont
  {Szorkovszky}, \citenamefont {Doherty}, \citenamefont {Harris},\ and\
  \citenamefont {Bowen}}]{PhysRevLett.107.213603}%
  \BibitemOpen
  \bibfield  {author} {\bibinfo {author} {\bibfnamefont {A.}~\bibnamefont
  {Szorkovszky}}, \bibinfo {author} {\bibfnamefont {A.~C.}\ \bibnamefont
  {Doherty}}, \bibinfo {author} {\bibfnamefont {G.~I.}\ \bibnamefont {Harris}},
  \ and\ \bibinfo {author} {\bibfnamefont {W.~P.}\ \bibnamefont {Bowen}},\
  }\href {\doibase 10.1103/PhysRevLett.107.213603} {\bibfield  {journal}
  {\bibinfo  {journal} {Phys. Rev. Lett.}\ }\textbf {\bibinfo {volume} {107}},\
  \bibinfo {pages} {213603} (\bibinfo {year} {2011})}\BibitemShut {NoStop}%
\bibitem [{\citenamefont {Rabl}\ \emph {et~al.}(2009)\citenamefont {Rabl},
  \citenamefont {Cappellaro}, \citenamefont {Dutt}, \citenamefont {Jiang},
  \citenamefont {Maze},\ and\ \citenamefont {Lukin}}]{PhysRevB.79.041302}%
  \BibitemOpen
  \bibfield  {author} {\bibinfo {author} {\bibfnamefont {P.}~\bibnamefont
  {Rabl}}, \bibinfo {author} {\bibfnamefont {P.}~\bibnamefont {Cappellaro}},
  \bibinfo {author} {\bibfnamefont {M.~V.~G.}\ \bibnamefont {Dutt}}, \bibinfo
  {author} {\bibfnamefont {L.}~\bibnamefont {Jiang}}, \bibinfo {author}
  {\bibfnamefont {J.~R.}\ \bibnamefont {Maze}}, \ and\ \bibinfo {author}
  {\bibfnamefont {M.~D.}\ \bibnamefont {Lukin}},\ }\href {\doibase
  10.1103/PhysRevB.79.041302} {\bibfield  {journal} {\bibinfo  {journal} {Phys.
  Rev. B}\ }\textbf {\bibinfo {volume} {79}},\ \bibinfo {pages} {041302}
  (\bibinfo {year} {2009})}\BibitemShut {NoStop}%
\bibitem [{\citenamefont {Rabl}\ \emph {et~al.}(2010)\citenamefont {Rabl},
  \citenamefont {Kolkowitz}, \citenamefont {Koppens}, \citenamefont {Harris},
  \citenamefont {Zoller},\ and\ \citenamefont {Lukin}}]{rabl2010quantum}%
  \BibitemOpen
  \bibfield  {author} {\bibinfo {author} {\bibfnamefont {P.}~\bibnamefont
  {Rabl}}, \bibinfo {author} {\bibfnamefont {S.~J.}\ \bibnamefont {Kolkowitz}},
  \bibinfo {author} {\bibfnamefont {F.}~\bibnamefont {Koppens}}, \bibinfo
  {author} {\bibfnamefont {J.}~\bibnamefont {Harris}}, \bibinfo {author}
  {\bibfnamefont {P.}~\bibnamefont {Zoller}}, \ and\ \bibinfo {author}
  {\bibfnamefont {M.~D.}\ \bibnamefont {Lukin}},\ }\href
  {https://www.nature.com/articles/nphys1679} {\bibfield  {journal} {\bibinfo
  {journal} {Nat. Phys.}\ }\textbf {\bibinfo {volume} {6}},\ \bibinfo {pages}
  {602} (\bibinfo {year} {2010})}\BibitemShut {NoStop}%
\bibitem [{\citenamefont {Childress}\ \emph {et~al.}(2006)\citenamefont
  {Childress}, \citenamefont {Gurudev~Dutt}, \citenamefont {Taylor},
  \citenamefont {Zibrov}, \citenamefont {Jelezko}, \citenamefont {Wrachtrup},
  \citenamefont {Hemmer},\ and\ \citenamefont {Lukin}}]{childress2006coherent}%
  \BibitemOpen
  \bibfield  {author} {\bibinfo {author} {\bibfnamefont {L.}~\bibnamefont
  {Childress}}, \bibinfo {author} {\bibfnamefont {M.}~\bibnamefont
  {Gurudev~Dutt}}, \bibinfo {author} {\bibfnamefont {J.}~\bibnamefont
  {Taylor}}, \bibinfo {author} {\bibfnamefont {A.}~\bibnamefont {Zibrov}},
  \bibinfo {author} {\bibfnamefont {F.}~\bibnamefont {Jelezko}}, \bibinfo
  {author} {\bibfnamefont {J.}~\bibnamefont {Wrachtrup}}, \bibinfo {author}
  {\bibfnamefont {P.}~\bibnamefont {Hemmer}}, \ and\ \bibinfo {author}
  {\bibfnamefont {M.}~\bibnamefont {Lukin}},\ }\href
  {https://www.science.org/doi/full/10.1126/science.1131871} {\bibfield
  {journal} {\bibinfo  {journal} {Science}\ }\textbf {\bibinfo {volume}
  {314}},\ \bibinfo {pages} {281} (\bibinfo {year} {2006})}\BibitemShut
  {NoStop}%
\bibitem [{\citenamefont {Taylor}\ \emph {et~al.}(2008)\citenamefont {Taylor},
  \citenamefont {Cappellaro}, \citenamefont {Childress}, \citenamefont {Jiang},
  \citenamefont {Budker}, \citenamefont {Hemmer}, \citenamefont {Yacoby},
  \citenamefont {Walsworth},\ and\ \citenamefont {Lukin}}]{taylor2008high}%
  \BibitemOpen
  \bibfield  {author} {\bibinfo {author} {\bibfnamefont {J.~M.}\ \bibnamefont
  {Taylor}}, \bibinfo {author} {\bibfnamefont {P.}~\bibnamefont {Cappellaro}},
  \bibinfo {author} {\bibfnamefont {L.}~\bibnamefont {Childress}}, \bibinfo
  {author} {\bibfnamefont {L.}~\bibnamefont {Jiang}}, \bibinfo {author}
  {\bibfnamefont {D.}~\bibnamefont {Budker}}, \bibinfo {author} {\bibfnamefont
  {P.}~\bibnamefont {Hemmer}}, \bibinfo {author} {\bibfnamefont
  {A.}~\bibnamefont {Yacoby}}, \bibinfo {author} {\bibfnamefont
  {R.}~\bibnamefont {Walsworth}}, \ and\ \bibinfo {author} {\bibfnamefont
  {M.}~\bibnamefont {Lukin}},\ }\href
  {https://www.nature.com/articles/nphys1075} {\bibfield  {journal} {\bibinfo
  {journal} {Nature Phys.}\ }\textbf {\bibinfo {volume} {4}},\ \bibinfo {pages}
  {810} (\bibinfo {year} {2008})}\BibitemShut {NoStop}%
\bibitem [{\citenamefont {Walker}(1957)}]{PhysRev.105.390}%
  \BibitemOpen
  \bibfield  {author} {\bibinfo {author} {\bibfnamefont {L.~R.}\ \bibnamefont
  {Walker}},\ }\href {\doibase 10.1103/PhysRev.105.390} {\bibfield  {journal}
  {\bibinfo  {journal} {Phys. Rev.}\ }\textbf {\bibinfo {volume} {105}},\
  \bibinfo {pages} {390} (\bibinfo {year} {1957})}\BibitemShut {NoStop}%
\bibitem [{\citenamefont {Walker}(1958)}]{doi:10.1063/1.1723117}%
  \BibitemOpen
  \bibfield  {author} {\bibinfo {author} {\bibfnamefont {L.~R.}\ \bibnamefont
  {Walker}},\ }\href {\doibase 10.1063/1.1723117} {\bibfield  {journal}
  {\bibinfo  {journal} {J. Appl. Phys.}\ }\textbf {\bibinfo {volume} {29}},\
  \bibinfo {pages} {318} (\bibinfo {year} {1958})}\BibitemShut {NoStop}%
\bibitem [{\citenamefont {Röschmann}\ and\ \citenamefont
  {Dötsch}(1977)}]{https://doi.org/10.1002/pssb.2220820102}%
  \BibitemOpen
  \bibfield  {author} {\bibinfo {author} {\bibfnamefont {P.}~\bibnamefont
  {Röschmann}}\ and\ \bibinfo {author} {\bibfnamefont {H.}~\bibnamefont
  {Dötsch}},\ }\href {\doibase https://doi.org/10.1002/pssb.2220820102}
  {\bibfield  {journal} {\bibinfo  {journal} {Phys. Status Solidi B}\ }\textbf
  {\bibinfo {volume} {82}},\ \bibinfo {pages} {11} (\bibinfo {year}
  {1977})}\BibitemShut {NoStop}%
\bibitem [{\citenamefont {Mills}(2006)}]{MILLS200616}%
  \BibitemOpen
  \bibfield  {author} {\bibinfo {author} {\bibfnamefont {D.}~\bibnamefont
  {Mills}},\ }\href {\doibase https://doi.org/10.1016/j.jmmm.2006.02.267}
  {\bibfield  {journal} {\bibinfo  {journal} {J. Magn. Magn. Mater.}\ }\textbf
  {\bibinfo {volume} {306}},\ \bibinfo {pages} {16} (\bibinfo {year}
  {2006})}\BibitemShut {NoStop}%
\bibitem [{\citenamefont {Kittel}(1948)}]{PhysRev.73.155}%
  \BibitemOpen
  \bibfield  {author} {\bibinfo {author} {\bibfnamefont {C.}~\bibnamefont
  {Kittel}},\ }\href {\doibase 10.1103/PhysRev.73.155} {\bibfield  {journal}
  {\bibinfo  {journal} {Phys. Rev.}\ }\textbf {\bibinfo {volume} {73}},\
  \bibinfo {pages} {155} (\bibinfo {year} {1948})}\BibitemShut {NoStop}%
\bibitem [{\citenamefont {Mamin}\ \emph {et~al.}(2012)\citenamefont {Mamin},
  \citenamefont {Rettner}, \citenamefont {Sherwood}, \citenamefont {Gao},\ and\
  \citenamefont {Rugar}}]{doi:10.1063/1.3673910}%
  \BibitemOpen
  \bibfield  {author} {\bibinfo {author} {\bibfnamefont {H.~J.}\ \bibnamefont
  {Mamin}}, \bibinfo {author} {\bibfnamefont {C.~T.}\ \bibnamefont {Rettner}},
  \bibinfo {author} {\bibfnamefont {M.~H.}\ \bibnamefont {Sherwood}}, \bibinfo
  {author} {\bibfnamefont {L.}~\bibnamefont {Gao}}, \ and\ \bibinfo {author}
  {\bibfnamefont {D.}~\bibnamefont {Rugar}},\ }\href {\doibase
  10.1063/1.3673910} {\bibfield  {journal} {\bibinfo  {journal} {Appl. Phys.
  Lett.}\ }\textbf {\bibinfo {volume} {100}},\ \bibinfo {pages} {013102}
  (\bibinfo {year} {2012})}\BibitemShut {NoStop}%
\bibitem [{\citenamefont {Li}\ \emph {et~al.}(2016)\citenamefont {Li},
  \citenamefont {Xiang}, \citenamefont {Rabl},\ and\ \citenamefont
  {Nori}}]{PhysRevLett.117.015502}%
  \BibitemOpen
  \bibfield  {author} {\bibinfo {author} {\bibfnamefont {P.-B.}\ \bibnamefont
  {Li}}, \bibinfo {author} {\bibfnamefont {Z.-L.}\ \bibnamefont {Xiang}},
  \bibinfo {author} {\bibfnamefont {P.}~\bibnamefont {Rabl}}, \ and\ \bibinfo
  {author} {\bibfnamefont {F.}~\bibnamefont {Nori}},\ }\href {\doibase
  10.1103/PhysRevLett.117.015502} {\bibfield  {journal} {\bibinfo  {journal}
  {Phys. Rev. Lett.}\ }\textbf {\bibinfo {volume} {117}},\ \bibinfo {pages}
  {015502} (\bibinfo {year} {2016})}\BibitemShut {NoStop}%
\bibitem [{\citenamefont {S\'anchez Mu\~noz}\ \emph {et~al.}(2018)\citenamefont
  {S\'anchez Mu\~noz}, \citenamefont {Lara}, \citenamefont {Puebla},\ and\
  \citenamefont {Nori}}]{PhysRevLett.121.123604}%
  \BibitemOpen
  \bibfield  {author} {\bibinfo {author} {\bibfnamefont {C.}~\bibnamefont
  {S\'anchez Mu\~noz}}, \bibinfo {author} {\bibfnamefont {A.}~\bibnamefont
  {Lara}}, \bibinfo {author} {\bibfnamefont {J.}~\bibnamefont {Puebla}}, \ and\
  \bibinfo {author} {\bibfnamefont {F.}~\bibnamefont {Nori}},\ }\href {\doibase
  10.1103/PhysRevLett.121.123604} {\bibfield  {journal} {\bibinfo  {journal}
  {Phys. Rev. Lett.}\ }\textbf {\bibinfo {volume} {121}},\ \bibinfo {pages}
  {123604} (\bibinfo {year} {2018})}\BibitemShut {NoStop}%
\bibitem [{\citenamefont {Zhou}\ \emph {et~al.}(2022)\citenamefont {Zhou},
  \citenamefont {Hu}, \citenamefont {L\"{u}}, \citenamefont {Li}, \citenamefont
  {Huang}, \citenamefont {Xiong},\ and\ \citenamefont {L\"{u}}}]{Zhou:22}%
  \BibitemOpen
  \bibfield  {author} {\bibinfo {author} {\bibfnamefont {Y.}~\bibnamefont
  {Zhou}}, \bibinfo {author} {\bibfnamefont {C.-S.}\ \bibnamefont {Hu}},
  \bibinfo {author} {\bibfnamefont {D.-Y.}\ \bibnamefont {L\"{u}}}, \bibinfo
  {author} {\bibfnamefont {X.-K.}\ \bibnamefont {Li}}, \bibinfo {author}
  {\bibfnamefont {H.-M.}\ \bibnamefont {Huang}}, \bibinfo {author}
  {\bibfnamefont {Y.-C.}\ \bibnamefont {Xiong}}, \ and\ \bibinfo {author}
  {\bibfnamefont {X.-Y.}\ \bibnamefont {L\"{u}}},\ }\href {\doibase
  10.1364/PRJ.459794} {\bibfield  {journal} {\bibinfo  {journal} {Photon.
  Res.}\ }\textbf {\bibinfo {volume} {10}},\ \bibinfo {pages} {1640} (\bibinfo
  {year} {2022})}\BibitemShut {NoStop}%
\bibitem [{\citenamefont {Li}\ \emph {et~al.}(2021)\citenamefont {Li},
  \citenamefont {Wang}, \citenamefont {Wu}, \citenamefont {Zhu},\ and\
  \citenamefont {You}}]{PRXQuantum.2.040344}%
  \BibitemOpen
  \bibfield  {author} {\bibinfo {author} {\bibfnamefont {J.}~\bibnamefont
  {Li}}, \bibinfo {author} {\bibfnamefont {Y.-P.}\ \bibnamefont {Wang}},
  \bibinfo {author} {\bibfnamefont {W.-J.}\ \bibnamefont {Wu}}, \bibinfo
  {author} {\bibfnamefont {S.-Y.}\ \bibnamefont {Zhu}}, \ and\ \bibinfo
  {author} {\bibfnamefont {J.}~\bibnamefont {You}},\ }\href {\doibase
  10.1103/PRXQuantum.2.040344} {\bibfield  {journal} {\bibinfo  {journal} {PRX
  Quantum}\ }\textbf {\bibinfo {volume} {2}},\ \bibinfo {pages} {040344}
  (\bibinfo {year} {2021})}\BibitemShut {NoStop}%
\bibitem [{\citenamefont {Norte}\ \emph {et~al.}(2016)\citenamefont {Norte},
  \citenamefont {Moura},\ and\ \citenamefont
  {Gr\"oblacher}}]{PhysRevLett.116.147202}%
  \BibitemOpen
  \bibfield  {author} {\bibinfo {author} {\bibfnamefont {R.~A.}\ \bibnamefont
  {Norte}}, \bibinfo {author} {\bibfnamefont {J.~P.}\ \bibnamefont {Moura}}, \
  and\ \bibinfo {author} {\bibfnamefont {S.}~\bibnamefont {Gr\"oblacher}},\
  }\href {\doibase 10.1103/PhysRevLett.116.147202} {\bibfield  {journal}
  {\bibinfo  {journal} {Phys. Rev. Lett.}\ }\textbf {\bibinfo {volume} {116}},\
  \bibinfo {pages} {147202} (\bibinfo {year} {2016})}\BibitemShut {NoStop}%
\bibitem [{\citenamefont {Tsaturyan}\ \emph {et~al.}(2017)\citenamefont
  {Tsaturyan}, \citenamefont {Barg}, \citenamefont {Polzik},\ and\
  \citenamefont {Schliesser}}]{tsaturyan2017ultracoherent}%
  \BibitemOpen
  \bibfield  {author} {\bibinfo {author} {\bibfnamefont {Y.}~\bibnamefont
  {Tsaturyan}}, \bibinfo {author} {\bibfnamefont {A.}~\bibnamefont {Barg}},
  \bibinfo {author} {\bibfnamefont {E.~S.}\ \bibnamefont {Polzik}}, \ and\
  \bibinfo {author} {\bibfnamefont {A.}~\bibnamefont {Schliesser}},\ }\href
  {https://www.nature.com/articles/nnano.2017.101.pdf?origin=ppub} {\bibfield
  {journal} {\bibinfo  {journal} {Nat. Nanotechnol.}\ }\textbf {\bibinfo
  {volume} {12}},\ \bibinfo {pages} {776} (\bibinfo {year} {2017})}\BibitemShut
  {NoStop}%
\bibitem [{\citenamefont {Ghadimi}\ \emph {et~al.}(2018)\citenamefont
  {Ghadimi}, \citenamefont {Fedorov}, \citenamefont {Engelsen}, \citenamefont
  {Bereyhi}, \citenamefont {Schilling}, \citenamefont {Wilson},\ and\
  \citenamefont {Kippenberg}}]{ghadimi2018elastic}%
  \BibitemOpen
  \bibfield  {author} {\bibinfo {author} {\bibfnamefont {A.~H.}\ \bibnamefont
  {Ghadimi}}, \bibinfo {author} {\bibfnamefont {S.~A.}\ \bibnamefont
  {Fedorov}}, \bibinfo {author} {\bibfnamefont {N.~J.}\ \bibnamefont
  {Engelsen}}, \bibinfo {author} {\bibfnamefont {M.~J.}\ \bibnamefont
  {Bereyhi}}, \bibinfo {author} {\bibfnamefont {R.}~\bibnamefont {Schilling}},
  \bibinfo {author} {\bibfnamefont {D.~J.}\ \bibnamefont {Wilson}}, \ and\
  \bibinfo {author} {\bibfnamefont {T.~J.}\ \bibnamefont {Kippenberg}},\ }\href
  {https://science.sciencemag.org/content/360/6390/764} {\bibfield  {journal}
  {\bibinfo  {journal} {Science}\ }\textbf {\bibinfo {volume} {360}},\ \bibinfo
  {pages} {764} (\bibinfo {year} {2018})}\BibitemShut {NoStop}%
\bibitem [{\citenamefont {Mamin}\ \emph {et~al.}(2007)\citenamefont {Mamin},
  \citenamefont {Poggio}, \citenamefont {Degen},\ and\ \citenamefont
  {Rugar}}]{mamin2007nuclear}%
  \BibitemOpen
  \bibfield  {author} {\bibinfo {author} {\bibfnamefont {H.}~\bibnamefont
  {Mamin}}, \bibinfo {author} {\bibfnamefont {M.}~\bibnamefont {Poggio}},
  \bibinfo {author} {\bibfnamefont {C.}~\bibnamefont {Degen}}, \ and\ \bibinfo
  {author} {\bibfnamefont {D.}~\bibnamefont {Rugar}},\ }\href
  {https://www.nature.com/articles/nnano.2007.105} {\bibfield  {journal}
  {\bibinfo  {journal} {Nat. Nanotechnol.}\ }\textbf {\bibinfo {volume} {2}},\
  \bibinfo {pages} {301} (\bibinfo {year} {2007})}\BibitemShut {NoStop}%
\bibitem [{\citenamefont {Poggio}\ and\ \citenamefont
  {Degen}(2010)}]{Poggio_2010}%
  \BibitemOpen
  \bibfield  {author} {\bibinfo {author} {\bibfnamefont {M.}~\bibnamefont
  {Poggio}}\ and\ \bibinfo {author} {\bibfnamefont {C.~L.}\ \bibnamefont
  {Degen}},\ }\href {\doibase 10.1088/0957-4484/21/34/342001} {\bibfield
  {journal} {\bibinfo  {journal} {Nanotechnology}\ }\textbf {\bibinfo {volume}
  {21}},\ \bibinfo {pages} {342001} (\bibinfo {year} {2010})}\BibitemShut
  {NoStop}%
\bibitem [{\citenamefont {Ishikawa}\ \emph {et~al.}(2012)\citenamefont
  {Ishikawa}, \citenamefont {Fu}, \citenamefont {Santori}, \citenamefont
  {Acosta}, \citenamefont {Beausoleil}, \citenamefont {Watanabe}, \citenamefont
  {Shikata},\ and\ \citenamefont {Itoh}}]{ishikawa2012optical}%
  \BibitemOpen
  \bibfield  {author} {\bibinfo {author} {\bibfnamefont {T.}~\bibnamefont
  {Ishikawa}}, \bibinfo {author} {\bibfnamefont {K.-M.~C.}\ \bibnamefont {Fu}},
  \bibinfo {author} {\bibfnamefont {C.}~\bibnamefont {Santori}}, \bibinfo
  {author} {\bibfnamefont {V.~M.}\ \bibnamefont {Acosta}}, \bibinfo {author}
  {\bibfnamefont {R.~G.}\ \bibnamefont {Beausoleil}}, \bibinfo {author}
  {\bibfnamefont {H.}~\bibnamefont {Watanabe}}, \bibinfo {author}
  {\bibfnamefont {S.}~\bibnamefont {Shikata}}, \ and\ \bibinfo {author}
  {\bibfnamefont {K.~M.}\ \bibnamefont {Itoh}},\ }\href
  {https://pubs.acs.org/doi/abs/10.1021/nl300350r} {\bibfield  {journal}
  {\bibinfo  {journal} {Nano Lett.}\ }\textbf {\bibinfo {volume} {12}},\
  \bibinfo {pages} {2083} (\bibinfo {year} {2012})}\BibitemShut {NoStop}%
\bibitem [{\citenamefont {Mamin}\ \emph {et~al.}(2013)\citenamefont {Mamin},
  \citenamefont {Kim}, \citenamefont {Sherwood}, \citenamefont {Rettner},
  \citenamefont {Ohno}, \citenamefont {Awschalom},\ and\ \citenamefont
  {Rugar}}]{mamin2013nanoscale}%
  \BibitemOpen
  \bibfield  {author} {\bibinfo {author} {\bibfnamefont {H.}~\bibnamefont
  {Mamin}}, \bibinfo {author} {\bibfnamefont {M.}~\bibnamefont {Kim}}, \bibinfo
  {author} {\bibfnamefont {M.}~\bibnamefont {Sherwood}}, \bibinfo {author}
  {\bibfnamefont {C.}~\bibnamefont {Rettner}}, \bibinfo {author} {\bibfnamefont
  {K.}~\bibnamefont {Ohno}}, \bibinfo {author} {\bibfnamefont {D.}~\bibnamefont
  {Awschalom}}, \ and\ \bibinfo {author} {\bibfnamefont {D.}~\bibnamefont
  {Rugar}},\ }\href {https://www.science.org/doi/full/10.1126/science.1231540}
  {\bibfield  {journal} {\bibinfo  {journal} {Science}\ }\textbf {\bibinfo
  {volume} {339}},\ \bibinfo {pages} {557} (\bibinfo {year}
  {2013})}\BibitemShut {NoStop}%
\bibitem [{\citenamefont {Bar-Gill}\ \emph {et~al.}(2013)\citenamefont
  {Bar-Gill}, \citenamefont {Pham}, \citenamefont {Jarmola}, \citenamefont
  {Budker},\ and\ \citenamefont {Walsworth}}]{bar2013solid}%
  \BibitemOpen
  \bibfield  {author} {\bibinfo {author} {\bibfnamefont {N.}~\bibnamefont
  {Bar-Gill}}, \bibinfo {author} {\bibfnamefont {L.~M.}\ \bibnamefont {Pham}},
  \bibinfo {author} {\bibfnamefont {A.}~\bibnamefont {Jarmola}}, \bibinfo
  {author} {\bibfnamefont {D.}~\bibnamefont {Budker}}, \ and\ \bibinfo {author}
  {\bibfnamefont {R.~L.}\ \bibnamefont {Walsworth}},\ }\href
  {https://www.nature.com/articles/ncomms2771} {\bibfield  {journal} {\bibinfo
  {journal} {Nat. Commun.}\ }\textbf {\bibinfo {volume} {4}},\ \bibinfo {pages}
  {1} (\bibinfo {year} {2013})}\BibitemShut {NoStop}%
\bibitem [{\citenamefont {Ovartchaiyapong}\ \emph {et~al.}(2014)\citenamefont
  {Ovartchaiyapong}, \citenamefont {Lee}, \citenamefont {Myers},\ and\
  \citenamefont {Jayich}}]{ovartchaiyapong2014dynamic}%
  \BibitemOpen
  \bibfield  {author} {\bibinfo {author} {\bibfnamefont {P.}~\bibnamefont
  {Ovartchaiyapong}}, \bibinfo {author} {\bibfnamefont {K.~W.}\ \bibnamefont
  {Lee}}, \bibinfo {author} {\bibfnamefont {B.~A.}\ \bibnamefont {Myers}}, \
  and\ \bibinfo {author} {\bibfnamefont {A.~C.~B.}\ \bibnamefont {Jayich}},\
  }\href {https://www.nature.com/articles/ncomms5429} {\bibfield  {journal}
  {\bibinfo  {journal} {Nat. Commun.}\ }\textbf {\bibinfo {volume} {5}},\
  \bibinfo {pages} {1} (\bibinfo {year} {2014})}\BibitemShut {NoStop}%
\bibitem [{\citenamefont {Rameshti}\ \emph {et~al.}(2022)\citenamefont
  {Rameshti}, \citenamefont {Kusminskiy}, \citenamefont {Haigh}, \citenamefont
  {Usami}, \citenamefont {Lachance-Quirion}, \citenamefont {Nakamura},
  \citenamefont {Hu}, \citenamefont {Tang}, \citenamefont {Bauer},\ and\
  \citenamefont {Blanter}}]{rameshti2022cavity}%
  \BibitemOpen
  \bibfield  {author} {\bibinfo {author} {\bibfnamefont {B.~Z.}\ \bibnamefont
  {Rameshti}}, \bibinfo {author} {\bibfnamefont {S.~V.}\ \bibnamefont
  {Kusminskiy}}, \bibinfo {author} {\bibfnamefont {J.~A.}\ \bibnamefont
  {Haigh}}, \bibinfo {author} {\bibfnamefont {K.}~\bibnamefont {Usami}},
  \bibinfo {author} {\bibfnamefont {D.}~\bibnamefont {Lachance-Quirion}},
  \bibinfo {author} {\bibfnamefont {Y.}~\bibnamefont {Nakamura}}, \bibinfo
  {author} {\bibfnamefont {C.-M.}\ \bibnamefont {Hu}}, \bibinfo {author}
  {\bibfnamefont {H.~X.}\ \bibnamefont {Tang}}, \bibinfo {author}
  {\bibfnamefont {G.~E.}\ \bibnamefont {Bauer}}, \ and\ \bibinfo {author}
  {\bibfnamefont {Y.~M.}\ \bibnamefont {Blanter}},\ }\href
  {https://www.sciencedirect.com/science/article/pii/S0370157322002460}
  {\bibfield  {journal} {\bibinfo  {journal} {Phys. Rep.}\ }\textbf {\bibinfo
  {volume} {979}},\ \bibinfo {pages} {1} (\bibinfo {year} {2022})}\BibitemShut
  {NoStop}%
\bibitem [{\citenamefont {Planat}\ \emph {et~al.}(2020)\citenamefont {Planat},
  \citenamefont {Ranadive}, \citenamefont {Dassonneville}, \citenamefont
  {Puertas~Mart\'{\i}nez}, \citenamefont {L\'eger}, \citenamefont {Naud},
  \citenamefont {Buisson}, \citenamefont {Hasch-Guichard}, \citenamefont
  {Basko},\ and\ \citenamefont {Roch}}]{PhysRevX.10.021021}%
  \BibitemOpen
  \bibfield  {author} {\bibinfo {author} {\bibfnamefont {L.}~\bibnamefont
  {Planat}}, \bibinfo {author} {\bibfnamefont {A.}~\bibnamefont {Ranadive}},
  \bibinfo {author} {\bibfnamefont {R.}~\bibnamefont {Dassonneville}}, \bibinfo
  {author} {\bibfnamefont {J.}~\bibnamefont {Puertas~Mart\'{\i}nez}}, \bibinfo
  {author} {\bibfnamefont {S.}~\bibnamefont {L\'eger}}, \bibinfo {author}
  {\bibfnamefont {C.}~\bibnamefont {Naud}}, \bibinfo {author} {\bibfnamefont
  {O.}~\bibnamefont {Buisson}}, \bibinfo {author} {\bibfnamefont
  {W.}~\bibnamefont {Hasch-Guichard}}, \bibinfo {author} {\bibfnamefont
  {D.~M.}\ \bibnamefont {Basko}}, \ and\ \bibinfo {author} {\bibfnamefont
  {N.}~\bibnamefont {Roch}},\ }\href {\doibase 10.1103/PhysRevX.10.021021}
  {\bibfield  {journal} {\bibinfo  {journal} {Phys. Rev. X}\ }\textbf {\bibinfo
  {volume} {10}},\ \bibinfo {pages} {021021} (\bibinfo {year}
  {2020})}\BibitemShut {NoStop}%
\bibitem [{\citenamefont {Grebel}\ \emph {et~al.}(2021)\citenamefont {Grebel},
  \citenamefont {Bienfait}, \citenamefont {Dumur}, \citenamefont {Chang},
  \citenamefont {Chou}, \citenamefont {Conner}, \citenamefont {Peairs},
  \citenamefont {Povey}, \citenamefont {Zhong},\ and\ \citenamefont
  {Cleland}}]{doi:10.1063/5.0035945}%
  \BibitemOpen
  \bibfield  {author} {\bibinfo {author} {\bibfnamefont {J.}~\bibnamefont
  {Grebel}}, \bibinfo {author} {\bibfnamefont {A.}~\bibnamefont {Bienfait}},
  \bibinfo {author} {\bibfnamefont {E.}~\bibnamefont {Dumur}}, \bibinfo
  {author} {\bibfnamefont {H.-S.}\ \bibnamefont {Chang}}, \bibinfo {author}
  {\bibfnamefont {M.-H.}\ \bibnamefont {Chou}}, \bibinfo {author}
  {\bibfnamefont {C.~R.}\ \bibnamefont {Conner}}, \bibinfo {author}
  {\bibfnamefont {G.~A.}\ \bibnamefont {Peairs}}, \bibinfo {author}
  {\bibfnamefont {R.~G.}\ \bibnamefont {Povey}}, \bibinfo {author}
  {\bibfnamefont {Y.~P.}\ \bibnamefont {Zhong}}, \ and\ \bibinfo {author}
  {\bibfnamefont {A.~N.}\ \bibnamefont {Cleland}},\ }\href {\doibase
  10.1063/5.0035945} {\bibfield  {journal} {\bibinfo  {journal} {Appl. Phys.
  Lett.}\ }\textbf {\bibinfo {volume} {118}},\ \bibinfo {pages} {142601}
  (\bibinfo {year} {2021})}\BibitemShut {NoStop}%
\end{thebibliography}%

\ \\\\
\noindent\textbf{\begin{large}Appendix A~~{Derivation of the coherent coupling between the magnon and the mechanical phonon}\end{large}}\label{appenA}
\\\\
Generally, the direct realization of the coherent coupling between the magnon and the mechanical phonon is difficult due to their large difference in the energy scales. We overcome this problem by choosing a proper parameter regime, i.e., $\Delta_{s}=\Delta_{K}$. To clarify this, we first give the system Hamiltonian (\ref{H_mms}) in the interaction picture after setting $\Delta_{s}=\Delta_{K}$,
\begin{equation}\label{H_mms_ap}
\begin{split}
\frac{\hat{H}_{\mathrm{mm}}^{\mathrm{S,I}}}{\hbar}=& g_{r}\hat{b}_{s}\hat{s}^\dagger-g_{c}\hat{b}_{s}^\dagger\hat{s}^\dagger e^{\mathrm{i}2\Delta_{s}t}+g_{r}\hat{b}_{s}^\dagger\hat{s}^\dagger e^{\mathrm{i}2(\omega_p+\Delta_{s})t}\\
&-g_{c}\hat{b}_{s} \hat{s}^\dagger e^{\mathrm{i}2\omega_pt}+\mathrm{H.c.}.
\end{split}
\end{equation}
Then, the contributions of each terms in the above Hamiltonian to the system dynamics can be easily identified by comparing the corresponding detunings. In the regime of $\Delta_{s}=\Delta_{K}$, the system parameters satisfy $2\sinh^2(r_{p})\omega_{p}=\cosh(2r_{p})\omega_{K}-\omega_{x}$, which is the exact condition for resonance. Based on this, we plot in Fig.~\ref{figA1} the ratios of $2|\Delta_{s}|$, $2(\omega_p+\Delta_{s})$ and $2\omega_p$ over $g_{\mathrm{max}}\equiv\max\{g_{r},g_{c}\}$ versus $r_p\in[0,3]$. The results show that we can safely neglect the contributions of the three terms that oscillate rapidly by the rotating-wave approximation (RWA), and keep only the slowly oscillating term $\propto(\hat{b}_{s}\hat{s}^\dagger+\mathrm{H.c.})$, at least for $r_{p}\le2.5$. Furthermore, in the weak parametric drive region of relatively small $r_{p}$, the resonance condition makes $\omega_{p}$ and the consequent detunings extremely large, which may exceed the range of experimentally feasible parameters. Nonetheless, this can be avoided when working with a strong parametric drive that is considered in this work for a large coupling enhancement.
The system Hamiltonian (\ref{H_mms}) is thus reduced to
\begin{equation}\label{H_mmsr_ap}
\begin{split}
\frac{\hat{H}_{\mathrm{mm,r}}^{\mathrm{S}}}{\hbar}\simeq&\Delta_{s} \hat{b}_{s}^{\dagger} \hat{b}_{s}+\Delta_{K} \hat{s}^{\dagger} \hat{s}+g_{r}(\hat{b}_{s}\hat{s}^\dagger+\mathrm{H.c.}),
\end{split}
\end{equation}
A direct demonstration of the validity of Eq.~(\ref{H_mmsr_ap}) is given in Fig.~\ref{fig2}(d), where the effective Hamiltonian (\ref{H_mmsr_ap}) reproduces exactly the system dynamics governed by the total Hamiltonian (\ref{H_mms}) for $r_{p}=2.5$.
\\
\begin{figure}
	\includegraphics[width=1\columnwidth]{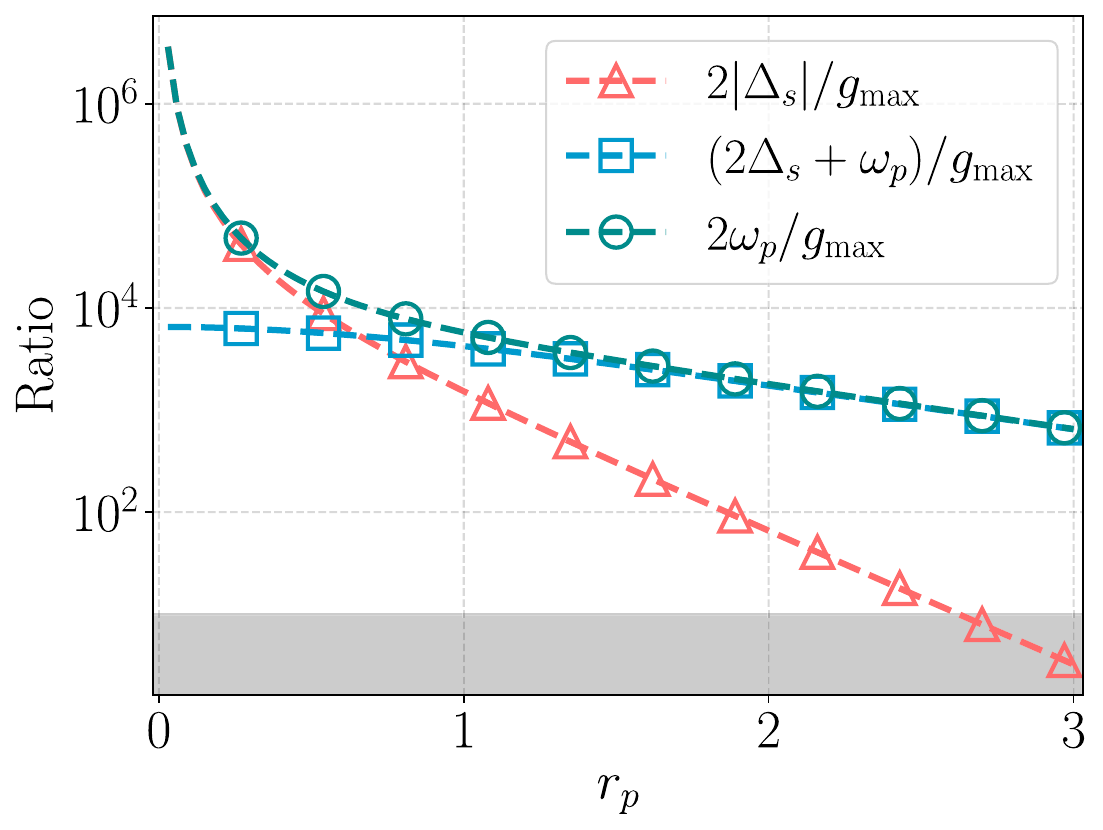}
	\caption{{Frequency detunings over coupling versus $r_p$. The shaded area shows the invalid region for the RWA with ratios less than 10. The parameters are the same as in Fig.~\ref{fig2}(a).}}
	\label{figA1}
\end{figure}
\\\\
\textbf{\begin{large}Appendix B~~{Master equation and squeezing-induced noise}\end{large}}\label{appenB}
\\\\
\noindent \textbf{1~~{Derivation of the master equation}}
\\\\
Considering the dissipation of the mechanical mode to a Markovian bath of thermal occupation $\bar{n}_{x}$ and damping rate $\gamma_{x}$, as well as the magnon mode to a bath of thermal occupation $\bar{n}_{s}$ and damping rate $\gamma_{s}$, the dynamics of the hybrid magnon-mechanical system, in the absence of the MPD, is described by the following master equation
\begin{equation} \label{me1}
\begin{split}
\dot{\hat{\rho}}=&-\frac{\mathrm{i}}{\hbar}[\hat{H}_{\mathrm{mm}}',\hat{\rho}]+\gamma_x\left(\bar{n}_{x}+1\right) L_{\hat{b}}\hat{\rho}+\gamma_x\bar{n}_{x} L_{\hat{b}^{\dagger}}\hat{\rho}\\&+\gamma_s\left(\bar{n}_s+1\right) L_{\hat{s}}\hat{\rho}+\gamma_s\bar{n}_s L_{\hat{s}^{\dagger}}\hat{\rho},
\end{split}
\end{equation}
where $\hat{H}_{\mathrm{mm}}'$ is given by Eq.~(\ref{H_mm}) with $\Omega_{p}=\omega_{p}=0$, $\hat{\rho}$ is the density operator, and $L_{\hat{o}} \hat{\rho}\equiv\hat{o} \hat{\rho} \hat{o}^{\dagger}-[\hat{o}^{\dagger}\hat{o} ,\hat{\rho}]/2$.

After introducing the MPD, the mechanical phonons can be squeezed, resulting in a squeezed-phonon mode $\hat{b}_{s}=\hat{b}\cosh(r_{p})+\hat{b}^\dagger\sinh(r_{p})$. In terms of the mode $\hat{b}_{s}$, the master equation (\ref{me1}) can be rewritten as
\begin{equation} \label{me2}
\begin{split}
\dot{\hat{\rho}}=&-\frac{\mathrm{i}}{\hbar}[\hat{H}_{\mathrm{mm}}^{\mathrm{S}},\hat{\rho}]+\gamma_x\left(\bar{n}_{x}^{s}+1\right) L_{\hat{b}_s}\hat{\rho}+\gamma_x\bar{n}_{x}^{s} L_{\hat{b}_s^{\dagger}}\hat{\rho}\\&-\gamma_xM(2\bar{n}_{x}+1)\mathscr{L}_{\hat{b}_s}\hat{\rho}-\gamma_xM(2\bar{n}_{x}+1)\mathscr{L}_{\hat{b}_s^{\dagger}}\hat{\rho}
\\&+\gamma_s\left(\bar{n}_s+1\right) L_{\hat{s}}\hat{\rho}+\gamma_s\bar{n}_s L_{\hat{s}^{\dagger}}\hat{\rho},
\end{split}
\end{equation}
where $\hat{H}_{\mathrm{mm}}^{\mathrm{S}}$ is given by Eq.~(\ref{H_mms}), $\bar{n}_{x}^{s}=\bar{n}_{x}\cosh(2r_{p})+\sinh^{2}(r_{p})$, $M=\cosh(r_{p})\sinh(r_{p})$, and
$\mathscr{L}_{\hat{o}} \hat{\rho}\equiv\hat{o} \hat{\rho} \hat{o}-[\hat{o}\hat{o} ,\hat{\rho}]/2$. Equation (\ref{me2}) shows that when the mechanical mode $\hat{b}$ is coupled to a Markovian bath
of thermal occupancy $\bar{n}_{x}$, the $\hat{b}_{s}$ mode is driven by a squeezed reservoir of thermal population $\bar{n}_{x}^{s}$.

In the limit of a weak mechanical dissipation, i.e., $\gamma_{x}\sinh(2r_{p})\ll\Delta_{s}$, the two terms (associated with $\mathcal{L}_{\hat{b}_s}$ and $\mathcal{L}_{\hat{b}_s^{\dagger}}$) describing the two-phonon correlation in Eq.~(\ref{me2}) are high-frequency oscillating terms and can thus be neglected. Therefore, the master equation (\ref{me2}) is reduced to 
\begin{equation} \label{me3}
\begin{split}
\dot{\hat{\rho}}=&-\frac{\mathrm{i}}{\hbar}[\hat{H}_{\mathrm{mm}}^{\mathrm{S}},\hat{\rho}]+\gamma_x\left(\bar{n}_{x}^{s}+1\right) L_{\hat{b}_s}\hat{\rho}+\gamma_x\bar{n}_{x}^{s} L_{\hat{b}_s^{\dagger}}\hat{\rho}
\\&+\gamma_s\left(\bar{n}_s+1\right) L_{\hat{s}}\hat{\rho}+\gamma_s\bar{n}_s L_{\hat{s}^{\dagger}}\hat{\rho},
\end{split}
\end{equation}
which is exactly the master equation (\ref{mas equ}) in the main text. From Eq.~(\ref{me3}), we find that the squeezed-phonon mode $\hat{b}_{s}$ can be considered as being coupled, with an unchanged vacuum damping rate $\gamma_{x}$, to a thermal bath but with an amplified thermal occupation $\bar{n}_{x}^{s}$. Apparently, the thermal noise with thermal occupation $\bar{n}_{x}^{s}$ increases drastically with increasing the MPD. In general, this amplified thermal noise could corrupt the coherent dynamics of the system even with an exponentially enhanced coupling, thus limiting the application of the system in the quantum regime. However, it can hardly affect the coherent state transfer in our system at least for $Q\ge10^5$, as the amplified thermal noise remains far less than the enhanced coupling strength even for a strong MPD (e.g., $r_{p}=2.5$). 
We note that although the intrinsic mechanical damping can be extremely low in state-of-the-art experiments, e.g., $Q_{x}\geq10^8$ in nanomechanical resonators \cite{PhysRevLett.116.147202,tsaturyan2017ultracoherent,ghadimi2018elastic}, our scheme presents a clear advantage in relaxing the high-$Q$ requirement of the mechanical resonator in realistic experiments. 
\\\\
\noindent \textbf{2~~{Elimination of the squeezing-induced noise}}
\\\\
To eliminate the amplified noise from the mechanical bath, a possible strategy is to use the squeezed-vacuum-reservoir technique, which has been widely used in previous schemes \cite{PhysRevLett.114.093602,PhysRevA.100.062501,PhysRevLett.120.093601,PhysRevLett.120.093602,PhysRevLett.126.023602,PhysRevLett.128.083604,PhysRevLett.125.153602,PhysRevApplied.17.024009}. The technique was first proposed in an optomechanical system to eliminate the noise of the squeezed cavity mode \cite{PhysRevLett.114.093602}. It employs an auxiliary, broadband squeezed-vacuum field to drive the cavity, which is phase matched with the parametric drive that squeezes the cavity mode. This ensures that the squeezed cavity mode is equivalently coupled to a vacuum bath, thereby allowing one to describe the system dynamics with a simplified master equation in the standard Lindblad form.

To demonstrate more explicitly how to eliminate the amplified mechanical noise in our scheme using the above auxiliary-reservoir technique, we now derive the master equation for the hybrid magnon-mechanical system. We first assume that the mechanical mode is coupled to a squeezed thermal reservoir with a squeezing parameter $r_{e}$ and a reference phase $\theta_{e}$. The dynamics of the system, in the absence of the MPD, is described by the master equation
\begin{equation} \label{me4}
\begin{split}
\dot{\hat{\rho}}=&-\frac{\mathrm{i}}{\hbar}[\hat{H}_{\mathrm{mm}}',\hat{\rho}]+\gamma_x\left(\mathcal{N}+1\right) L_{\hat{b}}\hat{\rho}+\gamma_x\mathcal{N} L_{\hat{b}^{\dagger}}\hat{\rho}\\&-\gamma_x\mathcal{M}\mathscr{L}_{\hat{b}}\hat{\rho}-\gamma_x\mathcal{M}^{\ast}\mathscr{L}_{\hat{b}^{\dagger}}\hat{\rho}
\\&+\gamma_s\left(\bar{n}_s+1\right) L_{\hat{s}}\hat{\rho}+\gamma_s\bar{n}_s L_{\hat{s}^{\dagger}}\hat{\rho},
\end{split}
\end{equation}
where $\mathcal{N}$ and $\mathcal{M}$ describe the thermal noise and the two-phonon correlations caused by the squeezed reservoir, which are given, respectively, by 
\begin{equation}
\begin{aligned}
\mathcal{N} & =\bar{n}_{x} \cosh \left(2 r_e\right)+\sinh ^2\left(r_e\right), \\
\mathcal{M} & =\left(2 \bar{n}_{x}+1\right) \cosh \left(r_e\right) \sinh \left(r_e\right) e^{\mathrm{i} \theta_e}.
\end{aligned}
\end{equation}
When squeezed, the master equation (\ref{me4}) can be rewritten, in terms of the mode $\hat{b}_{s}$, as
\begin{equation} \label{me5}
\begin{split}
\dot{\hat{\rho}}=&-\frac{\mathrm{i}}{\hbar}[\hat{H}_{\mathrm{mm}}^{\mathrm{S}},\hat{\rho}]+\gamma_x\left(\mathcal{N}_{s}+1\right) L_{\hat{b}_{s}}\hat{\rho}+\gamma_x\mathcal{N}_{s} L_{\hat{b}_{s}^{\dagger}}\hat{\rho}\\&-\gamma_x\mathcal{M}_{s}\mathscr{L}_{\hat{b}_{s}}\hat{\rho}-\gamma_x\mathcal{M}_{s}^{\ast}\mathscr{L}_{\hat{b}^{\dagger}_{s}}\hat{\rho}
\\&+\gamma_s\left(\bar{n}_s+1\right) L_{\hat{s}}\hat{\rho}+\gamma_s\bar{n}_s L_{\hat{s}^{\dagger}}\hat{\rho},
\end{split}
\end{equation}
where $\mathcal{N}_{s}$ and $\mathcal{M}_{s}$ are given, respectively, by
\begin{equation} \label{NsMs}
\begin{aligned}
\mathcal{N}_{s}=\! & \left[\bar{n}_{x} \cosh \left(2 r_e\right)\!+\!\sinh ^2\left(r_e\right)\right] \cosh \left(2 r_p\right)\!+\!\sinh ^2\left(r_p\right)\\
&+(\bar{n}_{x}\!+\!\frac{1}{2}) \sinh \left(2 r_e\right) \sinh \left(2 r_p\right) \cos \left(\theta_e\right),\\
\mathcal{M}_{s}=\! &-\left(2 \bar{n}_{x}\!+\!1\right)\big\{\frac{1}{2}\sinh\left(2 r_p\right) \cosh\left(2r_e\right)\!+\!\frac{1}{2} \sinh \left(2 r_e\right)\\
&\times\!\left[\exp \left(\mathrm{i}\theta_e\right) \cosh ^2\left(r_p\right)\!+\!\exp\left(-\mathrm{i}\theta_e\right) \sinh ^2\left(r_p\right)\right]\big\},
\end{aligned}
\end{equation}
corresponding to the effective thermal noise and two-phonon correlations of the squeezed phonon mode. These two terms describe the undesired noise induced by the phonon squeezing, which can be eliminated by setting proper parameters of the squeezed reservoir. 
When choosing $r_{e}=r_{p}$ and $\theta_{e}=\pm n\pi$ ($n=1,3,5,\dots$), $\mathcal{N}_{s}$ and $\mathcal{M}_{s}$ can be reduced to $\bar{n}_{x}$ and 0, respectively. The master equation (\ref{me5}) is thus simplified to 
\begin{equation} \label{me6}
\begin{split}
\dot{\hat{\rho}}=&-\frac{\mathrm{i}}{\hbar}[\hat{H}_{\mathrm{mm}}^{\mathrm{S}},\hat{\rho}]+\gamma_x\left(\bar{n}_{x}+1\right) L_{\hat{b}_{s}}\hat{\rho}+\gamma_x\bar{n}_{x} L_{\hat{b}_{s}^{\dagger}}\hat{\rho}
\\&+\gamma_s\left(\bar{n}_s+1\right) L_{\hat{s}}\hat{\rho}+\gamma_s\bar{n}_s L_{\hat{s}^{\dagger}}\hat{\rho}.
\end{split}
\end{equation}
Apparently, the squeezing-induced noises are completely eliminated, and the mode $\hat{b}_{s}$ is equivalently coupled to a thermal bath with the original thermal occupation $\bar{n}_{x}$. 

\begin{table*}
	\centering
	\footnotesize
	\caption{System parameters. Unless otherwise specified, these parameters are applied to the numerical calculations in the main text.}
	\label{tab1}
	\tabcolsep 40pt 
	\begin{tabular*}{\textwidth}{ccc}
		\toprule\\
		\hline
		Symbol&Name&Value\\\hline
		$\omega_{x}$&Mechanical resonance frequency&$2\pi\times3.8$ MHz\\
		$Q_{x}$&Mechanical quality factor&$10^{8}$\\
		$\gamma_{x}$&Mechanical linewidth&$2\pi\times0.038$ Hz\\
		$\bar{n}_{x}$&Thermal phonon occupation&55\\
		$(\ell,w,t)$&Dimensions of micromechanical cantilever&(4,0.1,0.02)$\mu$m\\
		$M$&Effective mass of cantilever&$7.0\times10^{-18}$ kg\\
		$x_{\mathrm{zpf}}$&Zero-point fluctuation amplitude&$2.69\times10^{-13}$ m\\
		$\omega_{p}$&Mechanical parametric drive frequency&$2\pi\times2.382$ GHz\\
		$\Omega_{p}$&Mechanical parametric drive amplitude&$2\pi\times2.378$ GHz\\
		$\omega_{K}$&Kittel magnon frequency&$2\pi\times2.35$ GHz\\
		$\gamma_{s}$&Magnon linewidth&$2\pi\times1.0$ MHz\\
		$\bar{n}_{s}$&Thermal magnon occupation&$1.3\times10^{-5}$\\
		$T$&Mechanical and magnon bath temperatures&10 mK\\
		$R$&Radius of YIG nanosphere&100 nm\\
		$V$&Volume of YIG nanosphere&$4.2\times10^{-21}$ m$^{3}$\\
		$M_{m}$&Effective mass of YIG nanosphere&$2.17\times10^{-18}$ kg\\
		$M_{t}$&Effective mass of Dy tip&$2.24\times10^{-18}$ kg\\
		$\omega_{\mathrm{NV}}$&NV spin frequency&$2\pi\times2.35$ GHz\\
		$\gamma_{z}$&Single NV spin dephasing rate&$2\pi\times1.0$ kHz\\
		$b_{0}$&Gradient of external magnetic field &$10^{7}$ T/m\\
		$b_{1}$&Magnetic field gradient produced by Dy tip&$4.5\times10^{7}$ T/m\\
		$B_{z}$&Homogeneous component of external magnetic field &0.084 T\\
		$g(\lambda)$&Magnon(Spin)-mechanical coupling rate&$2\pi\times0.69$ MHz\\
		\hline	
		\bottomrule
	\end{tabular*}
\end{table*}

Regarding the experimental implementation of the above approach, a possible arrangement is to integrate a microwave optomechanical system into the present hybrid setup, as discussed detailedly in Ref.~\cite{PhysRevApplied.17.024009}. A microwave resonator, which is electrically coupled to the mechanical resonator, is terminated by a
superconducting quantum interference device (SQUID) that provides squeezed microwave inputs to the mechanical mode.
The microwave drive with a sufficiently large squeezing bandwidth (experimental values of about gigahertz have been reported \cite{PhysRevX.10.021021,doi:10.1063/5.0035945}) can then be regarded as an effective squeezed reservoir of the mechanical mode. The squeezing parameter and the reference phase
of the auxiliary reservoir could be controlled by the amplitude and phase of the pump tone used to modulate the magnetic flux through the SQUID.	
\\\\
\textbf{\begin{large}Appendix C~~{System parameters}\end{large}}\label{appenC}\\
Table \ref{tab1} shows the symbols, names and values of the main system parameters that are used in our numerical calculations.
\\\\
\noindent \textbf{\begin{large}Appendix D~~{Derivation of the effective coupling between the magnon and the NV spin}\end{large}}\label{appenD}
\\\\
To begin with, we transform the Hamiltonian (\ref{H_mmnvs}) in the main text into the interaction picture, giving
	\begin{equation}\label{H_mmnvsI}
	\begin{split}
	\frac{\hat{H}_{\mathrm{Hybrid}}^{\mathrm{S,I}}}{\hbar}=&
	g_{r}\hat{b}_{s}\hat{s}^\dagger e^{\mathrm{-i}(\Delta_{s}-\Delta_{K})t}\!+\!
	g_{r}\hat{b}_{s}^\dagger\hat{s}^\dagger e^{\mathrm{i}(\Delta_{s}+\Delta_{K}+2\omega_p)t}\\&\!-
	\!g_{c}\hat{b}_{s}^\dagger\hat{s}^\dagger e^{\mathrm{i}(\Delta_{s}+\Delta_{K})t}\!-\!
	g_{c}\hat{b}_{s}\hat{s}^\dagger e^{\mathrm{-i}(\Delta_{s}-\Delta_{K}-2\omega_p)t}\\&\!+\!
	\lambda_{r}\hat{b}_{s}\hat{\sigma}_{+}e^{\mathrm{-i}(\Delta_{s}-\Delta_{\mathrm{NV}})t}\!+\!
	\lambda_{r}\hat{b}_{s}^\dagger\hat{\sigma}_{+}e^{\mathrm{-i}(\Delta_{s}+\Delta_{\mathrm{NV}}+2\omega_p)t}\\&\!-\!
	\lambda_{c}\hat{b}_{s}^\dagger\hat{\sigma}_{+}e^{\mathrm{i}(\Delta_{s}+\Delta_{\mathrm{NV}})t}\!-\!
	\lambda_{c}\hat{b}_{s}\hat{\sigma}_{+}e^{\mathrm{-i}(\Delta_{s}-\Delta_{\mathrm{NV}}-2\omega_p)t}\\&+\mathrm{H.c.},
	\end{split}
	\end{equation}
	In order to eliminate the undesired terms that oscillate rapidly in the Hamiltonian above and simultaneously eliminate the mechanical degree of freedom, we assume the following large detuning conditions 
	\begin{equation}\label{large detunings}
	\begin{split}
	\min\big\lbrace &\left|\Delta_{s}+\Delta_{K}\right|,\left|\Delta_{s}+\Delta_{K}+2\omega_p\right|,\\&\left|\Delta_{s}-\Delta_{K}-2\omega_p\right|\big\rbrace  \gg \left|\Delta_{s}-\Delta_{K}\right| \gg g_{r},\\
	\min\big\lbrace &\left|\Delta_{s}+\Delta_{\mathrm{NV}}\right|,\left|\Delta_{s}+\Delta_{\mathrm{NV}}+2\omega_p\right|,\\&\left|\Delta_{s}-\Delta_{\mathrm{NV}}-2\omega_p\right|\big\rbrace  \gg \left|\Delta_{s}-\Delta_{\mathrm{NV}}\right| \gg \lambda_{r}.
	\end{split}
	\end{equation}
	In this case, the mode $\hat{b}_{s}$ is only virtually excited and can then be eliminated adiabatically. Moreover, we assume that the phonon mode is initially in the vacuum state, and can always remain in the vacuum due to the effective decoupling from the system dynamics. Given above, we obtain the effective Hamiltonian involving only the magnon and spin degrees of freedom
	\begin{equation}\label{H_ms}
	\begin{split}
	\frac{\hat{H}_{\mathrm{ms}}^{\mathrm{S}}}{\hbar}=&\frac{g_{r}^2}{\Delta_{K}-\Delta_{s}}\hat{s}^\dagger\hat{s}+\frac{\lambda_{r}^2}{\Delta_{\mathrm{NV}}-\Delta_{s}}\hat{\sigma}_{+}\hat{\sigma}_{-}\\
	&+\frac{g_{r}\lambda_{r}}{2}(\frac{1}{\Delta_{K}-\Delta_{s}}+\frac{1}{\Delta_{\mathrm{NV}}-\Delta_{s}})(\hat{s}^\dagger\hat{\sigma}_{-}+ \hat{s}\hat{\sigma}_{+}).
	\end{split}
	\end{equation}
	
	We recall the realistic parameters considered in the main text, $\omega_{x}\approx2\pi\times3.8$ MHz, $\omega_{K}=\omega_{\mathrm{NV}}\approx2\pi\times0.96$ GHz, and $g=\lambda\approx2\pi\times0.69$ MHz. Then, the Hamiltonian above can be further reduced, yielding the Hamiltonian (\ref{H_msr}) in the main text. Moreover, considering the MPD with, e.g., $r_{p}=1.54$ and $\Omega_{p}\approx2\pi\times1.05$ GHz, we have $\Delta_{s}\approx-55g_{r}$ and $\Delta_{K}\approx-45g_{r}$, which validates the large detuning conditions (\ref{large detunings}) and therefore the effective Hamiltonian (\ref{H_ms}).
	\maketitle

\end{document}